\documentclass[useAMS,usenatbib]{mnras}

\usepackage{graphicx,color}
\usepackage{aas_macros}
\usepackage{amssymb}

\usepackage{longtable,lscape}
\title[Multi-transition VY CMa SiO maser polarimetry]{Simultaneous VLBA polarimetric 
observations of the v=$\{$1,2$\}$ J=1-0 and v=1, J=2-1 SiO maser emission toward VY CMa II: component-level polarization 
analysis.}

\author[L. Richter et al.]
{
L.~Richter,$^{1,3}$
A.~Kemball,$^{2,3}$\thanks{Visiting Professor affiliation} and
J.~Jonas$^3$
\\
  $^1$SKA South Africa, 3rd Floor, The Park, Park Road, Cape Town, 7405, South Africa\\
  $^2$Department of Astronomy and National Center for Supercomputing Applications, University of Illinois 
  at\\ Urbana-Champaign, 1002 W. Green Street, Urbana, IL, 61801, USA\\
  $^3$Department of Physics and Electronics, Rhodes University, Drostdy Road, Grahamstown, 6139, South Africa\\
}

\date{Released 2016 Xxxxx XX}

\pagerange{\pageref{firstpage}--\pageref{lastpage}} \pubyear{2016}

\def\LaTeX{L\kern-.36em\raise.3ex\hbox{a}\kern-.15em
    T\kern-.1667em\lower.7ex\hbox{E}\kern-.125emX}

\begin{document}

\label{firstpage}

\maketitle


\begin{abstract}

This paper presents a component-level comparison of the polarized v=1 J =1-0, v=2 J=1-0 and v=1 J=2-1 SiO maser emission 
towards the supergiant star VY CMa at milliarcsecond-scale, as observed using the VLBA at $\lambda=7$mm and 
$\lambda=3$mm. An earlier paper considered overall maser morphology and constraints on SiO maser excitation and pumping 
derived from these data. The goal of the current paper is to use the measured polarization properties of individual 
co-spatial components detected across multiple transitions to provide constraints on several competing theories for the 
transport of polarized maser emission. This approach minimizes the significant effects of spatial blending. We present 
several diagnostic tests designed to distinguish key features of competing theoretical models for maser polarization. 
The number of coincident features is limited by sensitivity however, particularly in the v=1 J=2-1 transition at 86 GHz, 
and deeper observations are needed. Preliminary conclusions based on the current data provide some support for: i) 
spin-independent solutions for linear polarization; ii) the influence of geometry on the distribution of fractional linear 
polarization with intensity; and, iii) $\pi/2$ rotations in linear polarization position angle arising from transitions 
across the Van Vleck angle ($\sin^2{\theta}=2/3$) between the maser line-of-sight and magnetic field. There is weaker
evidence for several enumerated non-Zeeman explanations for circular polarization. The expected 2:1 ratio in circular 
polarization between J=1-0 and J=2-1 predicted by standard Zeeman theory cannot unfortunately be tested conclusively due 
to insufficient coincident components.

\end{abstract}




\section{Introduction}

The nature and role of magnetic fields is an important open question in asymptotic giant branch (AGB), supergiant, and other 
late stages of stellar evolution. In late-type evolved stars, magnetic fields have been invoked to explain asymmetric mass 
loss \citep{Garcia-Segura:05}, localised features such as arcs observed in the circumstellar material \citep{Soker:99a,Soker:00}, 
and the origin of circumstellar disks \citep{Matt:00}. Magnetic fields may also play a role in the presupernova collapse of 
massive stars \citep{Heger:03}.

The primary means to detect magnetic fields towards late-type evolved stars is through linear and circular polarisation 
observations of several distinct radiation emission mechanisms. Several factors favour longer-wavelength observations in such 
studies. These include: i) the high degree of visible obscuration for late-type evolved stars with high mass-loss rates; 
ii) the technical and sensitivity limitations of optical polarimetry for weakly-magnetic stars \citep{Donati:97}; and iii) the 
abundance of dust and molecular species in the circumstellar environment (CSE) of these objects with associated emission in the 
sub-millimeter or millimeter regime.

Magnetic fields have been detected towards planetary nebulae and protoplanetary nebulae by interferometric imaging of 
linearly-polarised continuum sub-millimetre emission, which traces dust alignment due to the magnetic field 
\citep{Greaves:02,Sabin:07}. The linear polarisation of molecular line emission can trace the magnetic field through the 
Goldreich-Kylafis effect \citep{Goldreich:81}, which has been observed in 620.701~GHz $5_{32} - 4_{41}$ H$_2$O maser emission 
towards VY~CMa \citep{Harwit:10} and detected in thermal CO (J=2-1) and v=0, J-5-4 SiO emission toward IK Tau by 
\citet{Vlemmings:12}.
Optical spectropolarimetry has been used to detect magnetic fields of magnitude $\sim100$~G in an active giant \citep{Auriere:08}, 
of order a few Gauss in several rapidly rotating giants \citep{Konstantinova-Antova:09}, $\sim1$~G in the supergiant Betelgeuse 
\citep{Auriere:10} and $\lesssim1$~G in eight massive late-type supergiants \citep{Grunhut:10}.
These detections used a least-squares deconvolution technique to find composite circular polarisation profiles 
from hundreds of observed optical spectral lines in order to mitigate sensitivity limitations noted above \citep{Donati:97}.
The first detection of a photospheric magnetic field toward a Mira variable, the S-type star $\chi$Cyg, was recently 
reported using this technique by \citet{Lebre:14}, who measured a longitudinal magnetic field component of 2-3 G for this star.
\citet{Sabin:15}, using similar techniques, report longitudinal magnetic fields of 0.6~G and 10.2~G for the post-AGB stars 
R~Scuti and U~Monocerotis respectively.

Towards the end of a star's lifetime, its mass loss increases and it can shed a considerable fraction of its mass through the 
stellar wind. This matter forms a dusty circumstellar envelope, which obscures the star at optical and infrared wavelengths 
\citep{Iben:83}. Maser emission becomes a particularly important observational tool during these final stages of stellar 
evolution, as it is visible within dusty circumstellar envelopes \citep{Cohen:89}.
OH, H$_2$O and SiO masers have all been used to measure magnetic fields in the circumstellar envelopes of late-type stars, 
where they sample the magnetic field over a range of distances from the star \citep{Reid:81b}.
OH maser observations of Zeeman patterns have been used to derive magnetic field estimates for evolved stars 
\citep[e.g.][]{Reid:79,Chapman:86,Szymczak:98,Etoka:04} and protoplanetary nebulae 
\citep[e.g.][]{Bains:03,Bains:04,Szymczak:04,Gomez:09}. Circular polarisation observations of H$_2$O masers have also been 
used to derive mangetic field estimates in the circumstellar envelopes of late-type evolved stars 
\citep[e.g][]{Vlemmings:02a,Vlemmings:05,Richards:04,Leal-Ferreira:13}, in protoplanetary nebulae \citep{Vlemmings:08}, and 
in the so-called water-fountain jet sources \citep{Vlemmings:06d}. Similarly, circular polarisation studies of SiO masers 
have also been used to derive magnetic field estimates in the circumstellar envelopes of evolved stars 
\citep[e.g.][]{Barvainis:87,Kemball:97,Amiri:12}.

The high brightness temperature of maser emission makes it detectable at compact spatial scales with Very Long Baseline 
Interferometry (VLBI). SiO maser emission, in particular, can be imaged at milliarcsecond angular resolution in the inner 
circumstellar envelopes of late-type evolved stars \citep[e.g.][]{Diamond:94,Diamond:03}, making it a promising probe of 
the magnetic field at a distance of only a few stellar radii from the surfaces of these stars. However, in order for magnetic 
field information to be derived from the SiO maser polarisation observations, a maser polarisation radiative transfer model is 
required. 


\subsection{SiO maser polarisation theory}

SiO is a non-paramagnetic molecule \citep{bk:Elit} and invariably falls into the weak-splitting Zeeman regime, where 
fully-separated Zeeman patterns are not observed \citep{Gray:12}.

The theory of polarized maser emission in this regime has been investigated in series of papers by Elitzur 
\cite[e.g.][and references therein]{Elitzur:02}, and by Watson and collaborators \cite[e.g.][and references therein]{Watson:02}.
Both of these works build on the foundational maser polarisation theoretical model developed by \citet[][hereafter GKK]{GKK}.

In the Elitzur model, circular polarisation of the maser emission is considered in the small-splitting regime primarily 
arising from the standard Zeeman effect, subject to the criteria that the maser polarisation has reached its stationary state and 
there is no significant magnetic field line curvature along the maser path \citep{Elitzur:02}; this model can be used to derive 
directly the magnetic field strength in the envelope from a measurement of circular polarization.
In the Watson et al. models, multiple causes of circular polarisation are considered, including: i) the standard Zeeman effect 
alone; ii) the Zeeman effect with modifications due to saturation; and, iii) a change in quantisation axis as the maser propagates 
leading to the inter-conversion of linear to circular polarisation, termed non-Zeeman circular polarisation \citep{Watson:02}.
The two bodies of theoretical work differ primarily in their foundational assumptions regarding the rate of establishment and stable 
propagation of stationary polarization modes; accordingly their predictions of maser polarization as a function of saturation frequently 
differ.

\citet{Gray:03} compared the Elitzur and Watson models to the multi-level maser model by \citet{Gray:95c}. He found that the latter 
model level population equations could be reduced to those used by \citet{Watson:94} thus anticipating similar model predictions. The 
theory of maser polarization is described in further detail in the monograph by \citet{Gray:12}. 
We note that \citet{Asensio_Ramos:05} have considered the effect of radiation anisotropy and the Hanle effect on circumstellar SiO maser 
polarization. In addition, a recent paper by \citet{Houde:14} considers anisotropic resonant scattering. In the current paper we focus
primarily on testing current data against the core theories of Watson and Elitzur; future work will consider observational tests against 
more recent theoretical developments in the area of anisotropy in further detail.

Circumstellar SiO masers are strongly linearly polarised \citep[e.g.][]{Troland:79, Clark:82, Kemball:97} and probably at least partially 
saturated \citep{Nedoluha:94}. In this parameter regime the observed circular polarisation could be created by either Zeeman or non-Zeeman 
effects. Furthermore, anisotropic pumping is a consideration for circumstellar SiO maser observations due to their proximity to the central 
star \citep{Nedoluha:90a}.

When observations of SiO maser circular polarisation are interpreted as due to the standard Zeeman effect alone, they imply 
circumstellar magnetic fields in the range of a few Gauss up to a few tens of Gauss in some cases \citep{Barvainis:87,Kemball:97}.
These magnetic field magnitudes imply a magnetic energy density of more than $\sim10^{-0.5}$~dyne.cm$^{-2}$ \citep{Reid:07}. In contrast, 
the thermal pressure is $\sim10^{-2.7}$dyne.cm$^{-2}$ and the ram pressure is $\sim10^{-2.5}$dyne.cm$^{-2}$ in the SiO maser region of 
a typical AGB star circumstellar envelope \citep{Reid:07}. The magnetic energy density derived from the standard Zeeman interpretation 
is thus much greater than the thermal and ram energy densities. If this is the case, then the magnetic field may play a dominant role in 
the mass loss from the star and shaping of the envelope if globally organised \citep[e.g.][]{Matt:00,Garcia-Segura:05}.
Alternatively, the SiO masers may be preferentially sampling strong localised magnetic fields with low global filling factors, 
caused by amplification of the tangential magnetic field due to shock compression around pulsating stars \citep{Hartquist:97,Kemball:09}.

Under a non-Zeeman interpretation, the observed levels of circular polarisation could be created in the presence of magnetic fields of 
only a few tens of milliGauss \citep{Nedoluha:94,Wiebe:98}. For a magnetic field of this order the magnetic energy density is less than 
the thermal and kinetic energy densities in the SiO maser region. These uncertainties underline the scientific importance of placing 
closer observational constraints on the theoretical interpretation of SiO maser polarization data.


\subsection{Overview}

In this paper we systematically evaluate a number of observational tests designed to differentiate between the maser polarisation 
models discussed above, in order to improve the robustness of inferred magnetic field measurements from SiO maser emission 
observations. These tests entail component-level comparison of SiO maser features from the 43~GHz J=1-0 and 86~GHz J=2-1 
transitions at milliarcsecond (mas) resolution, in order to minimize the effects of spatial blending. The supergiant star VY~CMa 
was chosen as the target of this investigation, due to its high SiO maser luminosity over a wide range of SiO transitions 
\citep{Cernicharo:93}. The v=1 J=1-0, v=2 J=1-0 and v=1 J=1-0 SiO maser emission toward VY CMa was observed in full polarisation 
in this study, using the Very Long Baseline Array (VLBA\footnote{The VLBA is operated by the National Radio Astronomy Observatory 
(NRAO). The NRAO is a facility of the National Science Foundation operated under cooperative agreement by Associated Universities, 
Inc.}\textsuperscript{,}\footnote{science.nrao.edu/facilities/vlba/docs/manuals/oss}). The first paper based on these data primarily 
addressed the overall maser morphology and implications for pumping and excitation \citep[][hereinafter Paper I]{Richter:13}. The 
current paper concerns a detailed component-level analysis of the data and resulting observational constraints on theories of maser 
polarization propagation. The current component-level analysis is confined to the epoch 2 data in Paper I, which have a significantly 
higher signal-to-noise ratio (SNR) than the earlier epoch 1 data.

The VLBA observations and their reduction are described in Section~\ref{sec:Observations} and the results are presented in 
Section~\ref{sec:Results}. Six observational tests of the polarisation models are proposed in Section~\ref{sec:Discussion}, followed 
by their evaluation against the SiO data presented here. The conclusions are summarised in Section~\ref{sec:Conclusions}.


\section{Observations and Data Reduction}
\label{sec:Observations}

Full-polarisation VLBA observations were performed on 
15 and 19 March 2007, under project code BR123. The transitions $^{28}$SiO v=$\{$0,1,2$\}$ J=1-0 and \mbox{J=2-1}, $^{29}$SiO v=1 
\mbox{J=1-0} and $^{30}$SiO v=0 \mbox{J=1-0} were observed. We consider here only the $^{28}$SiO v=$\{$1,2$\}$ J=1-0 and v=1 J=2-1 
data, which were observed at adopted rest frequencies of 43122.03 GHz, 42820.48 GHz and 86243.37~GHz respectively \citep{Muller:05}. 
For each transition the spectral windows were centred in frequency assuming a systemic LSR (Local Standard of Rest) velocity of 
+18 km.s$^{-1}$ for the target source VY~CMa. 

These obsevations are described in detail in Paper I, which presents the total intensity and linear polarisation maps 
for each transition. The current paper presents the component-level linear and circular polarisation properties of the 
emission. A brief observational summary is provided below followed by a discussion of polarization calibration relevant 
to the current paper.
 
The target source VY~CMa was observed in conjunction with continuum extragalactic sources 3C454.3, J0423-0120, J0609-1542 
and 3C273, which were used as bandpass and continuum phase calibrators for all frequency bands. The data were sampled 
using two-bit quantisation, and correlated in full cross-polarisation over 128 frequency channels per spectral window.
The 43 GHz lines were recorded in \mbox{8-MHz} spectral windows and those in the 86~GHz band in \mbox{16-MHz} spectral 
windows. The nominal velocity channel width in all spectral windows is therefore $\approx0.43$~km.s$^{-1}$. 

The data were reduced following methods outlined in \citet{Kemball:11}; these methods extend the calibration techniques 
described in \citet{Kemball:95} and \citet{Kemball:97} to allow high-accuracy Stokes~$V$ measurement for millimeter-wavelength 
VLBI spectral-line observations. Data reduction refinements presented in \citet{Kemball:11}, and utilised here, include:
\begin{itemize}
   \item Solving for and correcting the bandpass phase response offset between RCP (right circular polarisation) and LCP 
         (left circular polarisation) receptors at the reference antenna.
   \item Applying an aliasing correction to the autocorrelation bandpass amplitude response, 
         before re-use as part of the cross-correlation complex bandpass correction.
   \item The use of autocorrelation polarisation self-calibration;
         a coupled iterative solution for instrumental polarization and amplitude calibration using autocorrelation 
         template spectral fitting.
   \item The use of a multi-antenna composite template spectrum during the latter amplitude calibration of the RCP and 
         LCP receptor systems.
   \item The use of a global fit to continuum calibrator data to determine the differential R/L ampliude gain offset 
         between the RCP and LCP receptor systems.
\end{itemize}


\begin{table}
\begin{tabular}{|lcccc|}
\hline 
SiO & Clean beam & N$_{ant}$ & I peak  & $\sigma_I$\\
transition & [$\mu$as] & & [Jy/beam] & [Jy/beam]\\
\hline
\hline
v=1 J=1-0 & $460\times150$ & 10$^a$ & 22.15 & 0.089 \\
v=2 J=1-0 & $430\times140$ & 9$^b$ & 12.16 & 0.134 \\ 
v=1 J=2-1 & $420\times90$ & 8$^c$ & 46.96 & 0.200 \\ 
\hline
\end{tabular}
\caption{Summary of the VLBA observations. 
For each SiO maser transition imaged, this table lists the major and minor angular dimensions of the 
CLEAN restoring beam, the number, N$_{ant}$, and configuration of VLBA antennas, the peak Stokes I brightness
in the image cube, 
and the broadened thermal noise estimate $\sigma_I$ (as defined in Paper I). 
\newline
$^a$VLBA antennas: BR, HN, KP, LA, MK, NL, OV, PT, SC, and a single VLA antenna;
$^b$VLBA antennas: BR, HN, KP, LA, MK, NL, OV, PT, SC;
$^c$VLBA antennas: BR, FD, KP, LA, MK, NL, OV, PT.
(VLBA antenna abbreviations given in science.nrao.edu/facilities/vlba/docs/manuals/oss).
}
\label{table:observation_summary}
\end{table}


The data reduction was performed using a customised version of the Astronomical Image Processing System 
(\textsc{AIPS}\footnote{\textsc{AIPS} is developed and maintained by the NRAO (http://www.aips.nrao.edu)}). 
The observations are summarised in Table~\ref{table:observation_summary}. 
For each SiO maser transition the table lists the CLEAN restoring beam major- and minor-axis angular dimensions, the 
antenna configuration, the peak Stokes I brightness (Jy/beam) in the resultant image cube, and the highest broadened 
thermal noise estimate $\sigma_I$ (Jy/beam) across all frequency channels of the imaged cube.
The noise from all four Stokes parameters were broadened as described in Paper I, to account for un-modeled residual 
calibration and deconvolution errors in the off-source noise estimate. All of the resulting Stokes parameter errors 
presented in this paper are similarly broadened. The total time on the target source VY CMa was 150~minutes for each 
transition.

The linear polarisation absolute electric vector position angle (EVPA) was determined using ancillary Very Large Array 
(VLA\footnote{science.nrao.edu/facilities/vla/docs/manuals/oss}).
observations. The VLA observed the primary polarisation calibrator J0521+166 (3C138) and secondary polarisation calibrators 
J0646+448, J0609-1542, J0423-013 and J0542+498 in Q-band on 17 March 2007. At this time the array was in D configuration.
The absolute EVPA of the primary polarisation calibrator 3C138 was adopted to be $-14^\circ$ \citep{bk:VLAmanual}, and 
was used to calibrate the absolute EVPA of the secondary VLA polarisation calibrators. The secondary polarisation 
calibrators were then included in the VLBA observations to establish by reference the absolute EVPA of all remaining VLBA 
sources (equivalently, the residual unknown R-L phase difference at the reference antenna, assumed constant \citep{Kemball:99}).
It was not possible to use this method to perform absolute EVPA calibration of the 86~GHz data, as the VLA is not 
equipped to observe at this frequency.


\subsection{Circular polarisation calibration}
\label{section:V-calibration}

The VY~CMa SiO maser emission is expected to be only weakly circularly polarised, at a level of a few percent 
\citep[e.g.][]{McIntosh:94,Herpin:06}. Accurate calibration is therefore required for the circular polarisation 
measurements, as described in \citet{Kemball:11}.

As outlined in the previous section, the calibration method employed in this work solves for the differential R/L amplitude 
gains from a global fit to the continuum calibrator data. The calibrators J0423-1020 and J0609-1542 were used in this fit. 
No continuum circular polarisation has been detected towards J0423-1020 \citep{Homan:01,Homan:06,Vitrishchak:08,Agudo:10}. 
Continuum circular polarisation has been detected towards J0609-1542, at frequencies $\le8~$GHz \citep{Homan:01, Homan:03, Aller:03}, 
and a $-0.23\%$ detection has been reported at 15~GHz \citep{Homan:03}, as well as a more recent 15~GHz non-detection with 
an upper limit of $0.21\%$ \citep{Homan:06}. A non-detection with an upper limit of $0.53\%$ has been reported at 86~GHz 
\citep{Agudo:10}. Circular polarisation at a level of $\lesssim0.5\%$ will not be significant relative to the noise in the 
work reported here, so J0609-1542 was also considered a suitable calibrator for the global continuum calibrator fit.

The errors $\sigma_{g_0^{RL}}$ in the R/L amplitude gains estimated using this method were determined using jackknife 
subsampling \citep{bk:Davison}. Table \ref{table:BR123-gRL-error} lists for each observed SiO transition, the reference 
differential R/L amplitude gain $g_0^{RL}$, associated jacknife error estimate $\sigma_{g_0^{RL}}$, and an independent 
measure of the error $\sigma_{m_c}$ in fractional circular polarization derived from continuum calibrator imaging of net 
residual Stokes $V$, described in further detail below.


\begin{table}
\begin{center}
\begin{tabular}{|lllll|}
\hline 
SiO        & Frequency & $g_0^{RL}$ & $\sigma_{g_0^{RL}}$ & $\sigma_{m_c}$ \\
transition & band      &            &         & \\
\hline
\hline
v=1 J=2-1 & 86~GHz & 0.990 & $8.5\times10^{-3}$ & $0.61\%$ \\ 
v=2 J=1-0 & 43~GHz & 1.024 & $7.5\times10^{-3}$ & $1.8\%$ \\ 
v=1 J=1-0 & 43~GHz & 0.993 & $7.4\times10^{-3}$ & $0.00064\%$ \\ 
\hline
\end{tabular}
\caption[Circular polarisation accuracy parameters]
{{\small The circular polarisation accuracy parameters for the BR123 data sets.
The columns from left to right are: 
the spectral line observed, the frequency band of the observation, 
the reference R/L amplitude gain $g_0^{RL}$,
the jackknife error estimate $\sigma_{g_0^{RL}}$, and an independent measure of 
error $\sigma_{m_c}$ in fractional circular polarization, derived from continuum 
calibrator imaging tests.
}}
\label{table:BR123-gRL-error}
\end{center}
\end{table}


\begin{figure*}
\hspace{-3.0cm}
\begin{minipage}[b]{0.32\linewidth}
   \begin{center}
      \includegraphics[width=3.5in]{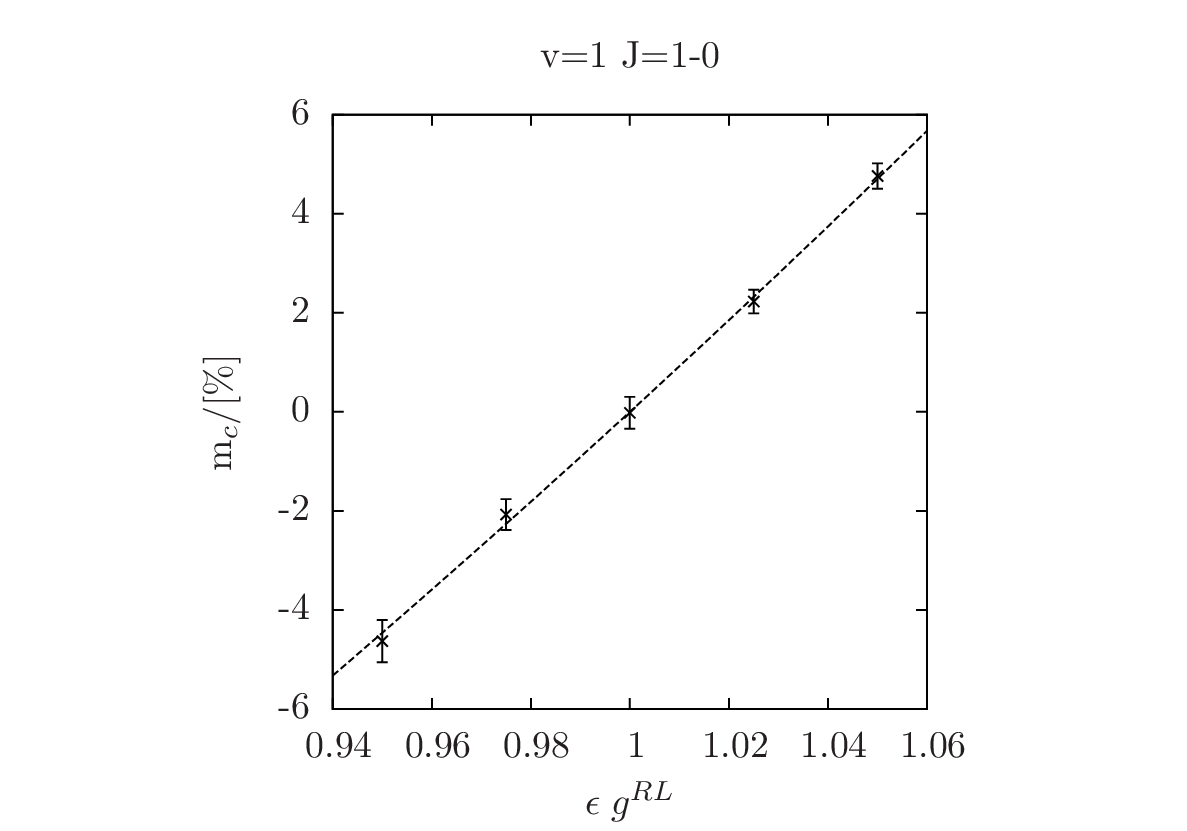}
   \end{center}
\end{minipage}
\begin{minipage}[b]{0.32\linewidth}
   \begin{center}
      \includegraphics[width=3.5in]{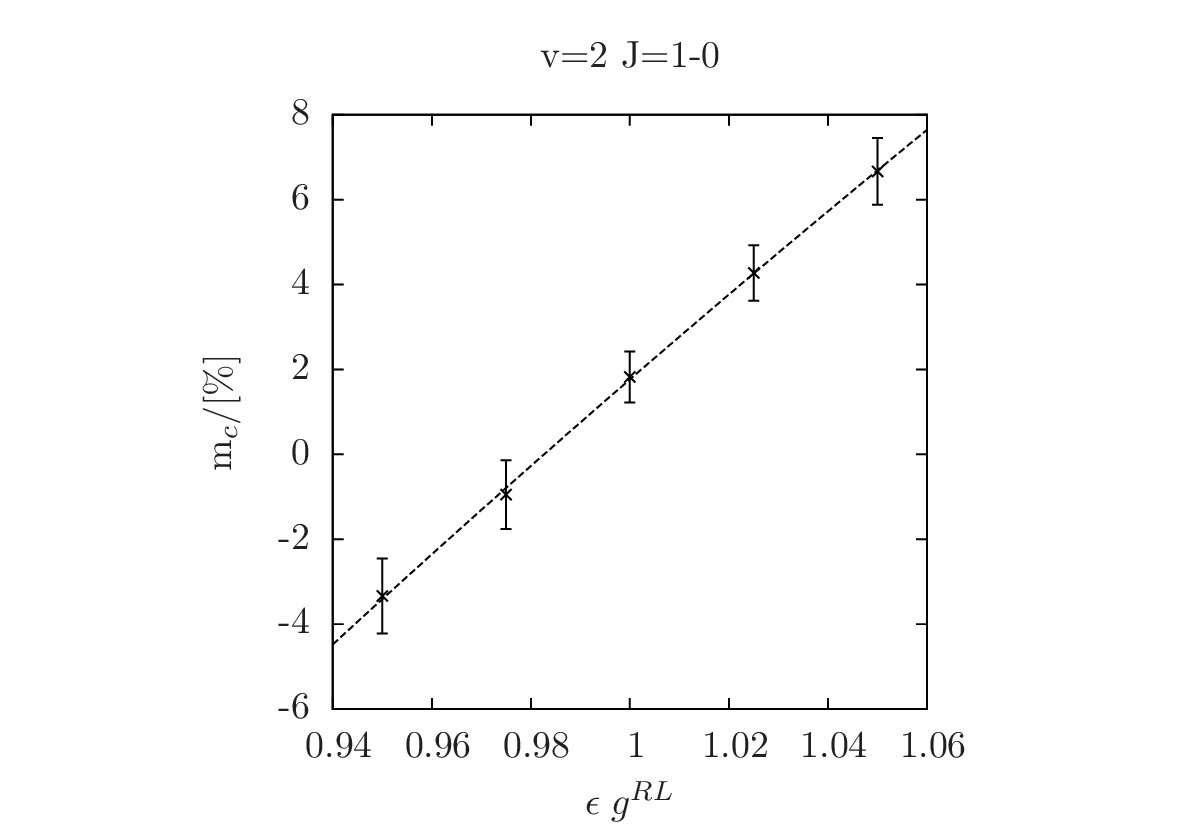}
   \end{center}
\end{minipage}
\begin{minipage}[b]{0.32\linewidth}
   \begin{center}
      \includegraphics[width=3.5in]{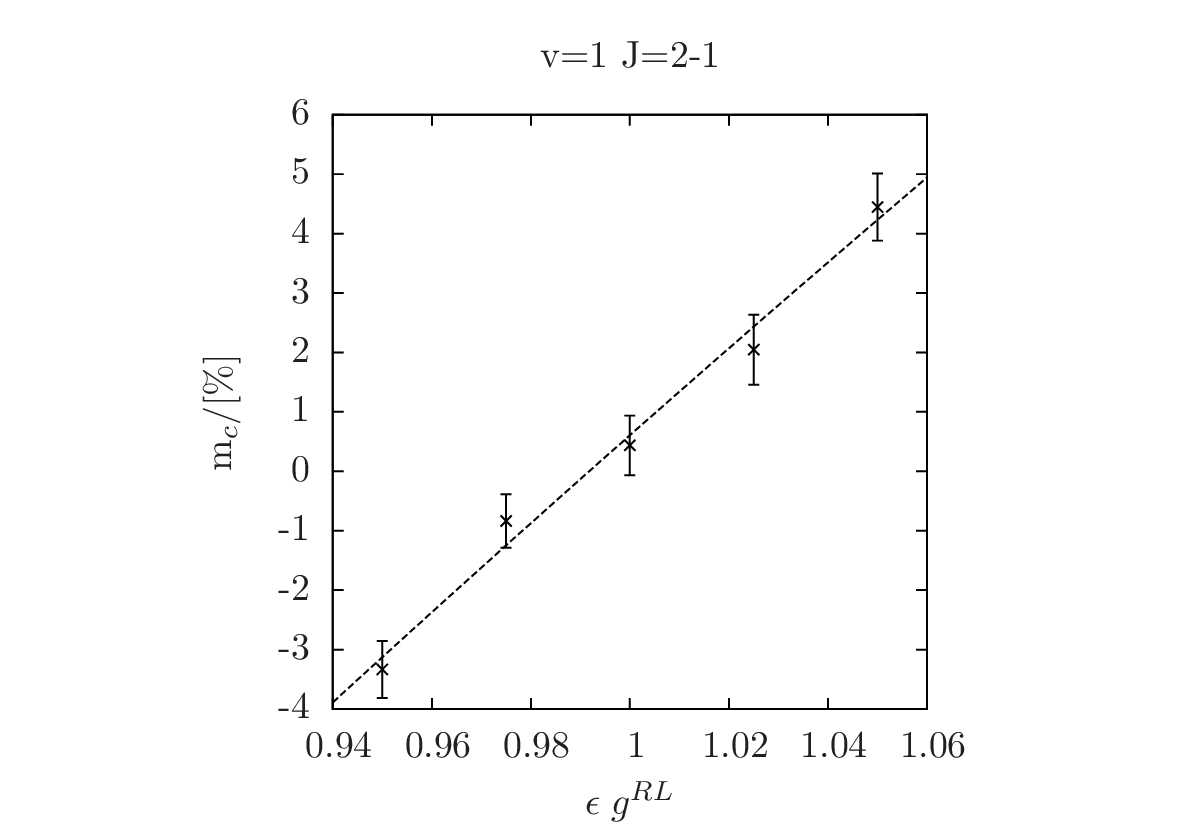} 
   \end{center}
\end{minipage} 
\caption[Calibrator imaging m$_c$ versus $\epsilon_{\mbox{g}^{RL}}$ offset]{{\small
Plots of continuum calibrator circular polarisation, m$_c$, versus multiplicative R-L amplitude gain offset $\epsilon_{\mbox{g}^{RL}}$ 
for calibrator J0423-0120, for 43~GHz v=1 J=1-0 data (left), 43 GHz v=2 J=1-0 data (middle), and the 86~GHz v=1 J=2-1 data (right). 
For each data set a second-order polynomial was fitted to the data, which is plotted as a dashed line.}}
\label{fig:Caibrator-imaging} 
\end{figure*}


In this latter test, the accuracy of the spectral-line circular polarization amplitude calibration was independently assessed 
by applying the line amplitude calibration derived for the SiO maser source to a continuum calibrator source in the data set; 
the calibrator was then imaged after application of an additional, multiplicative $\epsilon_{\mbox{g}^{RL}}$ offset before 
allowing only residual phase calibration. This cross-calibration was repeated independently for a range of values 
$\epsilon^{RL}$, spanning unity. Further details of this test can be found in \citet{Kemball:11}. Calibrator J0423-1020 
was chosen for this test because it is the brightest calibrator that was observed throughout the schedule. 

Ten iterations of phase-only self-calibration were performed for each $\epsilon_{\mbox{g}^{RL}}$ data set constructed in this 
manner, imaging down to a final deconvolution threshold of a few times the thermal noise limit. The J0423-1020 data from the 
\mbox{J=1-0} observation were imaged with pure uniform weighting. For the J0423-1020 data from the J=2-1 observation, Briggs 
weighting with a robustness parameter of zero \citep{Briggs:thesis} was found to produce superior imaging performance. This 
calibration and imaging procedure for J0423-1020 was performed independently for multiplicative $\epsilon_{\mbox{g}^{RL}}$ 
offsets of 0.95, 0.975, 1.0, 1.025 and 1.05.

However, J0423-0120 was only observed for nine scans of approximately seven minutes each over the course of the observations, 
so the data set does not contain a large number of visibilities. Furthermore, the number of unflagged visibilities is further 
reduced after interpolating the VY~CMa amplitude calibration gains onto J0423-0120, as there are time interval limits over 
which the line calibration solutions may reasonably be interpolated. A range of interpolation methods and flagging limits were 
investigated to determine the optimal interpolation parameters. These retain enough data to image while removing data where 
interpolation errors are most extreme. The optimal interpolation method was found to be a three-point median window filter 
with an interpolation limit of 14.25~minutes, half the length of a VY~CMa scan.

The measured image-plane calibrator circular polarisation percentages for each $\epsilon_{\mbox{g}^{RL}}$ offset value are 
plotted in Figure~\ref{fig:Caibrator-imaging}. The circular polarisation was calculated from average Stokes I and V values, 
measured in a tight image box enclosing the central Stokes I emission. If the relative amplitude gains between the RCP and 
LCP data are correct, we would expect a \mbox{m$_c = 0$} intercept for $\epsilon_{\mbox{g}^{RL}}=1$. The deviation of the 
m$_c$ intercept from zero provides a conservative upper bound $\sigma_{m_c}$ on the error in the $R/L$ line amplitude 
calibration, given the inherent interpolation errors between line and continuum scans in this test. A second-order polynomial 
was fitted to each measured sequence $m_c(\epsilon_{g^{RL}})$; values computed at $m_c(\epsilon_{g^{RL}}=1)$ are listed in 
the right-most column of Table~\ref{table:BR123-gRL-error}. 

For the v=2 J=1-0 data set, the calibrator imaging test gives a particularly poor result, with an estimated $\sigma_{m_c} = 1.8\%$ 
(Table~\ref{table:BR123-gRL-error}; Figure~\ref{fig:Caibrator-imaging}). Of the three calibrator data sets, the imaging artifacts 
were most extreme for the Stokes~$V$ v=2 J=1-0 data set J0423-1020 images, with deep off-source negatives around the central source 
region. If the Stokes~$V$ values for this data set are averaged over a larger box incorporating the negative regions around the 
Stokes $I$ source position, then the fitted $\epsilon_{\mbox{g}^{RL}}=1$ intercept occurs at $\|$m$_c\|\le0.5\%$. Thus, this 
independent estimate of error is at its limit of applicability for this transition, as discussed further below.

Outside of this discrepant v=2 J=1-0 calibrator imaging result, Table~\ref{table:BR123-gRL-error} shows that the errors in 
the R/L amplitude gain solutions are $\le1\%$. \citet{Kemball:11} estimate the accuracy of the circular polarisation calibration 
method applied to VLBA observations of SiO maser emission towards TX~Cam to be $\le0.5-1\%$ at 43~GHz and $\le1\%$ at 86~GHz.
These ranges are consistent with the jacknife error estimates $\sigma_{g_0^{RL}}$ in Table~\ref{table:BR123-gRL-error}. The poorer 
performance reported for the v=2 J=1-0 calibrator imaging test in the current work is due in part to the greater angular separation 
between the target source VY~CMa and calibrators J0423-1020 and J0609-1542, compared to the angular separation between the source 
and calibrators used in \citet{Kemball:11}, as well as the low elevation of VY CMa. Both effects heighten interpolation errors in 
this test. The current data set also contained fewer calibrator observations than that used in \citet{Kemball:11}.


\section{Results}
\label{sec:Results}

\begin{figure}
   \begin{center}
      \includegraphics[width=3.5in, angle=0]{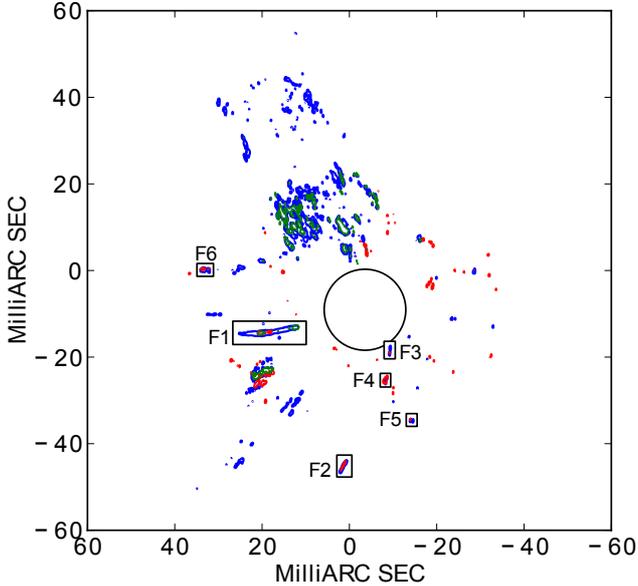}
      \caption{{\small Overlaid single-contour total intensity plots of the peak Stokes $I$ 
      brightness over frequency for the SiO transitions
      v=1 J=1-0 (blue), 
      v=2 J=1-0 (green), and 
      v=1 J=2-1 (red). 
      The contour level is $5\sigma_I$. The relative astrometric alignment was determined using a cross-correlation method, 
      as described in the text. The features F1 to F6 defined here are discussed in the text and plotted in Figures~\ref{fig:F1} 
      through \ref{fig:F6}.
      A circle of diameter 18.7~mas representing the estimated stellar diameter is plotted at the adopted stellar position 
      (described further in Section~\ref{section:poln_testing}).
      }}
      \label{fig:BR123-boxes}
   \end{center}
\end{figure}

For each transition, the peak intensity (over frequency channel) is plotted as a single-contour plot in Figure~\ref{fig:BR123-boxes}, 
colour-coded and overlaid by transition, at a contour level of 5$\sigma_I$. This follows Figure 8 in Paper I and defines features 
F1-F6, and includes a circle denoting the estimated stellar diameter (see Section~\ref{sec:Discussion} below). Absolute astrometric 
positions are lost during the data reduction process, due to the use of phase self-calibration \citep{Thompson:04}, so the relative 
alignment of the transition maps is unknown a priori. As described in Paper I the relative alignment was instead determined using a 
cross-correlation method and this alignment is used in Figure~\ref{fig:BR123-boxes}. The uncertainty in the map alignment is estimated 
to be $< 0.05$ mas (Paper I). The maps of the v=2 J=1-0 and v=1 J=2-1 SiO maser transitions were restored with the same beam size as 
the v=1 \mbox{J=1-0} SiO maser map, to allow component-level comparison of maser features.


\subsection{Maser feature parameters}
\label{sec:result_features}

Component-level parameters of the individual features in the full Stokes I image cubes were extracted using the three-dimensional 
source detection software Duchamp \citep{Whiting:12}. The detection threshold used in Duchamp was set to five times the broadened 
noise $\sigma_I$ in the channel with the highest root mean square (RMS) noise (Table~\ref{table:observation_summary}). The minimum 
channel width for feature detection was set to two channels, as the narrowest line widths of SiO maser features are typically 
$\sim0.5$~km/s \citep{Glenn:03}. 

The catalogue of maser features detected with Duchamp in Stokes $I$ is presented in Appendix~\ref{appendixA}. The table lists the 
mean velocity $v$ and velocity extent $\Delta v$ of each feature. The Stokes $Q$, $U$ and $V$ brightness values for each feature 
were taken to be the associated value of the emission at the pixel position of maximum Stokes $I$ in the feature. The quoted errors 
$\{\sigma_I$, $\sigma_Q$, $\sigma_U$, $\sigma_V\}$ in the Stokes parameters use broadened off-source noise estimates, as described 
above.

The fractional circular polarisation m$_c$, fractional linear polarisation m$_l$, and the EVPA $\chi$ were derived from the measured 
Stokes $I$, $Q$, $U$ and $V$ brightness values. The uncertainties in m$_c$, m$_l$ and $\chi$ are also included in the tables, calculated 
through propogation of the Stokes parameter errors.

The positions in the table are given as offsets ($\Delta\alpha$, $\Delta\delta$) on the projected plane of the sky, measured in 
milliarcseconds and increasing in the direction of increasing right ascension and declination. The offset is measured from the adopted 
centre of the map after relative alignment of the maser maps in each transition.


\begin{figure*}
\begin{minipage}[b]{0.48\linewidth}
   \begin{center}
      \includegraphics[width=3.0in]{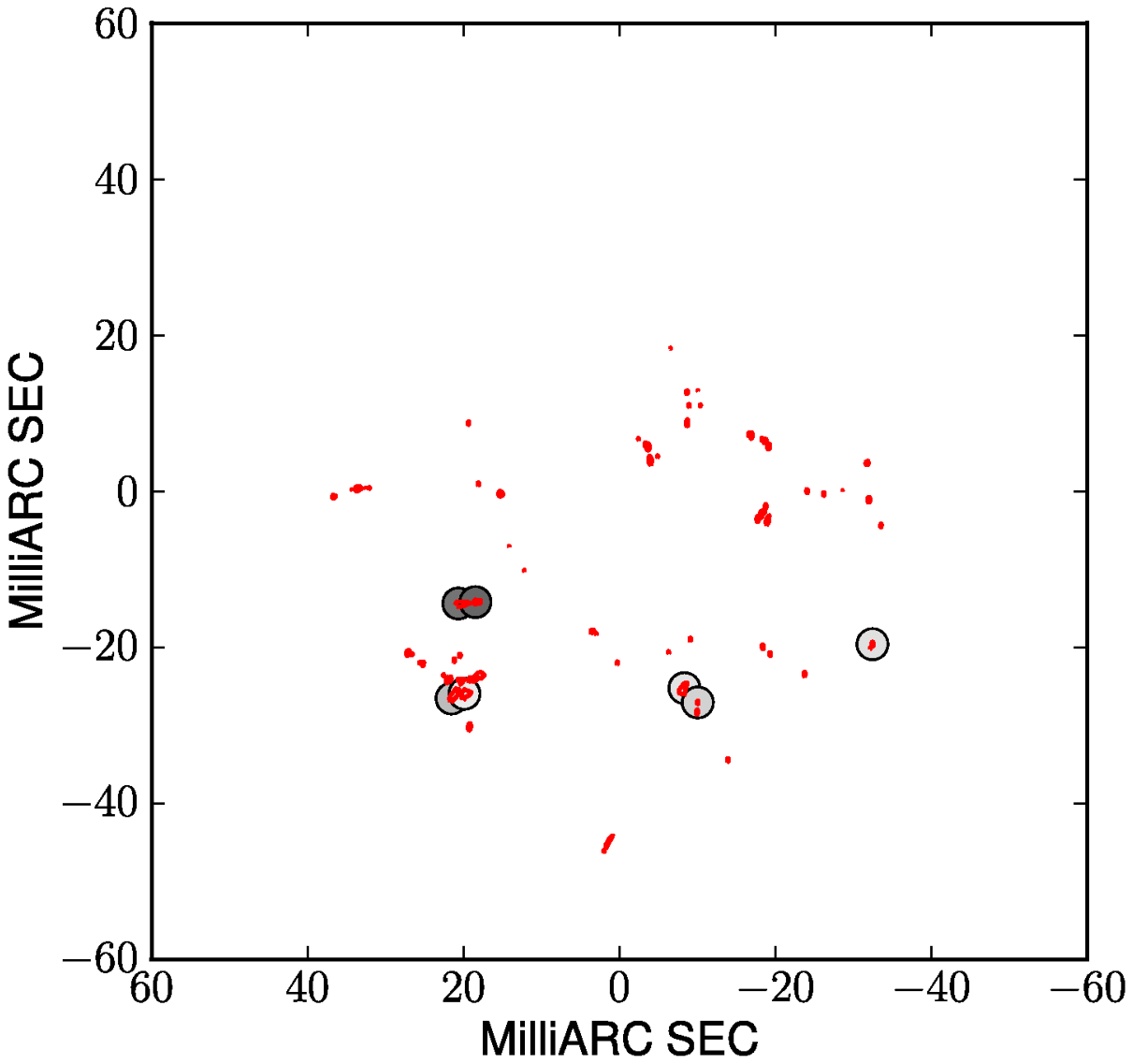}     
   \end{center}
\end{minipage}
\begin{minipage}[b]{0.48\linewidth}
   \begin{center}
      \includegraphics[width=3.0in]{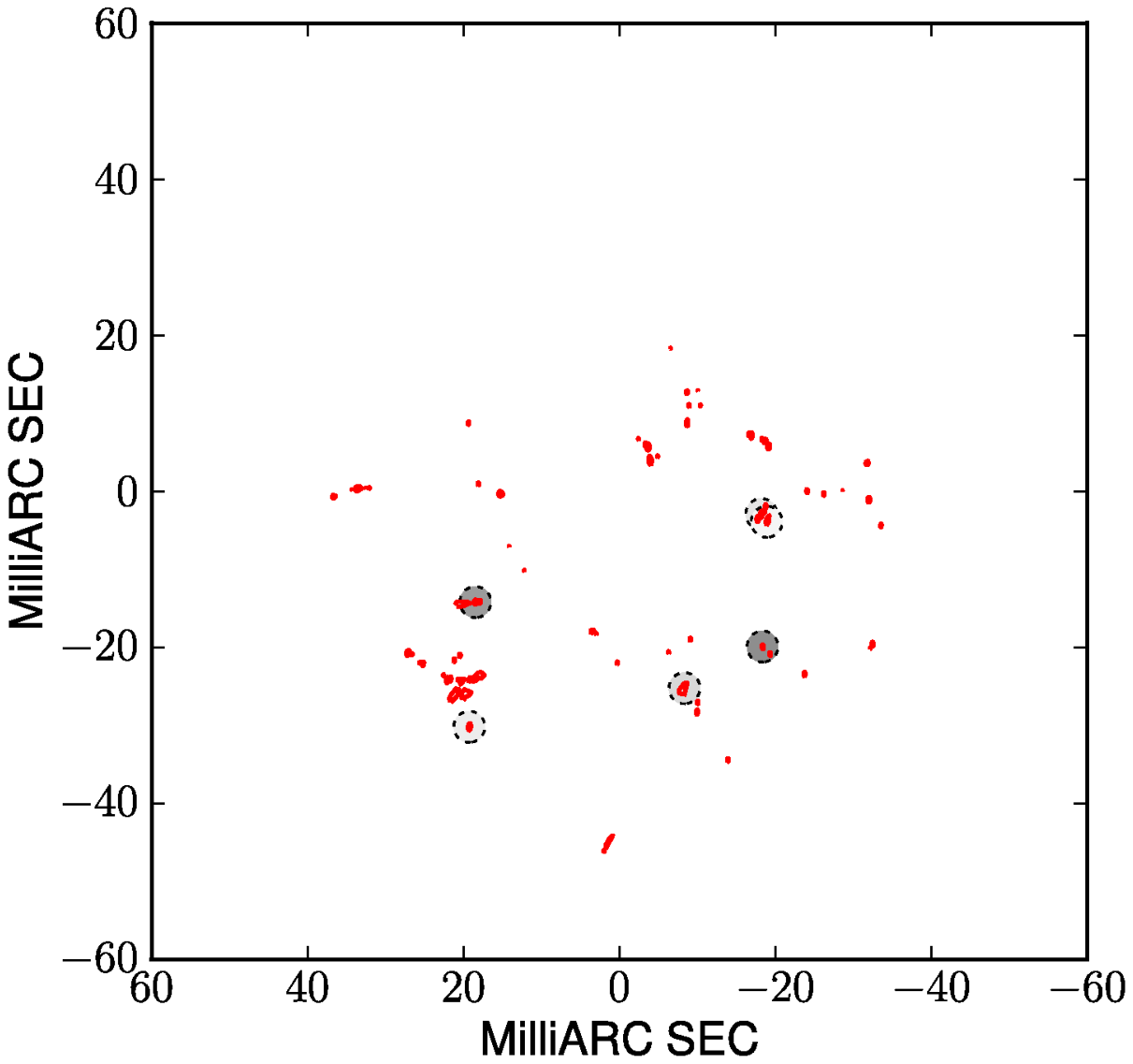}     
   \end{center}
\end{minipage}
\begin{minipage}[b]{0.48\linewidth}
   \begin{center}
      \includegraphics[width=3.0in]{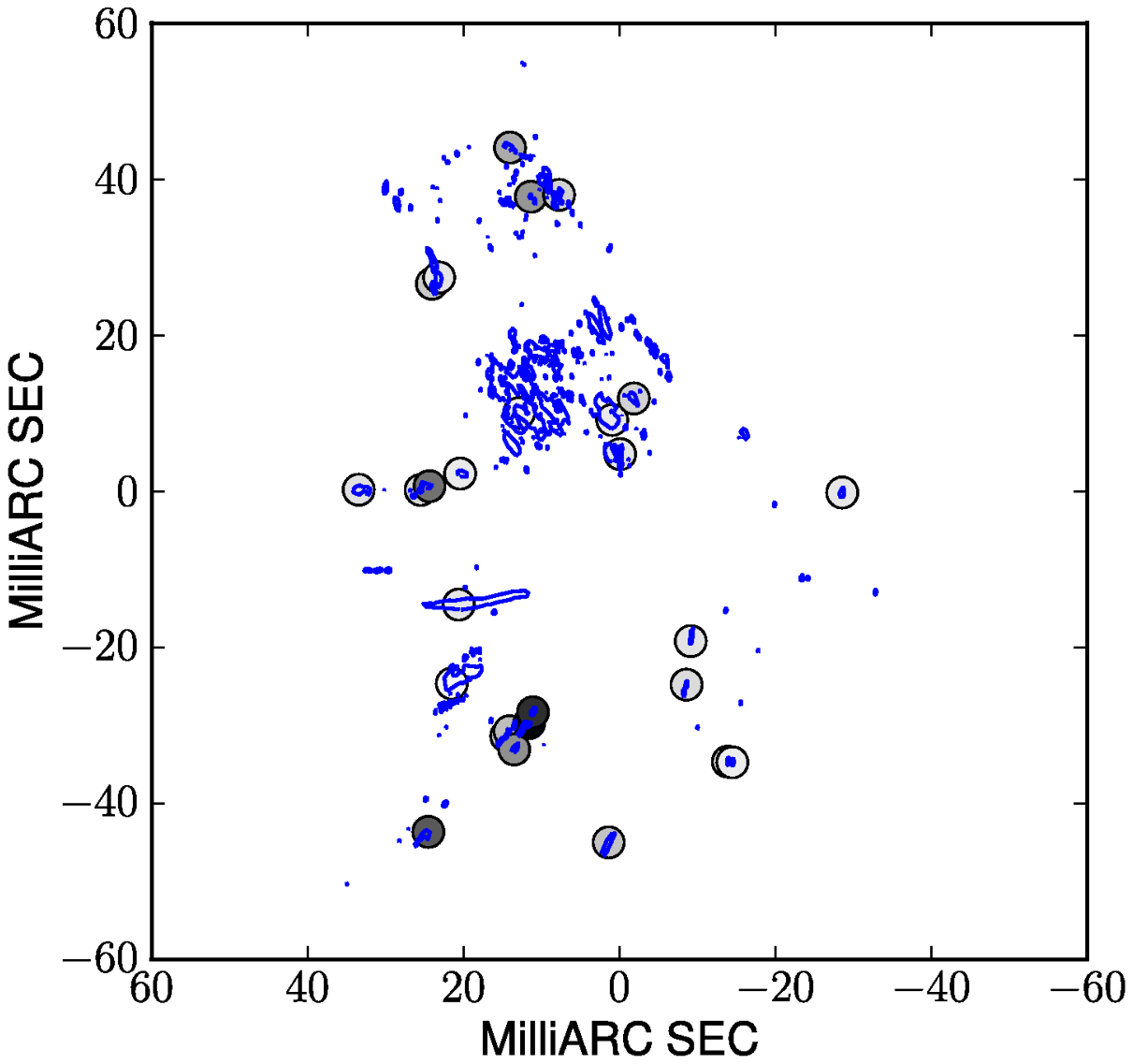}     
   \end{center}
\end{minipage}
\begin{minipage}[b]{0.48\linewidth}
   \begin{center}
      \includegraphics[width=3.0in]{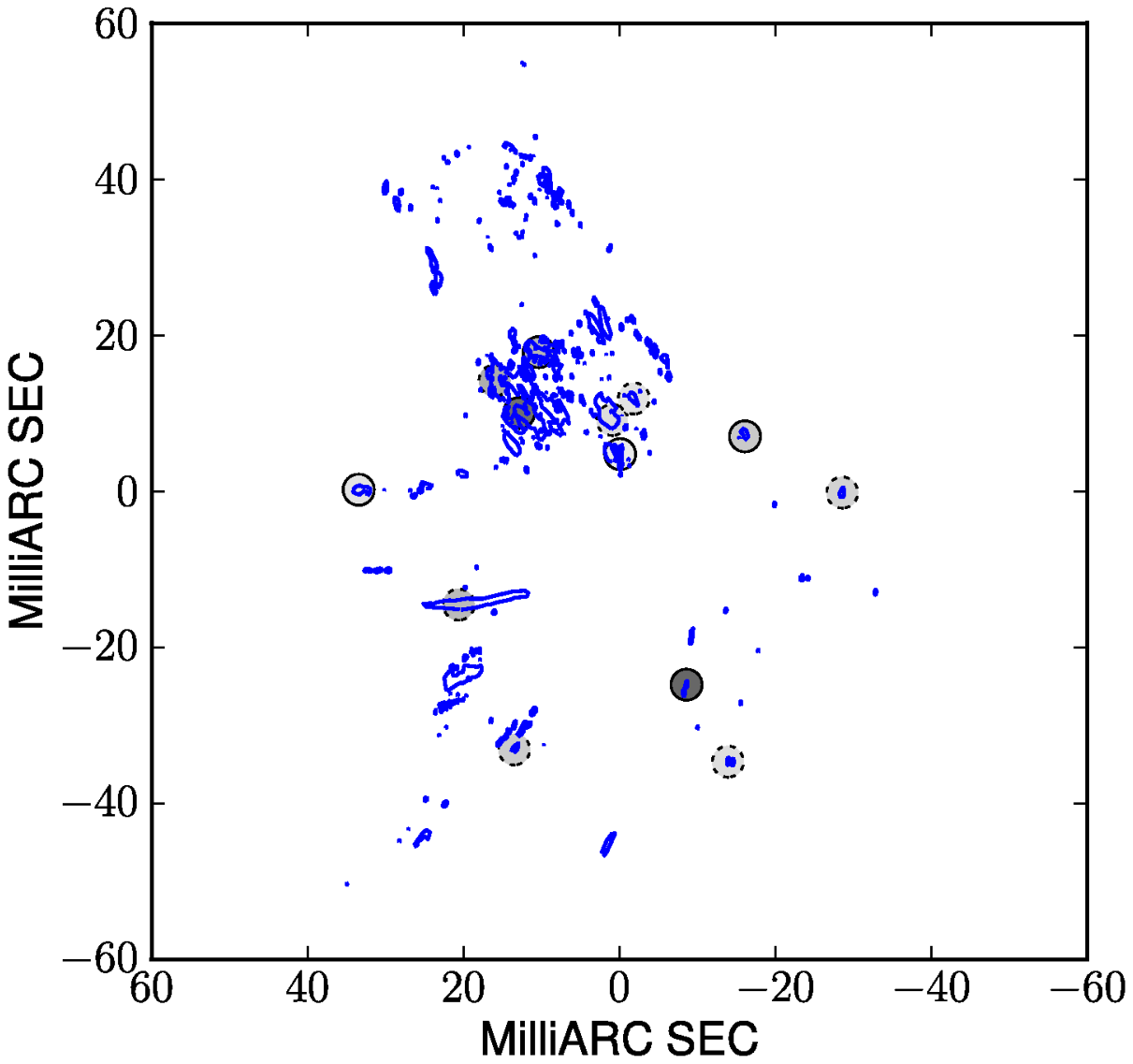}     
   \end{center}
\end{minipage}
\begin{minipage}[b]{0.48\linewidth}
   \begin{center}
      \includegraphics[width=3.0in]{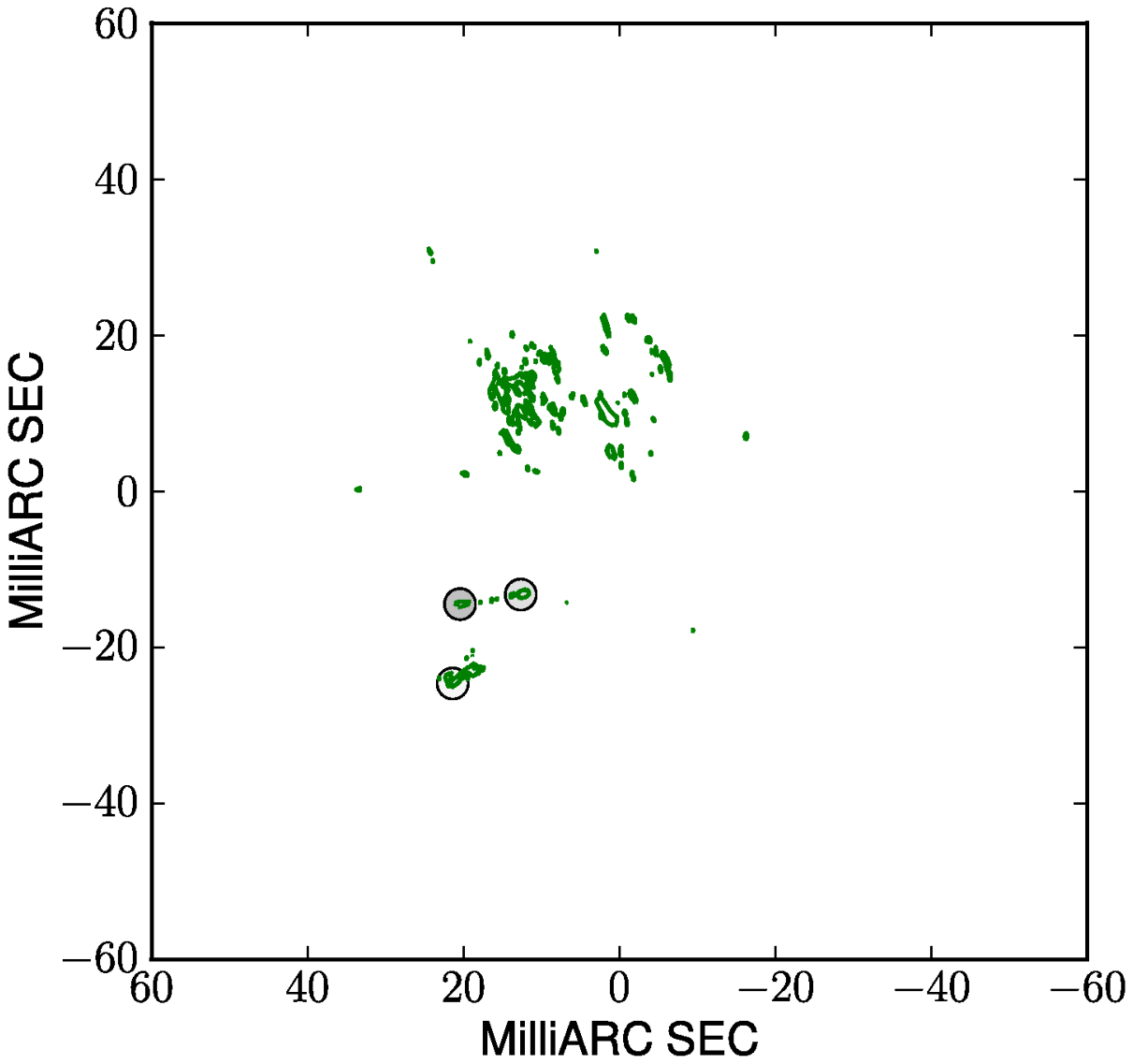}     
   \end{center}
\end{minipage}
\begin{minipage}[b]{0.48\linewidth}
   \begin{center}
      \includegraphics[width=3.0in]{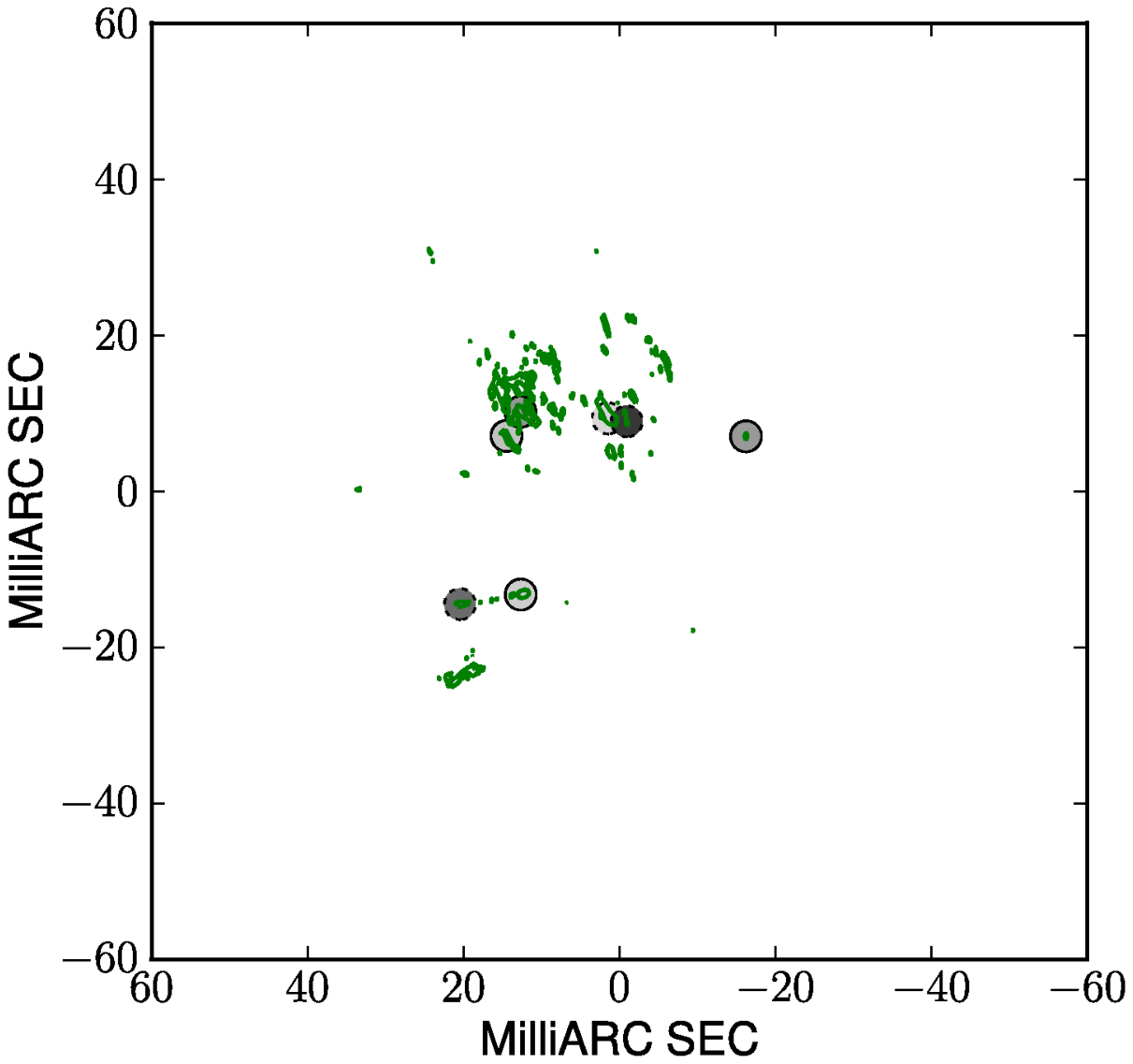}     
   \end{center}
\end{minipage} 
\caption[Linearly-polarisaed featues]{{\small
Linearly (left) and circularly (right) polarised features, from Tables~\ref{tab:BR123A-components} to \ref{tab:BR123D1-components},
for v=1 J=2-1 (red), v=1 J=1-0 (blue) and v=2 J=1-0 (green).
The fractional polarisation of each feature is represented by a circle centered on the position of the feature,
with the gray scale level of the circle increasing in proportion to the absolute fractional polarization of the feature. 
For the circularly-polarised features, the black border of the circle denotes whether $m_c > 0$ (solid) or $m_c < 0$ (dashed). 
The $|m_c|$ values fall in the range [$1.2\%$,$7.1\%$] and the $m_l$ values fall in the range [$1.6\%$,$46.5\%$].
}}
\label{fig:ml-mc-spots} 
\end{figure*}


Calculation of the measured linearly-polarized intensity $P$ are intrinsically biased due to the Ricean probability distribution 
of the non-negative $P = \sqrt{Q^2+U^2}$. This bias is taken into account by using the correction $P' = \sqrt{ P^2 - \sigma_{QU}^2 }$
\citep{Wardle:74}. The noise levels in the Stokes $Q$ and $U$ maps are similar, so the assumption is made that 
$\sigma_Q \sim \sigma_U$ in this analysis, and the geometric mean denoted as $\sigma_{QU} = \sqrt{\sigma_Q \sigma_U}$. The 
fractional linear polarisation values $m_l$ listed in the appendix have this correction taken into account. There is no bias 
correction needed for the position angle $\chi$ \citep{Wardle:74}.

The non-Gaussian probability density functions of $m_l$ and $m_c$ must also be taken into account when assessing statistical 
significance of a polarisation detection. In each case a detection limit was established by considering the null hypothesis that
the fractional polarisation is equal to zero. The detection limit was set to the upper threshold of the $95\%$ probability interval 
for zero fractional polarisation. These values can be determined through numerical intergration of the probability density 
functions (PDF) for $m_l$ and $m_c$ \citep{Kemball:thesis}, but have well-behaved limit approximations.

The detection limit $u_l$ for the fractional linear polarisation can be approximated by a range estimator 
\begin{equation}
 u_l = \frac{1.65}{\sqrt{2}} \left[ \frac{P'+\sigma_{P'}}{I-\sigma_I} - \frac{P'-\sigma_{P'}}{I+\sigma_I} \right]
\end{equation}
where $\sigma_{P'} = \frac{\sqrt{Q^2 \sigma_Q^2 + U^2 \sigma_U^2}}{{P'}}$ \citep{Kemball:thesis}. 
This prior work found that for m$_l$ values up to $5\%$ the range estimator approximation $u_l$ is an underestimate of the 
detection limit by up to $10\%$. However, when the fractional linear polarisation is large the range estimator may overestimate 
$u_l$ by $35\%$. In the catalogues in Appendix~\ref{appendixA}, only values of $m_l$ exceeding the detection limit $u_l$ are listed.

The $95\%$ probability interval for the fractional circular polarisation $m_c$ can be calculated using the Geary-Hinkley 
transformation, as described by \citet{Hayya:75}, to yield upper and lower limits of
\begin{equation}
 \mbox{u}_c = \pm 1.96 \frac{\sigma_V}{\sqrt{ \mu_I^2 - 1.96^2 \sigma_I^2}} 
\end{equation}
This approximation is good to within $5\%$ when $\sigma_I < 0.39 \mu_I$ and $\sigma_V > 0.005 \mu_V$, where $\mu_I$ and $\mu_V$ 
are the mean values of Stokes parameters $I$ and $V$ \citep{Hayya:75}. These conditions are met for features listed in the 
Appendix catalogue. Only circular polarisations greater than the detection limit $u_c$ are listed in the Appendix.

The features which display statistically-significant linear and circular polarisation are shown in Figure~\ref{fig:ml-mc-spots}.


\subsection{Sub-feature level parameter extraction}
\label{sec:detailed_features}

Six maser features extended across angular position and frequency were chosen for more detailed polarisation analysis. 
The features were chosen based on their spatial coincidence, or near spatial coincidence, in multiple SiO maser transions and 
are labelled F1 to F6 in Figure~\ref{fig:BR123-boxes}.

\begin{figure*}
\includegraphics[width=6.6in, angle=0]{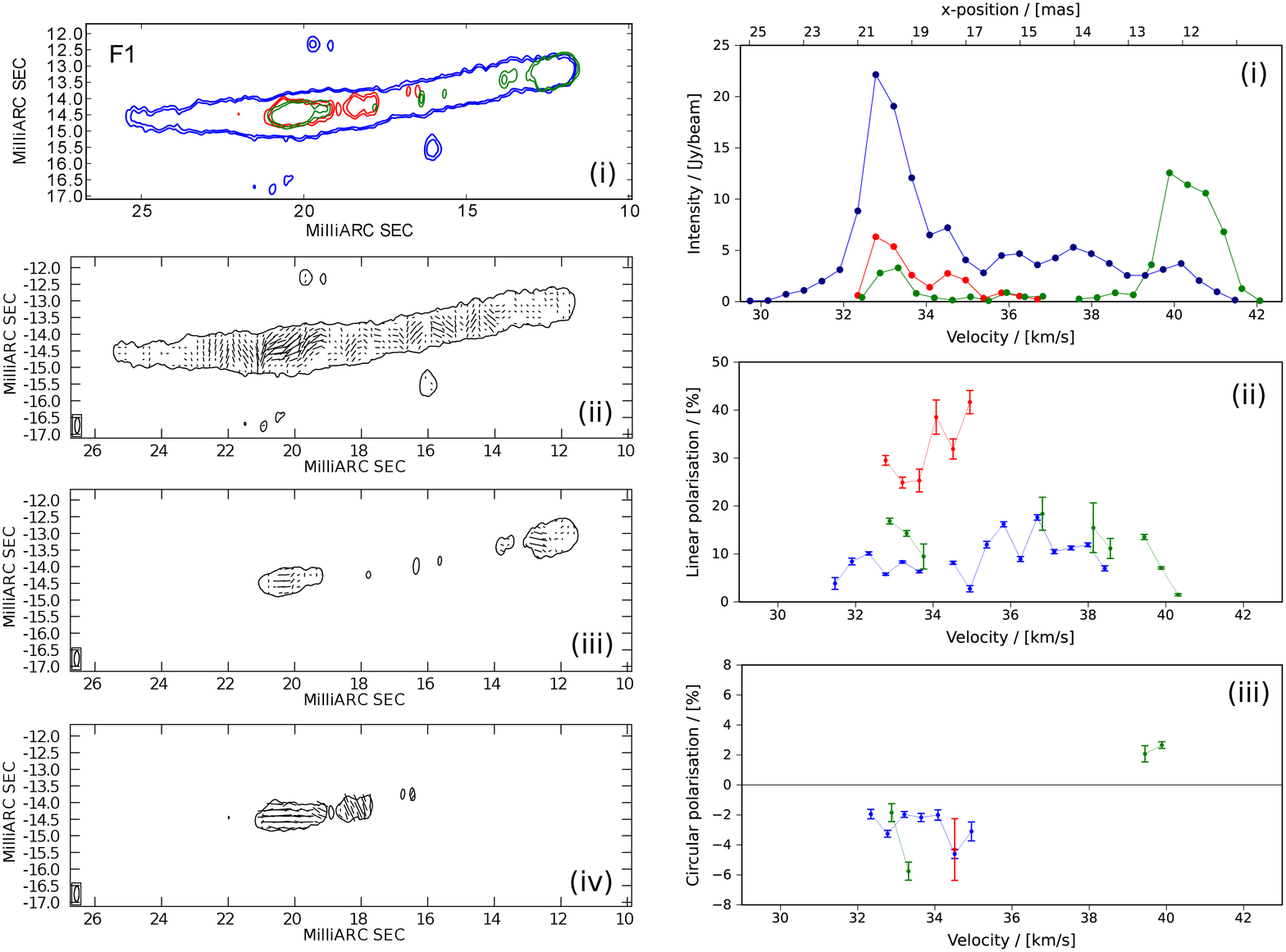} 
\caption[Emission properties of feature F1]
{{\small Emission properties of feature F1, which is defined in Figure~\ref{fig:BR123-boxes}. \\ 
Left, top to bottom: 
i) Overlaid contour plot of the v=1 J=1-0 (blue), v=2 J=1-0 (green) and v=1 J=2-1 (red) maser
emission drawn at contour levels of $\{3\sigma_I,\ 5\sigma_I\}$; 
ii-iv) contour plots of the v=1 J=1-0, v=2 J=1-0 and v=1 J=2-1 emission at a total-intensity contour level of $\{3\sigma_I\}$,
overlaid with vectors proportional in length to the underlying linearly polarized intensity on a scale where 
1~mas~=~$27.78\times10^{-3}$~Jy/beam. 
The vector orientation is in the direction of absolute EVPA for the J=1-0 lines (ii-iii; see text for J=2-1 EVPA alignment). 
The synthesised beam is drawn in lower-left of frames (ii) to (iv), and is $0.46 \times 0.15$~mas in half-power at a position angle of 
$-1.80^\circ$.\\
Right, top to bottom:
i) Feature spectra of Stokes $I$ intensity; ii) percentage linear polarisation, and iii) percentage circular polarisation
over line-of-sight LSR velocity (in km/s).
The upper axis of the Stokes $I$ spectrum shows the x-position at each velocity channel across the feature
where the v=1 J=1-0 Stokes parameters were measured.
}}
\label{fig:F1}
\end{figure*}

\begin{figure}
   \begin{center}
       \includegraphics[height=1.6in, angle=0]{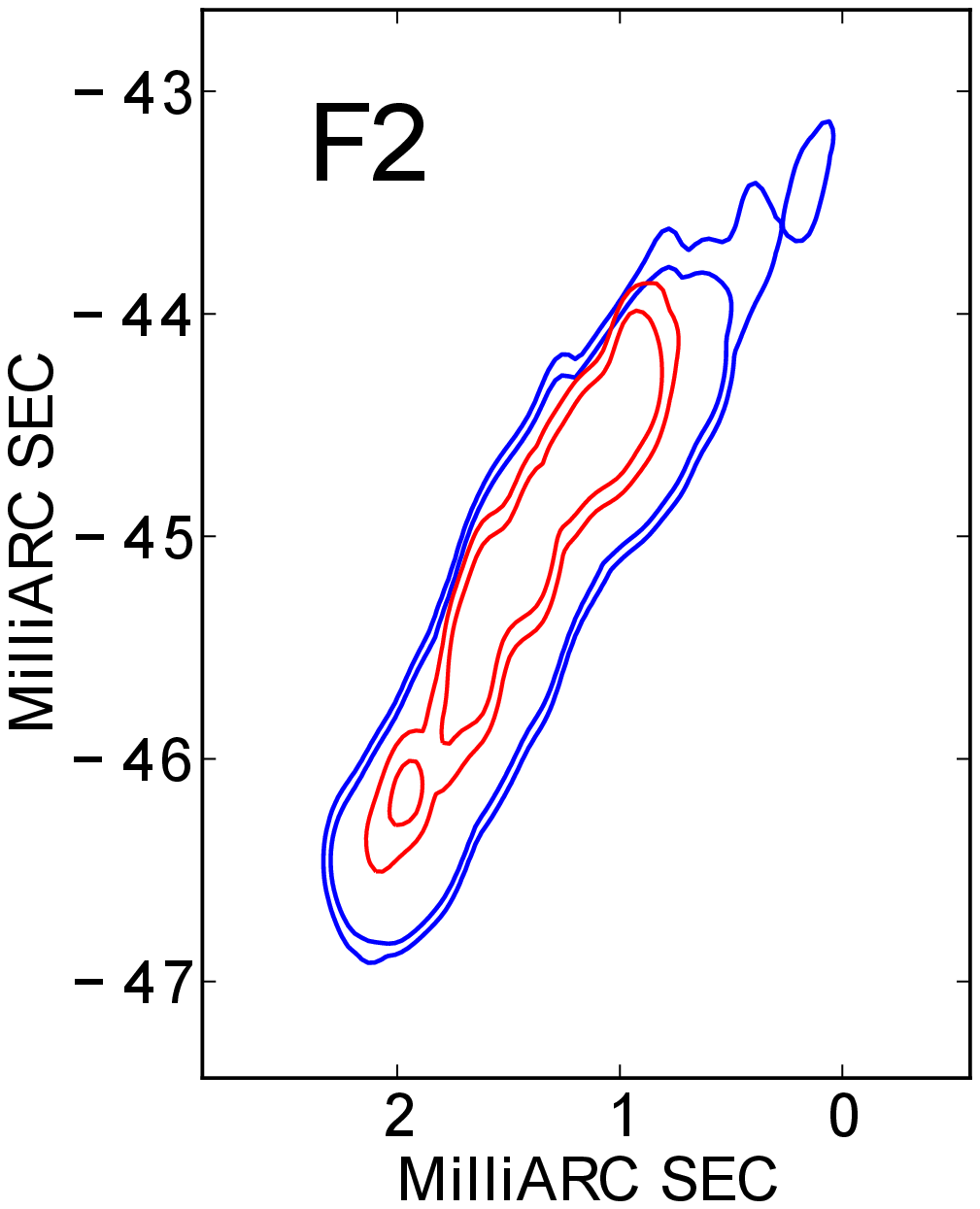}
       \includegraphics[height=1.6in, angle=0]{BR123D-12-F2-PANG.PS}
   \end{center}
\vspace{0.05cm}
   \begin{center}
       \includegraphics[width=3.2in, angle=0]{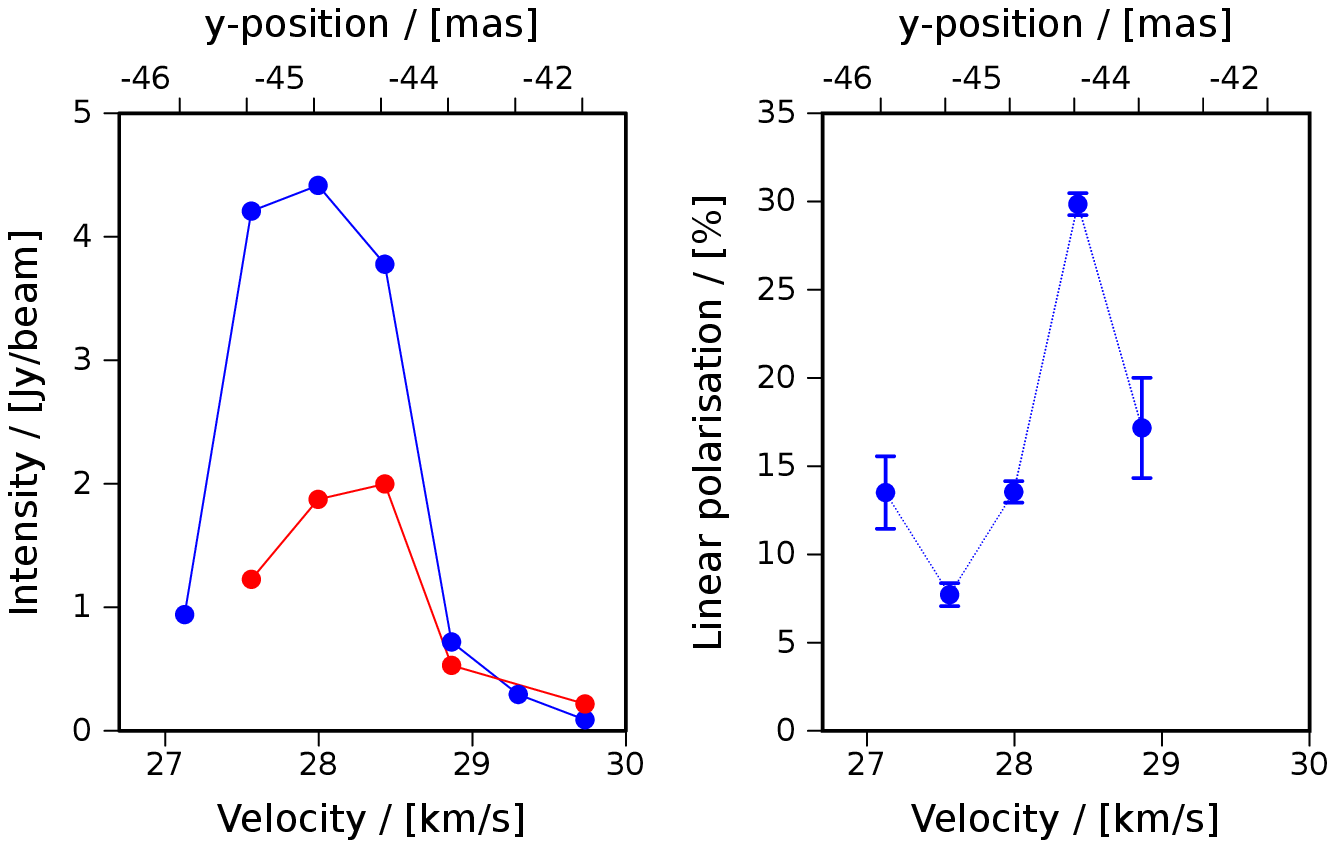}   
   \end{center}  
\caption[Emission properties of feature F2]
{{\footnotesize Emission properties of feature F2, which is defined in Figure~\ref{fig:BR123-boxes}. \\ 
Top, left to right:
Overlaid contour plot of the v=1 J=1-0 (blue) and v=1 J=2-1 (red) maser
emission, as for Figure~\ref{fig:F1};
Contour plot of the v=1 J=1-0 emission overlaid with linear polarisation EVPA vectors,
as for Figure~\ref{fig:F1}. 
Bottom, left to right: 
Stokes $I$ spectrum of the feature and percentage linear polarisation spectrum of the feature,
as for Figure~\ref{fig:F1}.
}}
\label{fig:F2}
\end{figure}

\begin{figure}
   \begin{center}
       \includegraphics[height=1.8in, angle=0]{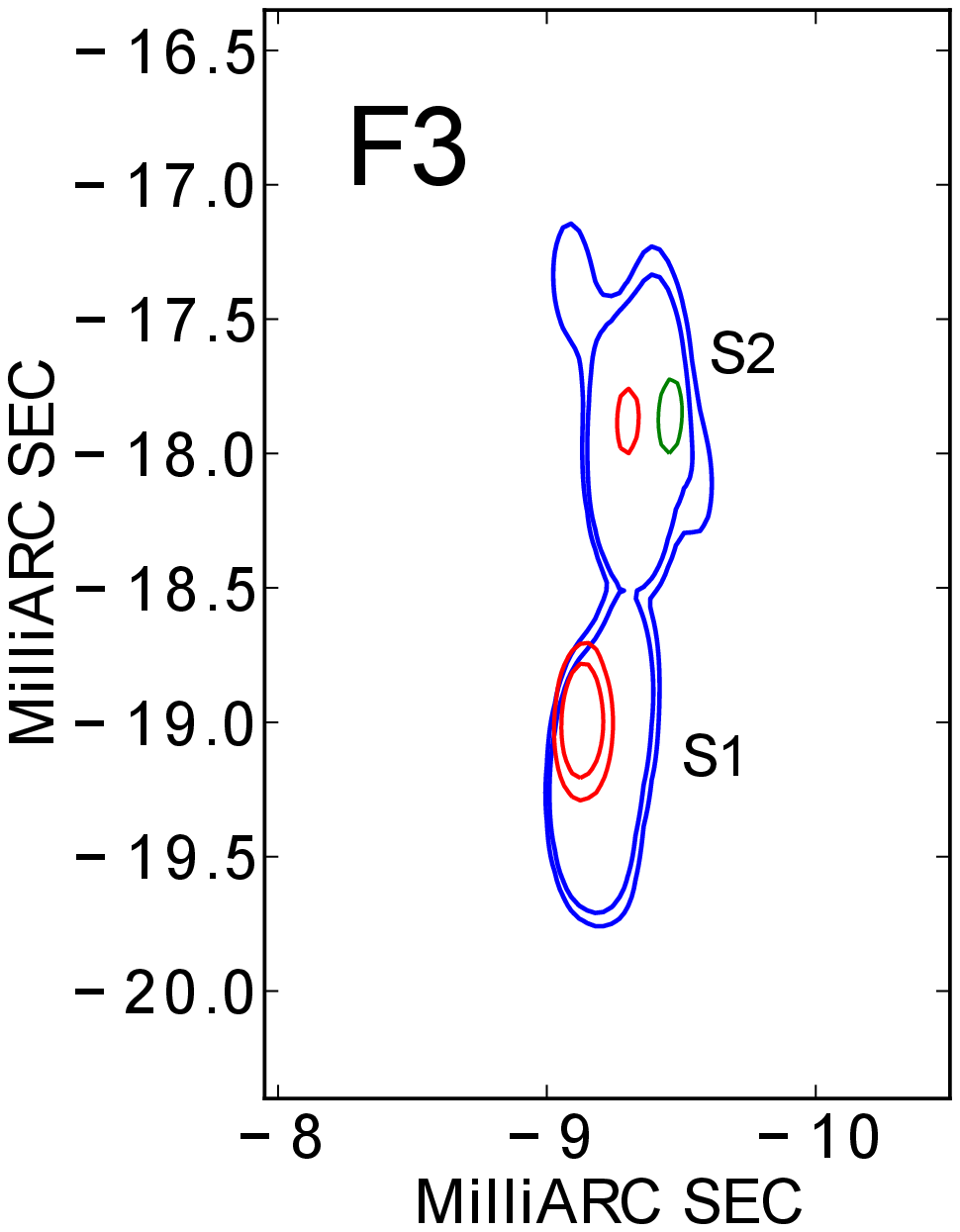}
       \includegraphics[height=1.8in, angle=0]{BR123D-12-F3-PANG.PS}
   \end{center}
\vspace{0.05cm}
   \begin{center}
       \includegraphics[width=3.0in, angle=0]{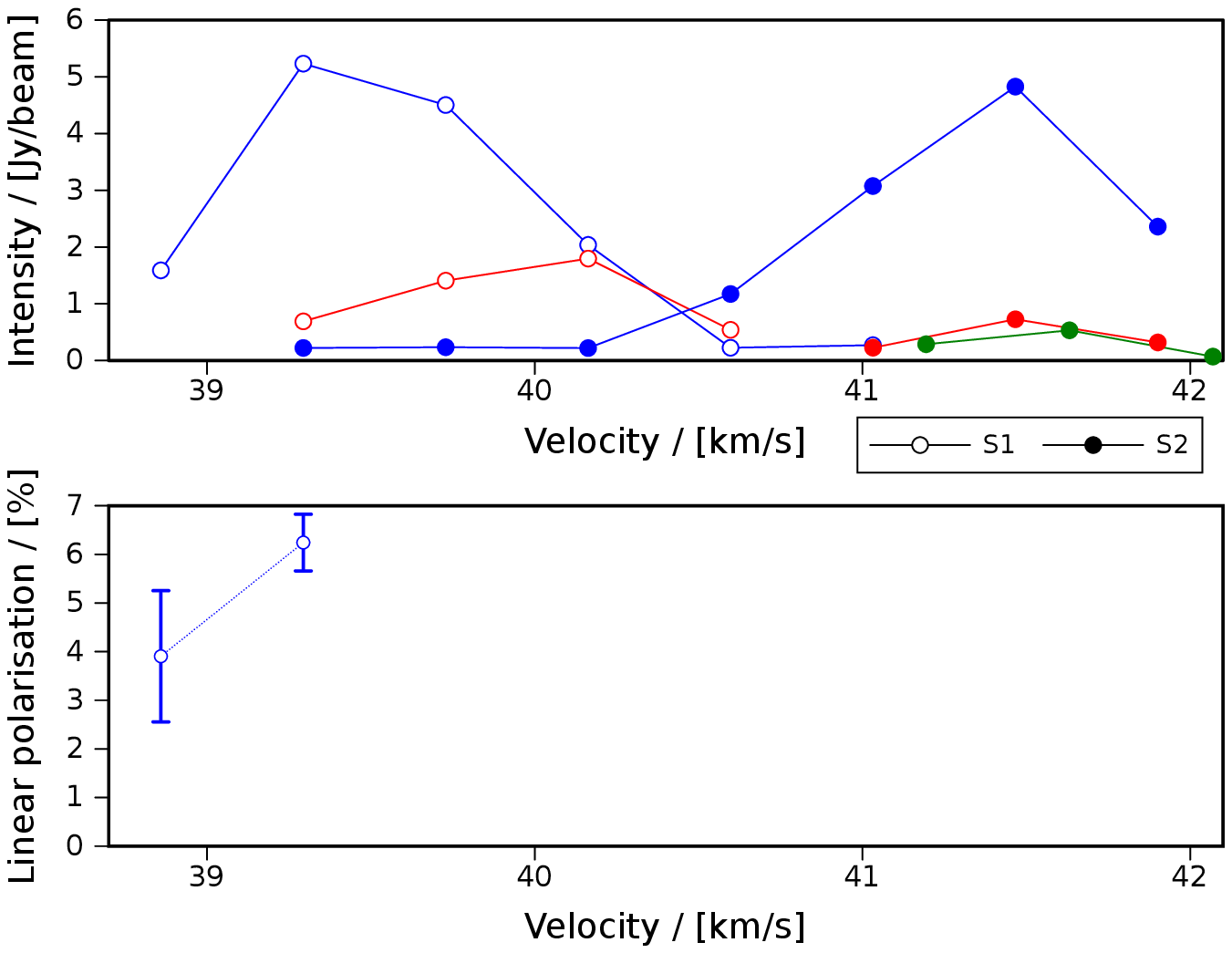}     
   \end{center}
\caption[Emission properties of feature F3]
{{\footnotesize Emission properties of feature F3, which is defined in Figure~\ref{fig:BR123-boxes}. \\ 
Top to bottom:
Overlaid contour plot of the v=1 J=1-0 (blue), v=2 J=1-0 (green) and v=1 J=2-1 (red) maser
emission, as for Figure~\ref{fig:F1};
Contour plot of the v=1 J=1-0 emission overlaid with linear polarisation EVPA vectors,
as for Figure~\ref{fig:F1}. 
Stokes $I$ intensity and percentage linear polarisation spectra of the feature,
as for Figure~\ref{fig:F1}.
}}
\label{fig:F3}
\end{figure}

\begin{figure}
\begin{minipage}{0.32\linewidth}
   \begin{center}
   \includegraphics[width=1.12in, angle=0]{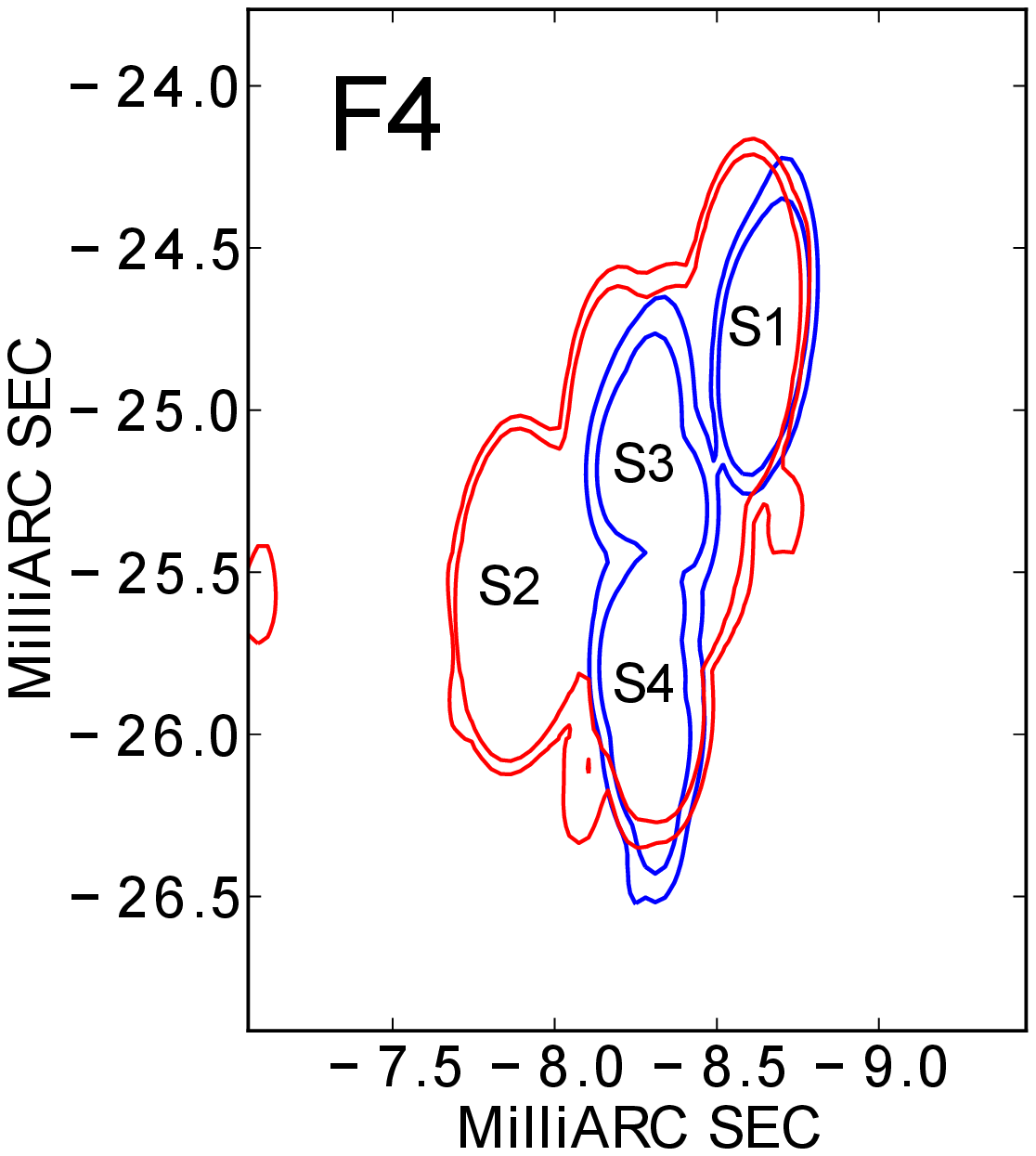}
   \end{center}
\end{minipage}
\begin{minipage}{0.32\linewidth}
   \begin{center}
   \includegraphics[width=1.2in, angle=0]{BR123D-12-F4-PANG.PS}
   \end{center}
\end{minipage}
\begin{minipage}{0.32\linewidth}
   \begin{center}
   \includegraphics[width=1.2in, angle=0]{BR123A-11-F4-PANG.PS}
   \end{center}
\end{minipage}
\vspace{0.2cm}
   \begin{center}
       \includegraphics[width=2.8in, angle=0]{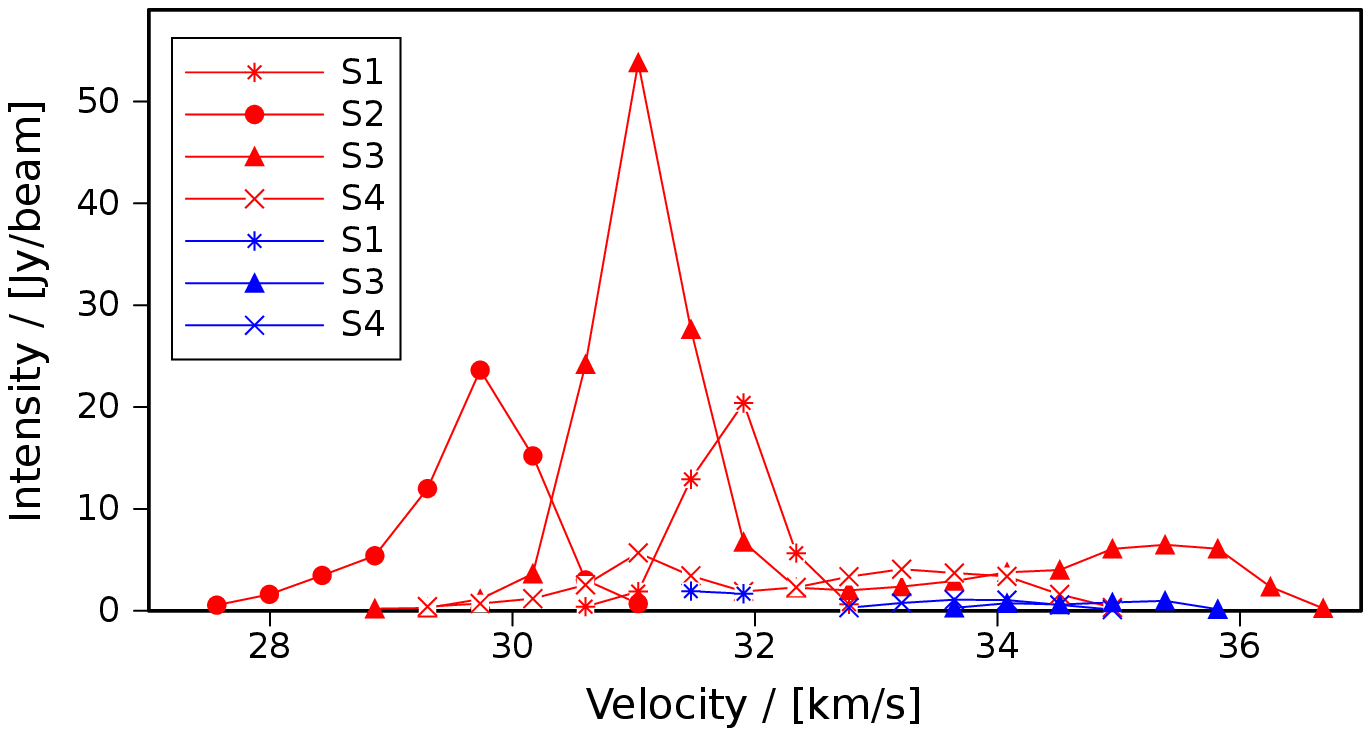}
       \includegraphics[width=2.8in, angle=0]{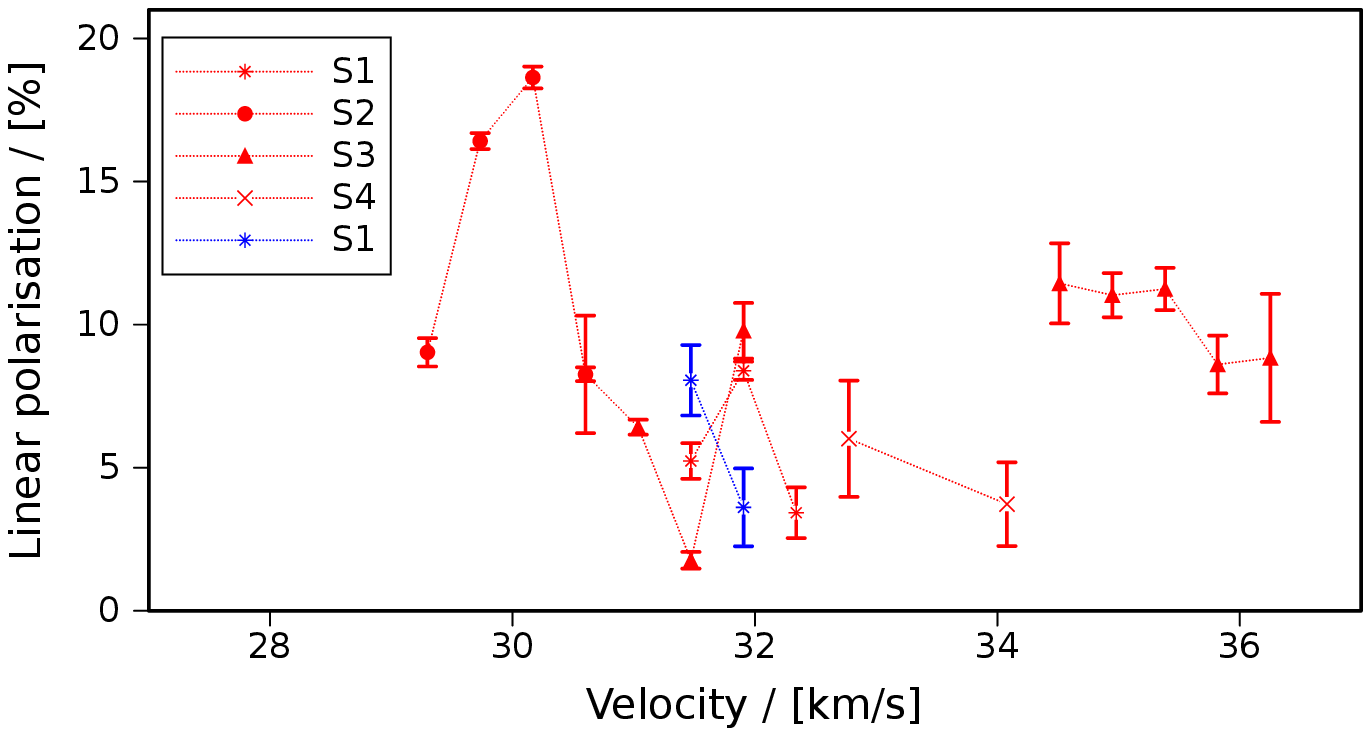}
       \includegraphics[width=2.8in, angle=0]{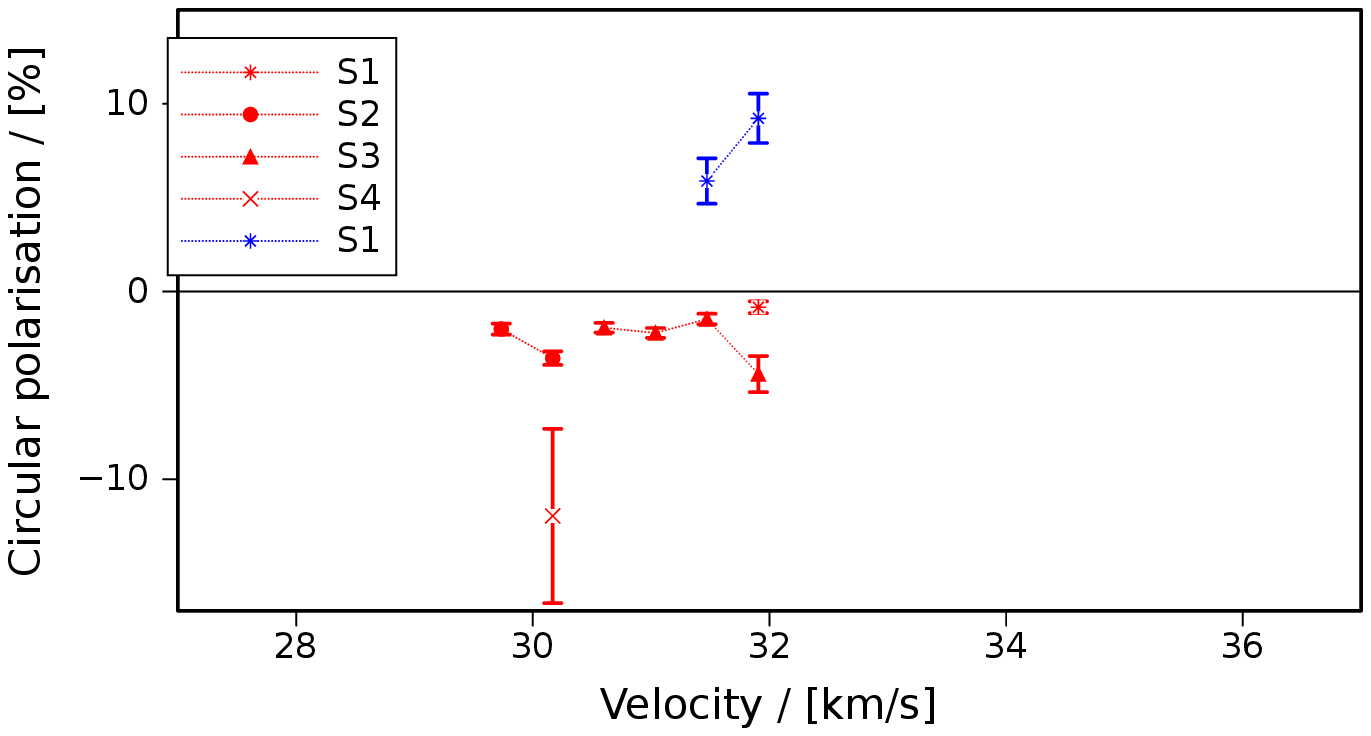}   
   \end{center}
%
%
\caption[Emission properties of feature F4]
{{\footnotesize Emission properties of feature F4, which is defined in Figure~\ref{fig:BR123-boxes}. \\ 
Top to bottom:
Overlaid contour plot of the v=1 J=1-0 (blue) and v=1 J=2-1 (red) maser
emission, as for Figure~\ref{fig:F1};
Contour plot of the v=1 J=1-0 emission overlaid with linear polarisation EVPA vectors,
as for Figure~\ref{fig:F1}. 
Contour plot of the v=1 J=2-1 emission overlaid with linear polarisation EVPA vectors,
as for Figure~\ref{fig:F1}, except that the vector length scale is 1~mas~=~$83.33\times10^{-3}$~Jy/beam.
Stokes $I$ intensity, percentage linear polarisation and percentage circular polarisation 
spectra for each of the spots in the feature, as for Figure~\ref{fig:F1}.
}}
\label{fig:F4}
\end{figure}

\begin{figure}
\begin{center}
    \includegraphics[height=1.4in, angle=0]{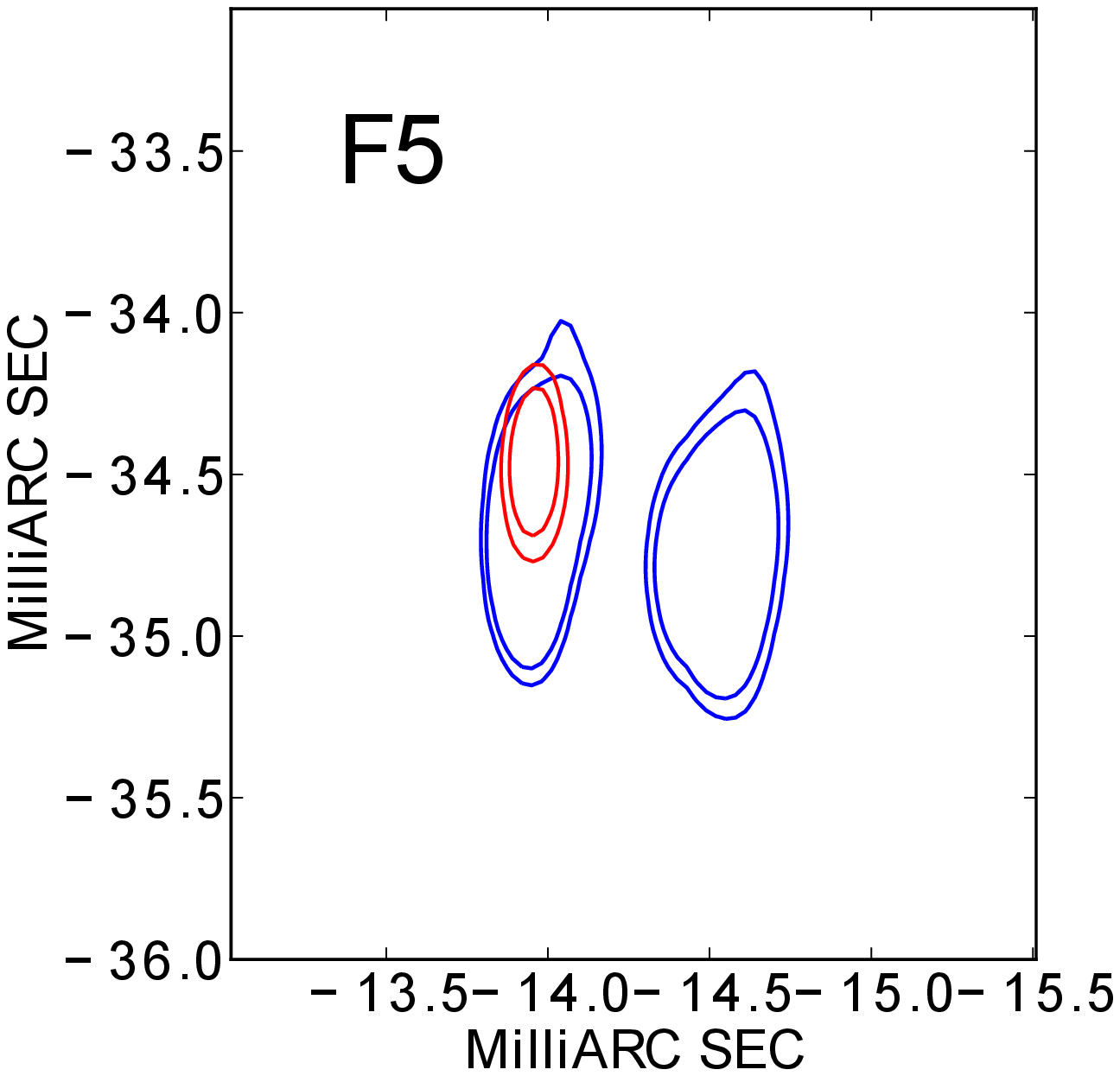}
    \includegraphics[height=1.5in, angle=0]{BR123D-12-F5-PANG.PS}
\end{center}
\vspace{0.02cm}
\begin{center}
    \includegraphics[width=2.5in, angle=0]{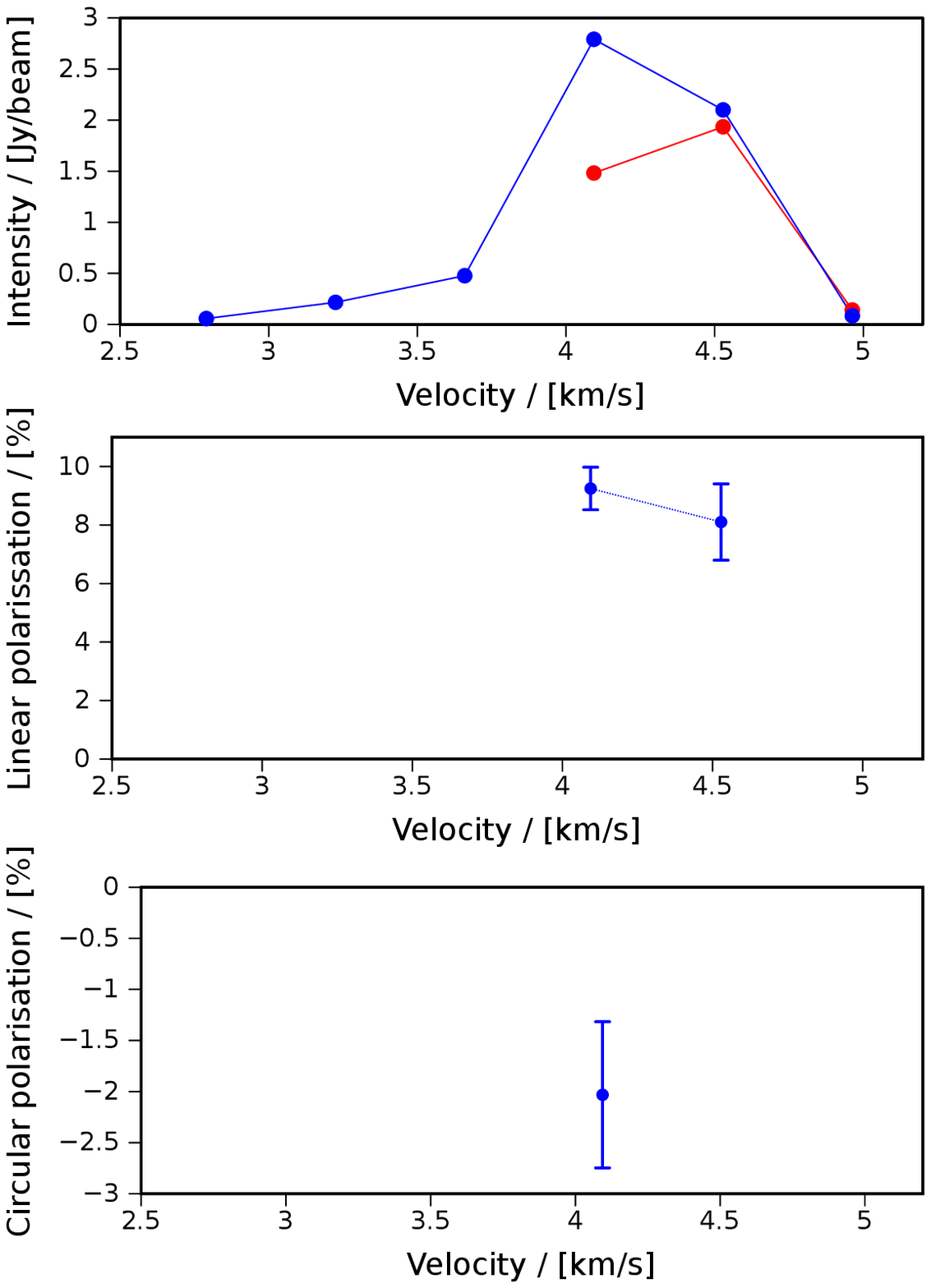}     
\end{center}
\caption[Emission properties of feature F5]
{{\footnotesize Emission properties of feature F5, which is defined in Figure~\ref{fig:BR123-boxes}. \\ 
Top to bottom:
Overlaid contour plot of the v=1 J=1-0 (blue) and v=1 J=2-1 (red) maser
emission, as for Figure~\ref{fig:F1};
Contour plot of the v=1 J=1-0 emission overlaid with linear polarisation EVPA vectors,
as for Figure~\ref{fig:F1}. 
Stokes $I$ intensity, percentage linear polarisation and percentage circular polarisation 
spectra for each of the spots in the feature, as for Figure~\ref{fig:F1}.
}}
\label{fig:F5}
\end{figure}

\begin{figure}
\begin{center}
    \includegraphics[height=1.2in, angle=0]{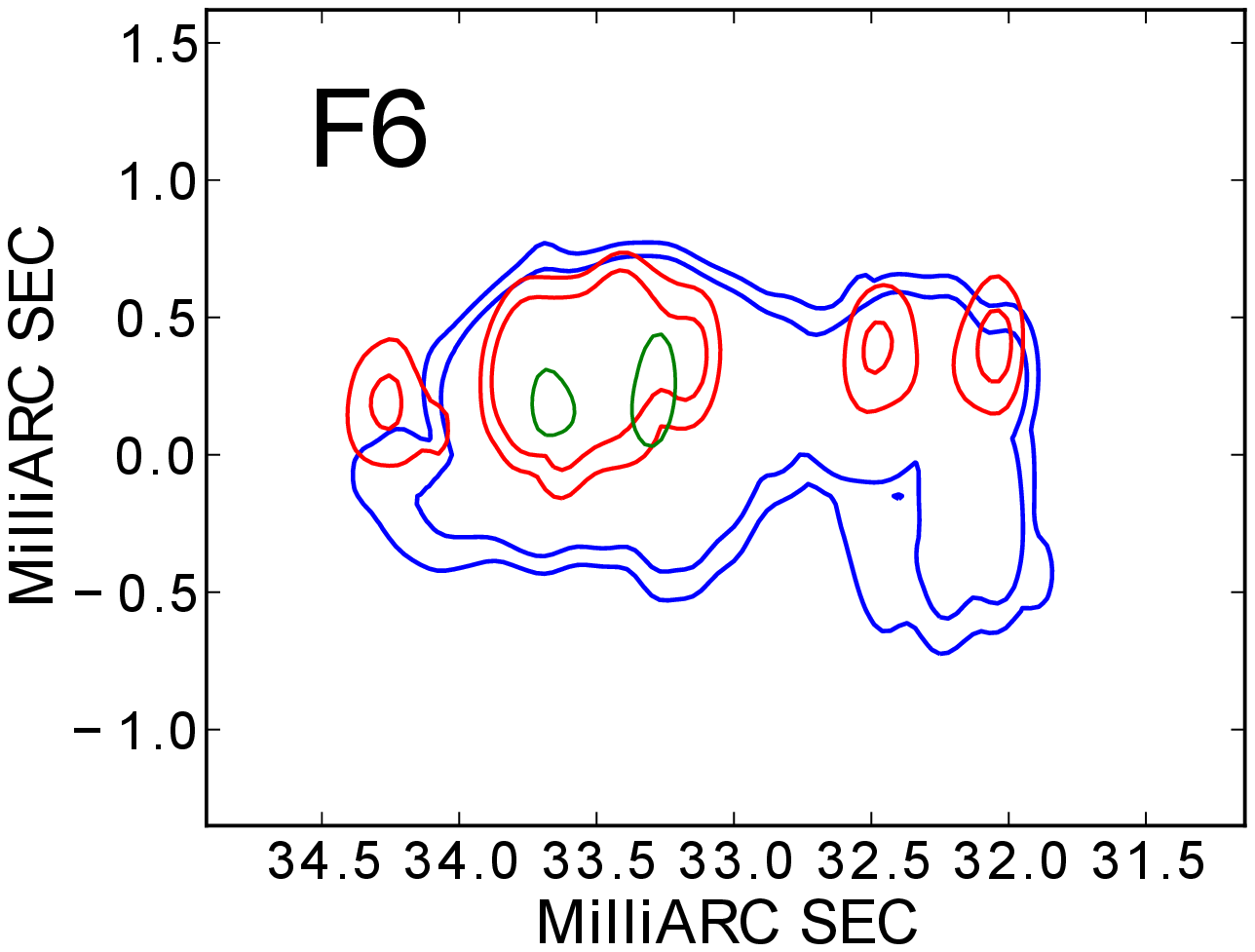}
    \includegraphics[height=1.3in, angle=0]{BR123D-12-F6-PANG.PS}
\end{center}
\vspace{0.02cm}
\begin{center}
    \includegraphics[width=2.5in, angle=0]{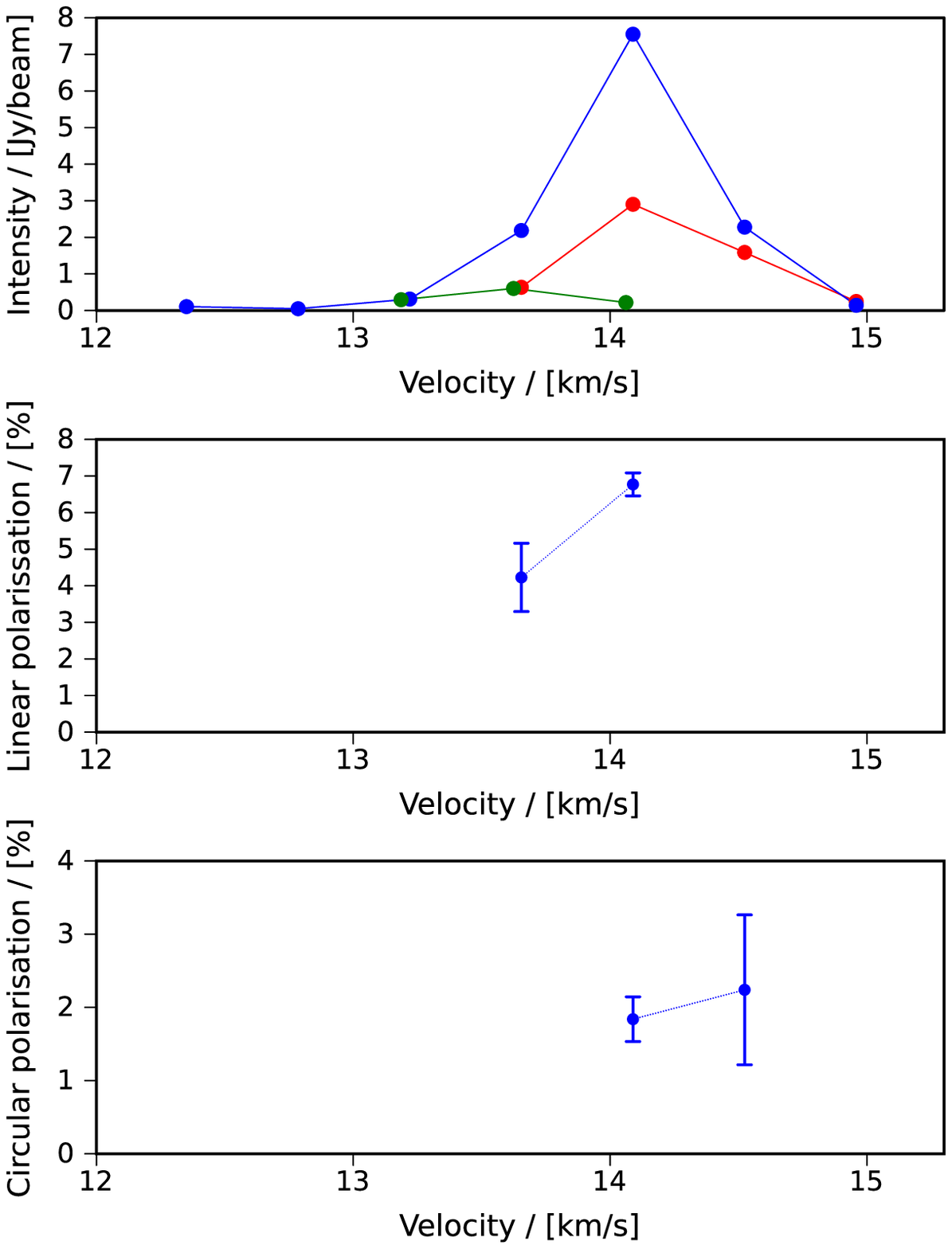}     
\end{center}
\caption[Emission properties of feature F6]
{{\footnotesize Emission proporties of feature F6, which is defined in Figure~\ref{fig:BR123-boxes}. \\ 
Top to bottom:
Overlaid contour plot of the v=1 J=1-0 (blue), v=2 J=1-0 (green) and v=1 J=2-1 (red) maser
emission, as for Figure~\ref{fig:F1};
Contour plot of the v=1 J=1-0 emission overlaid with linear polarisation EVPA vectors,
as for Figure~\ref{fig:F1}. 
Stokes $I$ intensity, percentage linear polarisation and percentage circular polarisation 
spectra for each of the spots in the feature, as for Figure~\ref{fig:F1}.
}}
\label{fig:F6}
\end{figure}

The features and their spectra are plotted in Figures~\ref{fig:F1} through \ref{fig:F6}. Where the features consist 
of multiple distinct maser spots, separated in position or frequency, the spots are labelled separately in the figures 
and separate spectra are plotted for each.

In these Figures, contour plots of overlapping v=1 \mbox{J=1-0} (blue), v=2 J=1-0 (green) and v=1 J=2-1 (red) Stokes $I$ emission 
are shown for each feature. The contours are drawn at levels $\{3\sigma_I, 5\sigma_I\}$, in terms of the broadened off-source noise 
limits described above. If contours for a particular transition are absent, the emission from that transition is weaker than the 
lowest contour. Associated EVPA plots overlaid on total intensity contour plots are also provided for those transitions with 
statistically-significant linear polarisation.

The accompanying Stokes $I$ spectra in these Figures were calculated from the peak Stokes $I$ pixel brightness values measured for 
each frequency channel across the feature. A threshold cutoff of three times the broadened noise in each channel was applied across 
the spectrum. The associated Stokes $Q$, $U$ and $V$ brightness values were measured at the pixel position of the Stokes $I$ maximum, 
and $m_l$ and $m_c$ computed accounting for statistical bias and the detection limits described above.
For features where statistically-significant linear or circular polarisation was measured, the percentage polarisation values are 
shown in separate spectra.

In all spectra, the v=2 J=1-0 frequency axes are shifted by two channels ($\sim0.86$~km s$^{-1}$) to account for an observed $\sim2$ 
channel frequency offset between the positions of maser features in the v=2 J=1-0 transition, and the v=1 J=1-0 and v=1 J=2-1 transitions.
As described in Paper I, the offset is likely due to errors in the assumed rest frequencies of these transitions.


\section{Discussion}
\label{sec:Discussion}

The foundational analysis of maser polarisation was provided by \citet{GKK} (GKK), who treated the maser emission 
semi-classically, with the molecules modeled in a quantum mechanical framework, and the radiation field modeled in 
a classical framework. The GKK polarisation solutions were derived in several limiting cases, defined by the relative 
values of the stimulated emission rate $R$ (s$^{-1}$), the decay rate $\Gamma$ (s$^{-1}$), the Zeeman splitting 
$g\Omega$ (Hz) and the spectral width of the line $\Delta \omega$ (Hz). Astrophysical SiO masers are in the 
weak-splitting regime $\Delta\omega \gg g\Omega$ \citep{Gray:12}.
As emphasized in a review by \citet{Watson:02}, the GKK solutions are an idealisation, as they deal with the specific 
case of a one-dimensional linear maser in a J=1-0 transition, weak continuum seed radiation, a constant magnetic field, 
$m$-isotropic pumping, a homogeneous environment, and derive the solutions at the line centre only. As further noted by 
\citet{Watson:02}, GKK solutions are not a continuous set of maser polarisation solutions over a range of maser 
intensities, or levels of saturation.
It is noted that the GKK solutions in the extreme limit of strong saturation are obtained by setting the derivatives 
of the Stokes parameters with respect to intensity (effectively equivalent to distance) to zero and solving the 
resulting algebraic equations for fractional polarisation.

Subsequent work extended the GKK solutions to higher-order $J$ transitions and to intermediate relative values of 
$(R, \Gamma, g\Omega, \Delta\omega)$, such as partial saturation. Additional effects such as $m$-anisotropic 
pumping and non-Zeeman mechanisms for producing circular polarisation have also been considered. In this paper we 
primarily consider the work by Elitzur \cite[][and references therein]{Elitzur:02} and by Watson and co-authors 
\cite[][and references therein]{Watson:02}. 
Elitzur has principally taken an analytic approach, finding stationary polarisation solutions. The work by Watson and 
collaborators makes use of numerical solutions of the polarised radiative transfer equations. A key foundational 
difference between the two approaches is that in the Watson approach the GKK solutions are considered applicable strictly 
under the asymptotic limits under which they were formulated \citep{Nedoluha:93}, as described above. In this view, the 
limiting solutions are not considered applicable to observational data \citep{Watson:09}. In the Elitzur approach, however, 
only stationary solutions are assumed to propagate, and the maser emission will rotate into the stationary solutions well 
before saturation \citep{Elitzur:96,Elitzur:02}.

Based on parameter estimates, circumstellar SiO maser emission most likely falls in the $g\Omega \gg R$ or $g\Omega > R$ 
regime with $R > \Gamma$ or $R \gg \Gamma$ \citep{Kemball:09,Watson:09,Assaf:13}; we consider this regime for the remainder 
of the paragraph. Here the GKK linear polarisation solutions have a position angle either parallel to or perpendicular to 
the projected magnetic field. The Elitzur model reproduces the GKK results in this regime \citep{Elitzur:91}.

Under the Watson model the linear polarisation solutions only asymptotically approach the GKK solution at high levels of 
saturation \citep{Western:84,Watson:01}. For high saturation the form of the linear polarisation as a function of the 
angle $\Theta$ between the magnetic field and the line of site is similar to the GKK solution, without the sharp cutoff 
at the $\Theta \approx 35^\circ$ break angle \citep[][GKK]{Watson:01}. In these solutions, if $g \Omega > R, \Gamma$ the 
EVPA will similarly be either parallel or perpendicular to the projected magnetic field \citep{Watson:02}.

In contrast, in the $g\Omega \simeq R$ regime, which is believed less likely based on parameter estimates, the linear 
polarisation position angle as a function of $\Theta$ will vary in form with intensity \citep{Nedoluha:90b}. At 
$g \Omega \sim R$ the fractional linear polarisation varies significantly with intensity \citep{Nedoluha:90a}.

Linear polarisation can also be created by $m$-anisotropic pumping of the masers, in the absence of a magnetic field in 
the medium \citep{Bujarrabal:81,Western:83c}, or in conjunction with a magnetic field \citep{Western:84,Nedoluha:90a}.

The primary cause of circular polarisation explored by the Elitzur model is standard Zeeman splitting in the presence of a 
magnetic field \citep{Elitzur:96}. The Watson models consider circular polarisation caused by standard Zeeman splitting, with 
modifications due to saturation effects \citep{Watson:01}, as well as non-Zeeman circular polarisation created by the the 
inter-conversion of linear to circular polarization in the intermediate intensity regime, due to intervening turbulent 
magnetic field directions or Faraday rotation \citep{Nedoluha:90b,Nedoluha:94,Wiebe:98}. 


\subsection{Faraday rotation}

Faraday rotation is a potential factor in any of the maser polarisation models, as Faraday rotation along the maser path may 
reduce the levels of integrated linear polarisation (GKK) and rotate the linear polarization EVPA \citep{Wallin:97}.

Faraday depolarisation becomes significant when the length of the region of plasma traversed by the radiation becomes close to 
the length scale $1.2\times10^{17} (\lambda^2 \; n \; B_{||})^{-1}$ (cm) corresponding to a Faraday rotation of $\pi$ radians, 
where $\lambda$ is the wavelength of the radiation (cm), $n$ is the electron density of the plasma (cm$^{-3}$) and $B_{||}$ is 
line of sight magnetic field (G) as derived using standard definitions of rotation measure \citep[e.g.][]{Draine:11}. The electron 
density in the SiO maser region of VY~CMa is unknown. 
Excess 8.4~GHz emission has been measured around VY CMa \citep{Knapp:95} and VLA continuum observations at 15, 22 and
43~GHz have been modelled as a radio photosphere extending out to 1.5-2~R$_\star$ \citep{Lipscy:05}. The extended atmospheres 
of late-type evolved stars are known to be complex in their layered structure, chemical composition, and kinematics 
\citep{Wittkowski:11,Ireland:11}, and similar complexity is likely in their ionization structure. An inner chromospheric 
ionized component is detected in the supergiant $\alpha$~Ori in UV spectroscopy, with a peak electron density 
n$_e \sim 10^8$~cm$^{-3}$ close to the photosphere and a low filling-factor at larger radii \citep{Harper:06}. The larger and 
cooler radio photosphere is believed ionized predominantly by photo-ionized metals, and lies predominantly interior to the 
SiO maser region \citep[][p.g. 149-245]{Reid:97,bk:Gustafsson}. The SiO masers in VY CMa lie at an approximate mean radius of
$\sim2.5$~R$_\star$. Number densities of order $\sim10^{10}$ cm$^{-3}$ and a mean temperature $1.4 \times 10^3$K are predicted 
at this radius by the AGB atmosphere models of \citet{Ireland:11}. A corresponding electron density estimate of order
$10^3$~cm$^{−3}$ is obtained at the SiO maser radius \citep{Assaf:13} using the ionization model of \citet{Reid:97}. This is 
consistent with an earlier independent estimate by \citet{Wallin:97}. A higher estimate of order
n$_e \sim 5 \times 10^5$~cm$^{-3}$ is obtained from the semi-empirical model of \citet{Harper:01} for the radio photosphere of 
$\alpha$~Ori if computed at the radius of the SiO maser emission. Significant uncertainties remain, specifically the fine-scale 
ionization conditions and spatial structure in the extended atmosphere, the relative abundance fractions of atomic or molecular 
hydrogen \citep{Glassgold:83,Wong:16}, and the neutral number density predictions of contemporary extended atmosphere models 
\citep{Wong:16}. 

If we assume a line of sight magnetic field in the range 0.5-1~G, somewhat higher than the mean level observed in the H$_2$O
maser region around AGB stars \citep{Vlemmings:07}, and an electron density of $10^3$~cm$^{-3}$, the Faraday depolarisation length 
scale is of order 2~R$_\star$ at 43~GHz and 9~R$_\star$ at 86~GHz, where R$_*$ is the stellar radius of VY CMa. Infrared-optical 
interferometric measurements of R$_*$ range from $9.88 \times 10^{11}$m \citep{Wittkowski:12} to $1.68 \times 10^{12}$m, the 
latter value derived from a stellar diameter measurement of 18.7 mas \citep{Monnier:04}; both  values of R$_*$ assume a distance 
of 1.2 kpc \citep{Zhang:12}. We conservatively adopt the larger value of R$_*$ in our order-of-magnitude estimates of depolarization 
length, but this choice does not affect our conclusion.
At the adopted electron density n$_e \sim 10^3$~cm$^{-3}$ Faraday depolarisation is not be a significant effect, even for the lower 
frequency 43~GHz \mbox{J=1-0} masers. However we note that this does not hold if electron densities approach the higher estimates 
of n$_e \sim 5 \times 10^5$~cm$^{-3}$. 

Several additional results support the conclusion of lower Faraday rotation. \citet{Wallin:97} performed numerical calculations of 
Faraday rotation in the weak-splitting regime, which showed that Faraday rotation in circumstellar SiO masers is not a dominant effect. 
\citet{Assaf:13} estimate a Faraday rotation of $\sim16^\circ$ for the \mbox{J=1-0} SiO masers toward R Cas. 
Faraday depolarisation of the maser emission would also result in higher levels of linear polarisation at higher $J$ levels 
\citep{Elitzur:91}; this pattern has not been unambiguously detected in prior single-dish studies (Section~\ref{section:poln_testing}, 
Test 1). Substantial Faraday depolarization would be accompanied by a significant rotation of the linear polarisation position 
angle, which is not observed in prior single-dish observations that suggest depolarization \citep{McIntosh:93}. Furthermore, 
numerous VLBI observations of SiO masers show the linear polarisation direction to be ordered, \citep[e.g.][]{Kemball:97,Cotton:06,Assaf:13}, 
with position angles predominantly tangential to the star, arguing against a large degree of Faraday rotation along the maser path.


\subsection{Observational tests}
\label{section:tests}

Six observational tests of the SiO maser polarisation models are discussed below, to be evaluated with the multi-transition SiO 
maser observations of VY~CMa presented in this paper. Tests that compare maser characteristics between different transitions are 
ideally performed using measurements of individual overlapping maser features, to ensure that the physical conditions of the 
masing gas are as similar as possible.

The tests are evaluated against observational data in the subsequent section, Section~\ref{section:poln_testing}. Several of the 
tests have been performed previously, using single dish and interferometric observations, and these prior results are discussed 
along with the application of the tests to the current observations.

\begin{itemize}
	\item[1.]
	\textit{Comparison of linearly-polarized intensity in the v=1 J=1-0 and J=2-1 transitions.}

	Under the Elitzur model, SiO maser transitions have spin-independent linear polarisation solutions \citep{Elitzur:91}.
	Under the Watson model the J=1-0 transition will have greater linearly-polarized intensity, if the transitions are 
	under comparable levels of saturation and degree of m-anisotropic pumping \citep{Western:84,Nedoluha:90a}. \\

	\item[2.] 
	\textit{Comparison of intensity and fractional linear polarisation.}

	The relationship between fractional linear polarisation and saturation level differs between maser 	polarisation models, 
	as described above, and observational evidence for the form of this relationship would provide a means to discriminate 
	between the models.  

	However, the saturation level depends on the unknown beaming angle as well as brightness, and the functional form of the 
	beaming angle varies with saturation \citep{bk:Elit}. In consequence, maser intensity is an imperfect proxy for the unknown 
	saturation level. Nonetheless, with this important disclaimer, the relationship between linear polarisation and intensity 
	is a potential 	diagnostic indicator of the relationship between linear polarisation and saturation.

	In the Elitzur model the polarisation solution is not dependent on the saturation level of the masers, as long as the maser 
	emission has evolved into the stationary solution \citep{Elitzur:91}. In the Watson model the fractional linear polarisation 
	level increases slowly with saturation, 	only asymptotically approaching the GKK solution \citep{Western:84,Nedoluha:90a}.
	For $g \Omega \gg R$, a correlation of fractional fractional linear polarisation with saturation level would therefore be 
	evidence for this model.

	Near $g \Omega \simeq R$, which is believed less physically likely as discussed above, 	fractional linear polarisation may 
	decrease with increasing maser saturation \citep{Nedoluha:90a}. In this regime, the addition of m-anisotropic pumping leads 
	to a reduction of fractional linear polarisation at higher saturation levels for the v=1 J=2-1 line than for the v=1 J=1-0 
	line \citep{Nedoluha:90a}. \\

	\item[3.]
	\textit{Comparison of fractional linear polarisation with distance from the star.}

	Circumstellar maser emission that is anisotropically pumped by stellar radiation will be strongest closest to the star 
	\citep{Western:83c,Desmurs:00}. The fractional linear polarisation created by anisotropic pumping is therefore expected to 
	be strongest closest 	to the star, where the anisotropy parameter is largest \citep{Kemball:09}. 	However, this trend 
	may also be influenced by shock compression of the magnetic field in the inner layers of the near-circumstellar envelope 
	\citep{Kemball:09}. \\

	\item[4.]
	\textit{Electric vector position angle rotation.}

	Circumstellar SiO masers often show 90$^\circ$ EVPA rotations across a single maser feature \citep[e.g.][]{Kemball:97}. 
	One natural explanation is a transition over the feature across the critical $55^\circ$ angle between the magnetic field 
	and the line of sight in the regime $g \Omega > R, \Gamma$, where the direction of the EVPA is predicted to change from
	parallel to perpendicular to the projected magnetic field direction \citep[GKK;][]{Elitzur:02,Watson:02}.

	Under the Watson model, EVPA rotation can also occur in the $g\Omega \simeq R$ regime, with degree of rotation dependent on 
	saturation \citep{Nedoluha:94}. 	This EVPA rotation is at most $\sim45^\circ$ over an order of magnitude in saturation
	level, except for very large magnetic fields ($100$G) directed almost perpendicular to the line of sight ($75^\circ$). It 
	is therefore unlikely that abrupt $\sim90^\circ$ changes in EVPA are caused by this mechanism.

	Linear polarisation EVPA rotation can also be explained by $m$-anisotropic radiative pumping, over a change in anisotropy 
	conditions \citep{Western:83c,Asensio_Ramos:05}. A change in the dominant anisotropy direction from radial to tangential 
	could result in a 	90$^\circ$ polarisation EVPA flip, but only if the magnetic field is not dynamically significant 
	\citep{Asensio_Ramos:05}. 	In the presence of a magnetic field of order 10-100 mG EVPA rotation of about 45$^\circ$ can occur, 
	however with considerable suppression of the masing effect \citep{Asensio_Ramos:05}.

	GKK-style EVPA flips can be modeled in features that contain a linear polarization EVPA position angle transition of 
	90$^\circ$ by modeling the fractional linear polarization and its dependence on the variation of the angle $\Theta$ between 
	the magnetic field and the line of sight $m_l(\Theta)$, along the maser feature \citep{Kemball:11b}. If the functional form 
	of $m_l(\Theta)$ across a 90$^\circ$ EVPA flip is well-described by the GKK model, this is more supportive of the Elitzur 
	models, because under the Watson model the fractional linear polarisation solutions approach the asymptotic GKK solution 
	only at very high levels of saturation \citep{Western:84,Watson:01}. \\

	\item[5.]
	\textit{Comparison of circular polarisation in the v=1 J=1-0 and J=2-1 transitions.}

	The ratio of standard Zeeman splitting in the presence of a magnetic field to the Doppler line width is proportional to the 
	wavelength of the transition \citep{Elitzur:96}. Standard Zeeman circular polarisation for the J=1-0 transition at 43~GHz 
	should therefore be double that of the J=2-1 transition at 86~GHz, all other effects being equal.

	The standard Zeeman circular polarisation can be increased by a factor of a few due to saturation under the Watson model 
	\citep{Watson:01}. Saturation has no effect on the predicted circular polarisation under the Elitzur model, so long as 
	J/J$_s \; > \frac{3}{4}$, where J is the maser intensity, and J$_s$ the saturation intensity \citep{Elitzur:96}.

	Under the Watson model, non-Zeeman circular polarisation can be created by a change in direction of linear polarisation. 
	This can occur when the maser emission falls in the $g\Omega \sim R$ regime, when the magnetic field direction changes 
	along the line of sight, or when Faraday rotation is significant \citep{Watson:09}.

	These non-Zeeman circular polarization mechanisms do not depend strongly on the angular momentum level of the transition. \\

	\item[6.]
	\textit{Correlation between circular and linear polarisation.}

	If the circular polarisation is caused by the non-Zeeman effects described in the previous sub-section, then the level of 
	circular polarisation will be correlated with the level of linear polarisation \citep{Watson:09}. This correlation can be 
	destroyed by statistical variations in the emission region, so the absence of this correlation is not necessarily evidence 
	against non-Zeeman circular polarisation \citep{Wiebe:98}. However, if many maser features display circular polarisation at 
	a level much greater than the average value of $m_l^2/4$, then the circular polarisation is unlikely to be caused by non-Zeeman 
	effects \citep{Wiebe:98}. 

\end{itemize}


\subsection{Observational test evaluation}
\label{section:poln_testing}

\begin{itemize}
	\item[1.]
	\textit{Comparison of linear polarisation in the v=1 J=1-0 and J=2-1 transitions.}

	Single-dish observations over a sample of sources by \citet{Barvainis:85} show strongly-correlated and comparable 
	spectrum-averaged fractional linear polarisation $\overline{m_l}_s$ in the v=1, J=1-0 and v=2, J=2-1 transitions. Similarly, 
	single-dish observations of these two transitions toward VY~CMa \citep{McIntosh:94} show broadly-comparable, or no clear 
	trend, in values for fractional linear polarisation averaged in velocity over coincident spectral features 	$\overline{m_l}_f$. 
	In contrast, a single-dish survey of late-type evolved stars in transitions v=1 J=1-0, v=1 J=2-1, and 	v=1, J=3-2 by \citet{McIntosh:91} 
	is consistent with increasing $\overline{m_l}_f$ in higher rotational transitions. Similar 	measurements in these three transitions 
	for features in the spectrum of Mira \citep{McIntosh:93} also generally favor a lower relative value $\overline{m_l}_f$ for the 
	J=1-0 transition. Collectively, these prior results do not allow an unambiguous rank ordering. 

	These previous observations were all performed with single dish telescopes, so spatial blending is a possible source of significant 
	systematic error, as noted by the authors. Ideally, the linear polarisation should be compared at the component-level in features 
	with spatially-coincident emission. In the current interferometric study, the only two features meeting these conditions that 
	showed statistically significant linear polarisation in both v=1, J=1-0 and v=1, J=2-1 transitions are feature F1 (Figure~\ref{fig:F1}) 
	and feature F4 	(Figure~\ref{fig:F4}). 

	In feature F1 the linear polarisation is considerably greater in the v=1 J=2-1 line, with an average percentage linear polarisation 
	of $32\%$ over the feature. The corresponding average in the v=1 J=1-0 line is $6\%$.

	In feature F4, only spot S1 shows significant linear polarisation in the v=1 J=1-0 line. In this spot, the linear polarisation level 
	is similar for both transitions. Significant linear polarisation is only measured in two frequency channels across the J=1-0 feature, 
	and three channels across the J=2-1 line. In both cases the peak linear polarisation is 8$\%$, measured at the brightest pixel of spot 
	S1 in the peak Stokes $I$ channel.

	For features F2, F3, F6 and the other F4 spots, linear polarisation is detected in the v=1 J=1-0 line, but not in the v=1 J=2-1 line, 
	possibly due to lower SNR. This was investigated by assuming that the v=1 J=2-1 masers are linearly polarised at the level of the v=1 
	J=1-0 masers, and checking if the J=2-1 linear polarisation would lie above the detection limits defined in Section~\ref{sec:result_features} 
	(and \textit{vice versa}). For F2, F5 and F4 S3, linear polarisation at the level of the v=1 J=1-0 feature would have been detected in 
	the J=2-1 line at the sensitivity of the current study. However, in features F3 and F4 S4 it would not have been detected.


	\begin{table}
	{\small
	\begin{center}
	\begin{tabular}{|lcccc|}
	\hline 
	Feature, & Ordinal relation in m$_l$  & $\overline{\mbox{m}}_l$ & $\overline{\mbox{m}}_l$ & Note \\
	spot && J=1-0 & J=2-1 & \\
	\hline
	\hline
	F1 & J=2-1 $>$ J=1-0 & 6.3$\%$ & 32.0$\%$ & \\
	F2 & J=1-0 $>$ J=2-1 & 16.4$\%$ & & $\star$ \\
	F3 & & 5.1$\%$ & &\\
	F4 S1 & J=1-0 $\sim$ J=2-1 & 5.8$\%$ & 6.8$\%$ & \\
	F4 S3 & J=2-1 $>$ J=1-0 & & 8.4$\%$ & $\star$ \\
	F4 S4 & & & 4.9$\%$ & \\
	F5 & J=1-0 $>$ J=2-1 & 8.7$\%$ & & $\star$\\
	F6 & & 5.5$\%$ & & \\
	\hline
	\end{tabular}
	\end{center}
	\caption[Fractional linear polarisation in v=1 J=1-0 and v=1 J=2-1]
	{\small Fractional linear polarisation comparison between the v=1 J=1-0 line and v=1 J=2-1 line.} 
	{\small $^\star$ In these cases, linear polarisation is only detected for the more linearly-polarised line.
	However, equal or greater fractional linear polarisation for the missing line would have been detected within the sensitivity 
	of the current observations.
	}
	\label{table:ml-comparison}
	}
	\end{table}


	A summary of the these linear polarisation results is given in Table~\ref{table:ml-comparison}. The average fractional linear 
	polarisation values $\overline{\mbox{m}}_l$ shown are computed over all detected linear polarisation values in the line or 
	feature, without any requirement for multi-transition detection. However, for features F1 and F4 S1 the values listed in the 
	table are the averages over the channel range where linear polarisation is detected in both lines.

	The average linear polarisation across all features in the source with linear polarisation detections, as enumerated in Appendix A, 
	is 13.7$\%$ and 7.7$\%$  for the 43~GHz v=1 J=1-0 and v=2 J=1-0 emission, and 15.0$\%$ for the 86~GHz v=1 J=2-1 emission. 

	The results in Table~\ref{table:ml-comparison} show that more cospatial components are needed than are detected in both rotational 
	transitions before a firm conclusion can be drawn regarding the relative magnitude of $\overline{m}_l$. However on the basis of the 
	net ordinal relation in Table~\ref{table:ml-comparison}, and secondarily the mean $\overline{m}_l$ over all features in Appendix A, 
	the current results are more consistent with the conclusion of comparable fractional linear polarization for both rotational transitions.

	Although component-level comparisons allow a more precise test of the dependence of fractional linear polarisation on rotational 
	transition than is possible in single-dish studies, even at VLBI resolution the physical conditions probed may not be exactly identical 
	in both transitions. For example, the large fractional linear polarisation difference in feature F1 could be explained through a small 
	positional offset between the maser emission from the J=1-0 and J=2-1 transitions, translating to a small difference in angle $\Theta$ 
	between the magnetic field and the line of sight. Different $\Theta$ implies a different fractional linear polarisation between the two 
	transitions in the GKK model for $g \Omega \gg R \gg \Gamma$, as plotted in Figure~\ref{fig:gkk_ml}. The offset in F1 fractional linear 
	polarisation between the v=1 J=1-0 emission (6.3$\%$ on average) and the v=1 J=2-1 emission (32.0$\%$ on average) can be accounted for 
	by the gradient over the region $Q/I \simeq -0.05$ to $-0.3$ which corresponds to a change in angle $\Theta < 10^\circ$.
	Alternatively, field line curvature along the path length of the maser emission may reduce the observed linear polarisation. If the path 
	length of the J=2-1 maser is a limited fraction of the path length of the J=1-0 maser, the effect of field line curvature will be 
	diminished for the J=2-1 maser, possibly explaining the higher J=2-1 polarisation.

	Faraday depolarization predicts lower fractional linear polarization in J=1-0 accompanied by a large linear polarization position 
	angle rotation between the two transitions; this EVPA rotation is not generally observed in the current data
	(Section~\ref{section:poln_testing}, Test 4). \\


	\begin{figure}
	\begin{center}
	  \includegraphics[width=3.3in, angle=0]{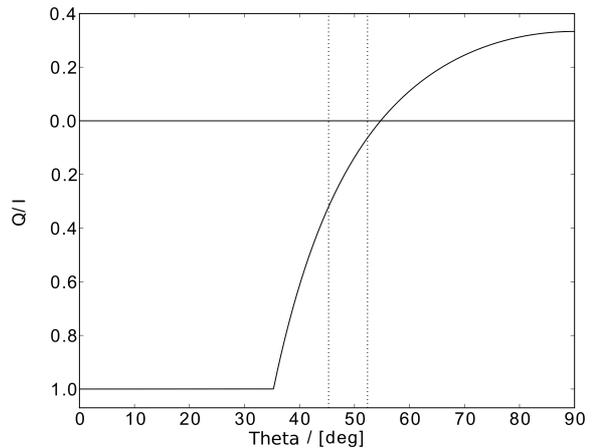}
	  \caption[The GKK fractional linear polarisation solution]
	  {{\small Plot of the GKK fractional linear polarisation solution.
	  Vertical lines are plotted through $Q/I = -0.32$ and $-0.063$ (see text for discussion).
	  }}
	  \label{fig:gkk_ml}
	\end{center}
	\end{figure}


	\item[2.] 
	\textit{Comparison of saturation and linear polarisation.}

	In Figure~\ref{fig:ml_I} the fractional linear polarisation is plotted against the total intensity for each of the maser features with 
	statistically significant linear polarisation, separately for each transition. The plots show a general trend of higher linear 
	polarisation for the weaker maser emission, particularly for the v=1 J=1-0 emission, for which the largest number of maser features 
	were detected. 


	\begin{figure*}
	\begin{center}
	  \includegraphics[width=6.5in, angle=0]{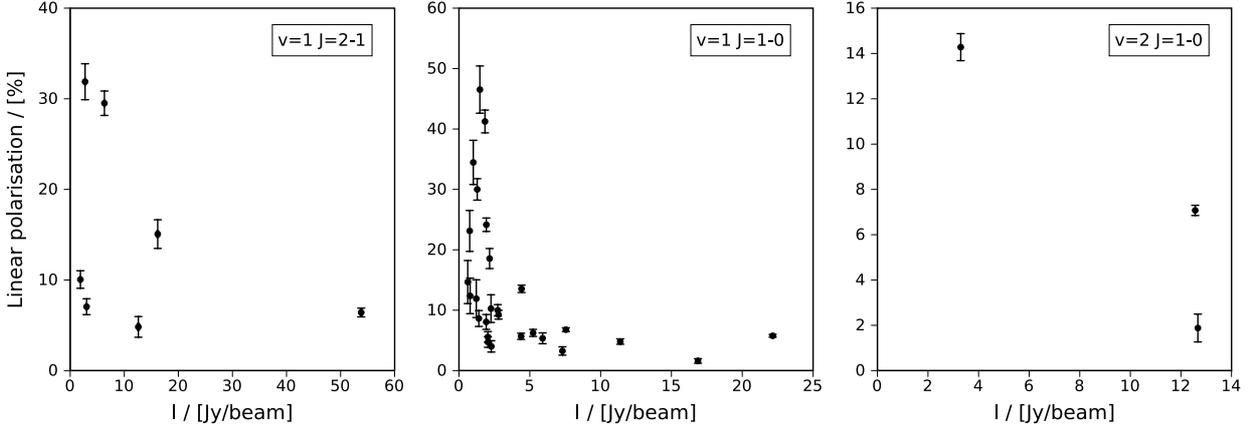}
	  \caption[Linear polarisation versus total intensity]
	  {{\small Plots of fractional linear polarisation versus total intensity, for the v=1 J=2-1 (left), v=1 J=1-0
	  (centre), and v=2 J=1-0 (right) maser features.
	  }}
	  \label{fig:ml_I}
	\end{center}
	\end{figure*}


	Trends of higher fractional linear polarisation for weaker SiO masers have previously been observed in the late-type evolved stars 
	R~Aquarii \citep{Allen:89,Hall:90,Boboltz:thesis}, R~Cassiopeia \citep{McIntosh:89,Assaf:13} and R~Leo \citep{Clark:84}.

	A trend of decreasing fractional linear polarisation with saturation is at odds with the predictions of both Watson and Elitzur models 
	in the regime $g \Omega \gg R \gg \Gamma$. However, in addition to the caveat noted above regarding the use of intensity as a proxy 
	for saturation, there are very likely to be additional variables in play. We note that the observed trend of higher linear polarisation 
	for the weaker masers has been previously hypothesized to result from relative saturation and m-anisotropic pumping near the regime 
	$g \Omega \simeq R$ \citep{Nedoluha:90a}. As discussed above, this regime is believed less likely to be applicable to circumstellar SiO 
	masers based on current parameter estimates. We also note that \citet{McIntosh:89} argue that stronger maser emission arises out of 
	longer maser path lengths, these components may suffer the greatest levels of Faraday depolarisation leading to the observed trend in 
	$m_l(I)$. However, for reasons discussed in earlier sections, there is no strong evidence of dominant Faraday depolarization effects 
	in the SiO maser region.


	\begin{figure}
	\begin{center}
	  \includegraphics[width=3.8in, angle=0]{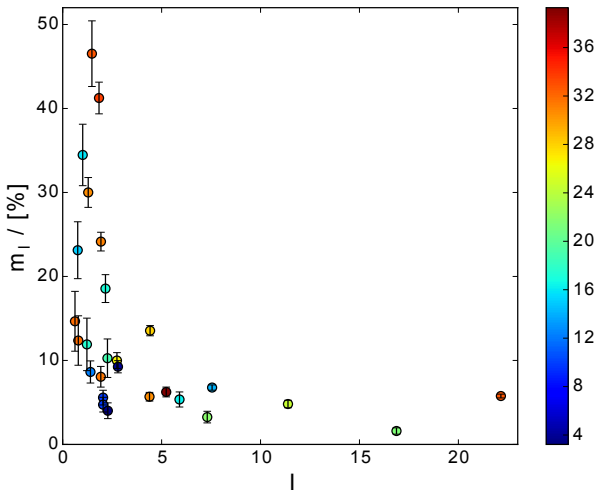}
	  \caption[Linear polarisation versus total intensity]
	  {{\small Plot of fractional linear polarisation versus total intensity, for the v=1 J=1-0
	  maser features. The colour scale is line-of-sight velocity in the LSR frame, in km/s, as shown in 
	  the color bar on the right. The adopted systemic LSR velocity for VY~CMa is +18~km~s$^{-1}$.
	  }}
	  \label{fig:ml_I_v}
	\end{center}
	\end{figure}


	\citet{Assaf:13} report higher fractional linear polarization in the inner shell of R Cas; by projection arguments the authors note 
	that inner-shell features are more likely to be at extreme velocities in the spectrum. In Figure~\ref{fig:ml_I_v} we plot the linear 
	polarization percentage against total intensity for all detected v=1, J=1-0 maser features, colour-coded by LSR velocity. As noted 
	earlier, a systemic stellar velocity $V_* = +18$ km s$^{-1}$ has been adopted for VY CMa in the current work. Figure~\ref{fig:ml_I_v} 
	shows a tendency of weak-$I$ high-m$_l$ features to fall further from the stellar velocity and \textit{vice versa} for high-$I$ weak-m$_l$ 
	features. Note that the highest $I$ feature on the figure, with m$_l~6\%$, is the unusual feature F1.

	In the model of tangential amplification for circumstellar SiO masers \citep[e.g.][]{Diamond:94} longer coherent path lengths and higher 
	intensities occur closer to the systemic stellar velocity $V_*$; to first order, maser features near this velocity are expected to lie 
	closer to the plane of the sky. If the magnetic field structure is such that masers at velocities $V$ further from the systemic velocity 
	are more likely to have smaller angles $\theta$ between the magnetic field and line-of-sight then under the GKK model for $m_l(\theta)$ 
	for $g \Omega \gg R \gg \Gamma$ (Figure~\ref{fig:gkk_ml}), or related functional forms for lower saturation in the models of \citet{Watson:01},
	a possible explanation is provided for the trends in $m_l(I)$ and $m_l(|V-V_*|)$ in Figure~\ref{fig:ml_I_v}. This $\theta(V)$ dependence 
	arises naturally with a radial magnetic field \citep{Assaf:13}, but may also arise from other local or global morphologies. These effects 
	may also be enhanced by lower gradients in magnetic field along the maser path for smaller $|V-V_*|$. \\


	\item[3.]
	\textit{Comparison of linear polarisation with distance from the star.}

	Time-series VLBA images of polarised SiO maser emission towards TX Cam and R Cas show strongest linearly-polarised intensity 
	\citep{Kemball:09} and fractional linear polarisation \citep{Assaf:13} respectively at the inner boundary of the projected maser 
	shell.

	Figure~\ref{fig:linpol-dist} shows the component-level fractional linear polarisation plotted against projected distance from the 
	assumed stellar position, for the three transitions observed in the current paper. The stellar position is unknown, but we adopt the 
	zeroth-order assumption that the central star is most likely located toward the centroid of the inner shell of SiO maser features. 
	The position was estimated through a grid search of the inner 30~mas of the image, to find the the position which maximises the minimum 
	projected distance of the candidate centroid position to the closest maser feature. The feature positions in Appendix~\ref{appendixA} 
	were used in this minimisation. A circle representing the star is shown on Figure~\ref{fig:BR123-boxes}, centred on the stellar 
	position estimated using this method, and adopting a diameter of 18.7~mas from \citet{Monnier:04}.


	\begin{figure}
	\begin{center}
	  \includegraphics[height=2.5in, angle=0]{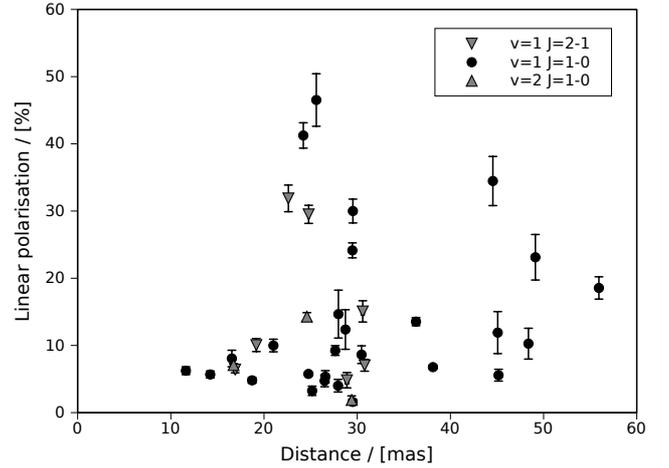}
	  \caption[Percentage linear polarisation versus projected radial distance]
	  {{\small Percentage linear polarisation versus projected radial distance from the assumed stellar position,
	  for the v=1 J=1-0, v=2 J=1-0 and v=1 J=2-1 maser features.
	  }}
	  \label{fig:linpol-dist}
	\end{center}
	\end{figure}


	\citet{Zhang:12} determined the VY~CMa stellar position relative to VLBA observations of the SiO masers through VLA observations of 
	the radio photosphere. The stellar position assumed in this paper is coincident with the \citet{Zhang:12} position within the 10~mas 
	uncertainty cited by these authors.

	There is no visible trend in Figure~\ref{fig:linpol-dist} of higher fractional linear polarisation closer to the star. However, we 
	stress that VY CMa is a supergiant, with a complex circumstellar environment and likely asymmetric mass loss (described in greater 
	detail in Paper I), so the absence of this correlation does not exclude anisotropic pumping. \\


	\item[4.]
	\textit{Electric vector position angle rotation.}

	Analysis of individual SiO maser features with 90$^\circ$ EVPA rotations were performed by \citet{Kemball:11b} and \citet{Assaf:13} 
	who found that the EVPA rotation and percentage linear polarisation were consistent with the GKK linear polarisation solution for 
	$g \Omega \gg R \gg \Gamma$. This solution for $m_l(\theta)$ is shown in Figure~\ref{fig:gkk_ml}, where $\theta$ is the angle between 
	the magnetic field and the line-of-sight.

	In the current data, features F1 and F2 are candidates for a similar analysis. \\ 


	\textit{Feature F2}

	Following \citet{Kemball:11b}, a fit was performed for the fractional linear polarisation in v=1, J=1-0 across feature F2 
	(Figure~\ref{fig:F2}), modeling the angle $\Theta$ as a second order polynomial in projected angular distance along the feature, 
	\begin{equation}
	\Theta = p \; (d - d_{55})^2 + q \; (d - d_{55}) + 55
	\label{eqn:theta_eqn}
	\end{equation}
	where $\Theta$ is in units of degrees and $d_{55}$ is the projected angular position of the minimum in the polarised emission, at 
	the mid-point of the $90^\circ$ EVPA rotation. The projected angular distances were measured radially from the assumed stellar position. 
	The measured fractional linear polarisation values were jointly fit to Equation~\ref{eqn:theta_eqn} and the GKK linear polarisation
	solution described earlier in this section, using a $\chi^2$-fit for parameters $p$ and $q$. The results of the fit for feature F2 are 
	shown in Figure~\ref{fig:thetafit}. 


	\begin{figure}
	\begin{center}
	  \includegraphics[height=2.6in, angle=0]{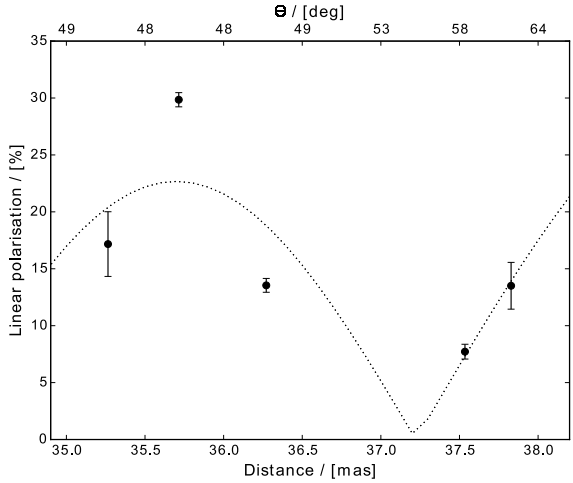}
	  \caption{{\small Fractional linear polarisation fit for feature F2 against the GKK model. 
	  The plot shows percentage linear polarisation versus projected angular distance from the assumed stellar position
	  (bottom) and fitted angle $\Theta$ between the magnetic field and the line of sight (top).
	  }}
	  \label{fig:thetafit}
	\end{center}
	\end{figure}


	The fit is poor relative to that shown in \citet{Kemball:11b} but it is not inconsistent with the GKK linear polarisation solution given 
	the uncertain $\Theta(d)$ relationship. Higher-order polynomial models could be used to provide a better fit to the data, but this is 
	not warranted given the small number of data points across the feature.

	If we assume that the $90^\circ$ linear polarisation flip of feature F2 is produced by a transition through the critical angle, it is 
	consistent with both the Elitzur model and the Watson model (in regime $g\Omega > R, \Gamma$ \citep{Watson:02}).

	Anisotropic pumping is not a likely explanation for the EVPA rotation in this feature. The magnitude of the rotation is too large to be 
	caused by the anisotropic pumping in the presence of a magnetic field as described by \citet{Asensio_Ramos:05}. Following similar arguments
	presented in \citet{Kemball:11b} we believe that it is less likely that an EVPA rotation of this magnitude results from a change in 
	anisotropy or pumping conditions \citep{Western:83c,Asensio_Ramos:05}. In addition the EVPA is radially directed closest to the star in 
	this feature, which further does not support this hypothesis. Feature F2 is also not close to the star, so a large change in anisotropy 
	parameter over the length of the maser feature is less likely. \\


	\textit{Feature F1}

	The elongated feature F1 (Figure~\ref{fig:F1}) is a second candidate feature for a $90^\circ$ EVPA rotation analysis, but has complex 
	structure.


	\begin{figure}
	\begin{center}
	  \includegraphics[width=3.2in, angle=0]{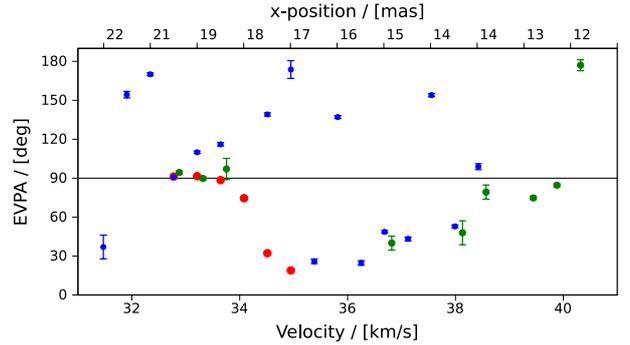}
	  \caption[Linear polarisation electric vector position angles for feature F1]
	  {{\small Linear polarisation EVPAs for feature F1,      
	  for the v=1 J=1-0 (blue), v=2 J=1-0 (green) and v=1 J=2-1 (red) SiO maser emission.
	  The lower x-axis shows the channel LSR velocity, and the upper x-axis shows the x-position where the v=1 J=1-0 Stokes 
	  parameters were measured, at each velocity channel across the feature. 
	  The formal statistical errors are shown for the v=1 J=1-0 and v=2 J=1-0 emission, and the error bars are 
	  smaller than the data points in some cases. Additional systematic errors are estimated to be $\le5\%$,
	  as discussed in the text.    
	  }}
	  \label{fig:F1chi}
	\end{center}
	\end{figure}


	The linear polarization EVPA variation across the feature is shown in Figure~\ref{fig:F1chi}. The two-channel shift discussed in 
	Section~\ref{sec:detailed_features} has been applied to the v=2 J=1-0 data in the Figure. The position angles have a 180$^\circ$ 
	ambiguity, so they were all rotated by integral multiples of 180$^\circ$ to fall within $[0^\circ,180^\circ]$. Error bars are 
	omitted from the v=1 \mbox{J=2-1} data points to indicate that the absolute values of these angles are uncertain to within a single 
	unknown additive constant offset in the 86 GHz band, as discussed in Section~\ref{sec:Observations}. This offset can be selected 
	prudently if it can be physically justified through, for example, alignment with the EVPAs of SiO maser emission from other 
	transitions. In the overlapping region of F1 the EVPAs of all three transitions agree relatively closely, within $\le20-30^\circ$ 
	(Figure~\ref{fig:F1chi}). This suggests that the v=1 J=2-1 EVPAs are close to their absolute values, within $\sim20-30^\circ$.

	The v=1 J=1-0 and v=2 J=1-0 error bars in Figure~\ref{fig:F1chi} were determined through propagation of the absolute error in the 
	EVPA calibration transfer and the Stokes parameter errors of each maser component, listed in Appendix~\ref{appendixA}. These formal 
	statistical errors do not include systematic errors due to the different angular scales sampled by the VLBA and VLA during the 
	absolute EVPA calibration transfer, or systematic errors arising from source variability during the time delay between the VLA and 
	VLBA observations. The VLBA and VLA observations were separated by 2 days (Section~\ref{sec:Observations}). The systematic errors 
	are estimated to be $\le5\%$. \\

	\textit{Feature F1: v=1 J=2-1}

	An EVPA rotation of $\sim 90^\circ$ is evident in Figure~\ref{fig:F1chi} for the v=1, J=2-1 data near an LSR velocity of 34~km~s$^{-1}$. 
	However, The rotation is not as abrupt as for feature F2 and neither is it accompanied by the minimum in $m_l(\Theta)$ predicted by the 
	GKK solution. \\

	\textit{Feature F1: v=1 J=1-0}

	The v=1 J=1-0 emission in feature F1 displays what appear to be multiple $\sim90^\circ$ EVPA rotations across the length of the feature. 

	These flips may be caused by multiple crossings of the $55^\circ$ critical angle, if the feature is elongated along a magnetic field 
	direction oriented close to the critical angle. The v=1 J=1-0 fractional linear polarisation of this feature is $\lesssim20\%$ 
	(Figure~\ref{fig:F1}). According to the GKK linear polarisation solution (Figure~\ref{fig:gkk_ml}), linear polarisation of less than 
	$20\%$ arises over a range of angles $\Theta = 48^\circ$ to $66^\circ$. The three-dimensional position of the feature in the 
	circumstellar envelope is unknown, but for tangential amplification in an accelerating shell we would expect maser emission arising
	further from the stellar velocity to arise in regions of gas moving at smaller angles to the line of sight. The line of sight velocity 
	of this feature is redshifted by $\sim 11-21$km/s relative to the stellar velocity, so it is possible that this feature is elongated 
	along an axis oriented near $55^\circ$ to the line of sight, possibly along a local magnetic field direction.

	Elongation of a feature in the direction of the magnetic field could be caused by ionised gas dragging the magnetic field along the 
	direction of outflow, or a stronger magnetic field may constrain the ionised gas to move along the field lines 
	\citep{Vlemmings:05,Cotton:06}. Circumstellar maser images often show radially extended features with polarisation position angles either
	parallel or perpendicular to the radial direction, which have been explained by such alignment with magnetic field lines 
	\citep{Cotton:06,Kemball:09}.

	The multiple $\sim90^\circ$ rotations visible in the v=1 J=1-0 emission may alternatively be caused by a helical magnetic field threading 
	the elongated maser feature. In this geometry the EVPA would rotate by 180$^\circ$ through the coils of the helix, and there is no need 
	to invoke a transition through the 55$^\circ$ critical angle to explain 90$^\circ$ flips. \\

	The similarity of the total intensity spectral shape in the v=1 J=1-0 and v=1 J=2-1 transitions across F1 shown in Figure~\ref{fig:F1} 
	suggests strorngly that the emission from these transitions arises from the same physical conditions.
	However, the measured EVPA values are integrated along the three-dimensional coherent path length of the maser emission, and the maser 
	excitation conditions and SiO density may vary locally in detail across this region. The measured EVPAs are also spatially filtered by 
	the different surface brightness sensitivity of the 43 GHz and 86 GHz arrays. These effects will introduce some level of variance in the 
	emission from different transitions across the feature, even for physically-coincident maser components. In this context, the v=1 
	\mbox{J=1-0} and v=1 J=2-1 fractional linear polarisation difference may be explained by modest differences in $\Theta$ between these two 
	transitions, as considered earlier, in Test~1. \\


	\item[5.]
	\textit{Comparison of circular polarisation in the v=1 J=1-0 and J=2-1 transitions.}

	A prior comparison of circular polarisation of SiO masers at v=1 J=1-0 and v=1 J=2-1 was performed with a single dish telescope, towards 
	VY~CMa, by \citet{McIntosh:94}. The circular polarisation was measured for four velocity features in the spectrum. For each velocity feature 
	the v=1 J=1-0 circular polarisation was significantly higher than that in the v=1 J=2-1 transition. In one case the v=1 J=1-0 circular 
	polarisation was measured to be double that of the v=1 J=2-1 circular polarisation, consistent with standard Zeeman splitting, as noted above.
	As single-dish observations, these prior results are subject to significant but unknown systematic errors arising from spatial blending of 
	individual SiO maser components. Our current study attempts to make this comparison for individual spatially-resolved SiO maser features. 

	The circular polarisation was compared for the six overlapping features in the current interferometric component-level study. Of the 
	overlapping features, only F1 and F4 show significant circular polarisation in both the v=1 J=1-0 and v=1 J=2-1 transitions.

	In feature F1, only a single channel displays statistically significant circular polarisation in both transitions: $-4.61\pm0.30\%$ for 
	v=1 J=1-0, and $-4.31\pm2.06\%$ for v=1 J=2-1. Standard Zeeman splitting cannot be ruled out for this feature due to the the large 
	uncertainty of the v=1 \mbox{J=2-1} circular polarisation measurement.

	In the Watson model, saturation effects can increase the standard Zeeman circular polarisation by factors of a few \citep{Watson:01} so 
	this result could be explained in this model by more highly saturated v=1 J=2-1 emission. This is possible, but not likely, due to the lower 
	intensity of the v=1 J=2-1 line evident in Figure~\ref{fig:F1}.

	Feature F4 is a group of four spots spanning almost 10~km/s (Figure~\ref{fig:F4}). Only spot S1 displays significant circular polarisation 
	in both the v=1 J=1-0 and v=1 J=2-1 lines. The measured fractional circular polarisation of S1 is completely different for the two lines, 
	with a maximum of $9.22\pm1.31\%$ in the v=1 J=1-0 line, and $-0.84\pm0.32\%$ in the v=1 J=2-1 line. Closer inspection of the feature shows 
	that the location of the circular polarisation peak of spot S1 is offset between the two lines. The circular polarisation measurements are 
	therefore unlikely to be probing the same region of gas, so this component is not included in the current test.

	The average fractional circular polarisation magnitude across all of the circularly polarised features in Appendix~\ref{appendixA} is 3.0$\%$
	for the v=1 J=1-0 line, 4.2$\%$ for the v=2 J=1-0 line, and 2.6$\%$ for the v=1 J=2-1 line. We note however that this is not a comparison between 
	individual coincident components, so is not highly dispositive. \\


	\item[6.]
	\textit{Correlation between circular and linear polarisation.}

	Evidence of a circular-linear polarisation correlation has been sought, but not detected, in single dish spectra of VY~CMa \citep{McIntosh:94} 
	as well as VLBA images of v=1 J=1-0 SiO maser emission towards R~Aquarii \citep{Boboltz:thesis} and TX~Cam \citep{Kemball:97}. In contrast, 
	\citet{Herpin:06} do report a preliminary correlation between integrated circular and linear polarization in a single-dish survey of late-type 
	evolved stars.

	Figure~\ref{fig:circ-lin} shows circular polarisation percentages versus linear polarisation percentages for maser features with statistically 
	significant linear and circular polarisation. There is no observed correlation between the fractional circular and linear polarisation values.

	There are a total of 37 maser features that display statistically significant linear polarisation in the v=1 J=1-0, v=2 J=1-0 and v=1 J=2-1 
	feature lists in Appendix~\ref{appendixA}. Of those 37 features, 14 display statistically significant circular polarisation. All of the 14 
	circularly-polarised features are circularly polarised at a level $>m_l^2/4$ \citep{Wiebe:98}. Another 12 features display circular polarisation 
	without significant linear polarisation. 
	This strongly suggests that the circular polarisation does not arise from non-Zeeman effects, which are described in Section~\ref{section:tests} 
	Test 6. \citet{Cotton:11} report a similar result from polarised VLBA observations of the AGB star IK~Taurii in v=1 J=1-0 and v=2 J=1-0 SiO maser 
	emission.


	\begin{figure}
	\begin{center}
	  \includegraphics[width=3.2in, angle=0]{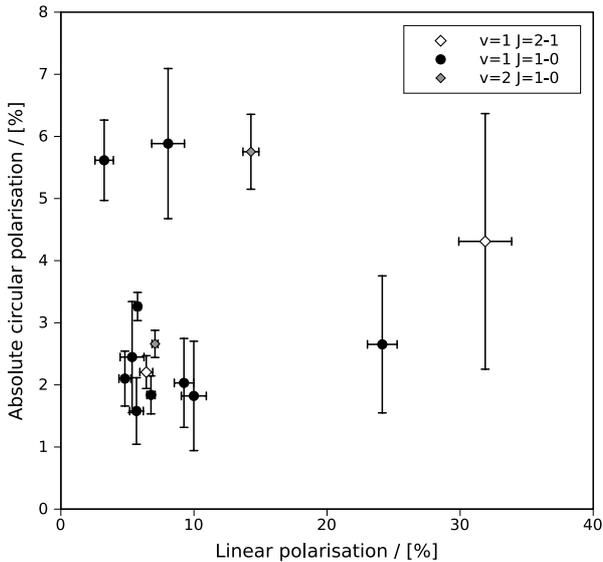}
	  \caption[Fractional circular polarisation versus fractional linear polarisation]
	  {{\small Plot of fractional circular polarisation magnitude versus fractional linear polarisation
	  magnitude, for the v=1 J=1-0, v=2 J=1-0 and v=1 J=2-1 maser features.
	  }}
	  \label{fig:circ-lin}
	\end{center}
	\end{figure}


\end{itemize}

 
\subsection{Magnetic field estimates}
\label{sec:bfield_discussion}

If the fractional circular polarisation is generated by the standard Zeeman mechanism, then magnetic field estimates can be derived from 
the circular polarisation levels. As mentioned previously, the average magnitude of the fractional circular polarisation is 3.0$\%$ for the 
v=1 J=1-0 line, 4.2$\%$ for the v=2 J=1-0 line, and 2.6$\%$ for the v=1 J=2-1 line. From \citet{Elitzur:96}, the magnetic field can be 
calculated for standard Zeeman circular polarisation, as $B = 2 \mbox{m}_c \Delta v_5$~G for J=1-0 SiO maser transitions, where $\Delta v_5$ 
is the Doppler width of the line in units of km/s, fractional circular polarisation $\mbox{m}_c$ is taken as a percentage, and adopting 
$\Theta = 45^\circ$. The magnetic field relation will be approximately double for the higher frequency J=2-1 transition, 
$B = 2 \times 2 \mbox{m}_c \Delta v_5$~G \citep{Elitzur:96}.

This relation predicts mean magnetic fields of 3.6 and 5.0 G for the v=1 and v=2 J=1-0 lines respectively, and 6.2 G for the v=1 J=2-1 line,
assuming a Doppler velocity line-width of 0.6~km/s. Magnetic field estimates from the Watson model will be similar, for standard Zeeman 
splitting, differing by only a factor of $\sim2-3$ due to saturation effects \citep{Watson:01,Watson:09}. 

Similar magnetic field estimates have been reported by \citet{Barvainis:87} and \citet{Kemball:97} for SiO maser emission towards a number 
of late-type evolved stars, assuming standard Zeeman splitting. \citet{Barvainis:87} estimate a magnetic field value of 65~G for VY CMa, 
which is considerably larger than the magnetic field reported here. The discrepancy is due to their use of a different scale factor in the 
Zeeman relation than that in the \citet{Elitzur:96} expression used here.

Several magnetic field estimates for VY CMa have also been published for regions at a larger projected radius from the star: $\sim2$~mG from 
satellite-line OH maser emission \citep{Cohen:87b}, 1~mG from main line OH maser emission \citep{Benson:82}, and $\sim175 - 200$~mG from 
H$_2$O maser emission \citep{Vlemmings:02a}. A magnetic field of order several Gauss is a plausible extrapolation for the SiO maser region, 
assuming a solar-type $B \propto r^{-2}$ law as a function of radius \citep{Sabin:15}.

If the standard Zeeman derivation of the magnetic field values above is appropriate, then the measured magnetic field values are either 
sampling a strong global magnetic field, or the masers are preferentially probing regions of locally-enhanced magnetic field with low global 
filling-factor \citep{Kemball:09}. Figure~\ref{fig:ml-mc-spots} shows that maser features displaying statistically significant circular 
polarisation are dispersed throughout the circumstellar envelope. If the circular polarisation is evidence of localised magnetic field 
enhancements, then the stronger fields appear to be randomly distributed throughout the maser region.

For the elongated feature F1, the v=1 J=1-0 emission displays circular polarisation at a level of $\sim-2$ to $-4\%$ across a region of 
$\sim4$~mas in angular extent over the region of the highest total intensity emission in the feature. Using the standard Zeeman interpretation 
described above, this equates to a magnetic field of $\sim2$ to 5~G. In that case, the magnetic energy density in the masing gas will exceed 
the thermal and kinetic energy densities and the magnetic field would have a dominant role in shaping the feature, as suggested by 
\citet{Vlemmings:05} and \citet{Cotton:06} for other late-type evolved stars. Feature F1 is a long lived maser feature, at least 18 months 
old \citep{Zhang:12}. If the magnetic field in this feature is as high as the estimates provided by the Zeeman interpretation, then the feature 
may possibly be located above a highly magnetised outflow above a magnetic cool spot or convective cell.


\section{Conclusions}
\label{sec:Conclusions}

Full-polarisation VLBA images of v=1 J=1-0, v=2 J=1-0 and v=1 J=2-1 SiO masers toward VY~CMa were presented. Component-level characteristics 
of six maser features that are spatially-coincident in more than one transition were used to test maser polarisation models.

The following summary conclusions were reached:
\begin{itemize}
	\item A comparison of the fractional linear polarisation in the v=1 J=1-0 and v=1 J=2-1 SiO maser emission of the six coincident features 
	showed no clear ordinal relationship with rotational quantum number $J$. This result is more consistent with models predicting 
	spin-independent polarization solutions \citep{Elitzur:96}.

	\item An analysis of the dependence of fractional linear polarization on intensity, for the transition v=1 J=1-0, containing the largest 
	number of detected components, proved difficult as a test on saturation effects due to the likely influence of geometric effects evident 
	in the data.

	\item A trend of stronger fractional linear polarisation closer to the star is not observed, in weak contradiction to the first-order 
	predictions of m-anisotropic pumping models.

	\item The form of the fractional linear polarisation variation across feature F2 is broadly consistent with a transition across the critical 
	55$^\circ$ angle $\Theta$ between the magnetic field and the line of sight, as in the GKK model for $m_l(\Theta)$ in the parameter regime 
	$g \Omega \gg R \gg \Gamma$ (GKK). However, the limited number of data points across this feature limit the quality of the fit relative to 
	prior results \citep{Kemball:11b}.
	Feature F1 is more complex, displaying multiple $\sim90^\circ$ EVPA rotations and considerably greater fractional linear polarisation in the 
	J=2-1 transition than the J=1-0 transition. The EVPA rotation of the feature may be caused by a helical magnetic field, or orientation close 
	to the $55^\circ$ critical 	angle.

	\item A component-level comparison of circular polarisation could only be performed for one maser feature (F1) believed to be sampling the 
	same physical conditions in both transitions. In this feature, statistically-significant circular polarisation in the v=1 J=1-0 and 
	v=1 J=2-1 transitions was only observed in one channel across the feature, where the fractional circular polarisation was the same for both 
	transitions, within the measured uncertainties. The large uncertainty in the v=1 J=2-1 fractional circular polarisation measurement means
	that the standard Zeeman circular polarisation model cannot be ruled out by this single component test.

	\item The fractional linear and circular polarisation of the maser features were found to be uncorrelated. A significant number of maser 
	features are circularly polarised at a level greater than $m_l^2/4$, which provides strong evidence against non-Zeeman circular polarisation 
	mechanisms \citep{Wiebe:98}.
\end{itemize}

This work shows that further full polarisation comparisons of J=1-0 and J=2-1 SiO masers are strongly warranted in order to increase the number 
of coincident components. A larger sample of multi-transition circular polarisation measurements will be especially valuable.  


\vskip 1cm

This material is based upon work supported by the National Science Foundation under Grant No. AST 0507473.


\bibliographystyle{mnras}
\bibliography{richter_references}

\begin{thebibliography}{}
\makeatletter
\relax
\def\mn@urlcharsother{\let\do\@makeother \do\$\do\&\do\#\do\^\do\_\do\%\do\~}
\def\mn@doi{\begingroup\mn@urlcharsother \@ifnextchar [ {\mn@doi@}
  {\mn@doi@[]}}
\def\mn@doi@[#1]#2{\def\@tempa{#1}\ifx\@tempa\@empty \href
  {http://dx.doi.org/#2} {doi:#2}\else \href {http://dx.doi.org/#2} {#1}\fi
  \endgroup}
\def\mn@eprint#1#2{\mn@eprint@#1:#2::\@nil}
\def\mn@eprint@arXiv#1{\href {http://arxiv.org/abs/#1} {{\tt arXiv:#1}}}
\def\mn@eprint@dblp#1{\href {http://dblp.uni-trier.de/rec/bibtex/#1.xml}
  {dblp:#1}}
\def\mn@eprint@#1:#2:#3:#4\@nil{\def\@tempa {#1}\def\@tempb {#2}\def\@tempc
  {#3}\ifx \@tempc \@empty \let \@tempc \@tempb \let \@tempb \@tempa \fi \ifx
  \@tempb \@empty \def\@tempb {arXiv}\fi \@ifundefined
  {mn@eprint@\@tempb}{\@tempb:\@tempc}{\expandafter \expandafter \csname
  mn@eprint@\@tempb\endcsname \expandafter{\@tempc}}}

\bibitem[\protect\citeauthoryear{{Agudo}, {Thum}, {Wiesemeyer}  \&
  {Krichbaum}}{{Agudo} et~al.}{2010}]{Agudo:10}
{Agudo} I.,  {Thum} C.,  {Wiesemeyer} H.,   {Krichbaum} T.~P.,  2010, \mn@doi
  [\apjs] {10.1088/0067-0049/189/1/1}, \href
  {http://adsabs.harvard.edu/abs/2010ApJS..189....1A} {189, 1}

\bibitem[\protect\citeauthoryear{{Allen}, {Hall}, {Norris}, {Troup}, {Wark}  \&
  {Wright}}{{Allen} et~al.}{1989}]{Allen:89}
{Allen} D.~A.,  {Hall} P.~J.,  {Norris} R.~P.,  {Troup} E.~R.,  {Wark} R.~M.,
  {Wright} A.~E.,  1989, \mnras, \href
  {http://adsabs.harvard.edu/abs/1989MNRAS.236..363A} {236, 363}

\bibitem[\protect\citeauthoryear{{Aller}, {Aller}  \& {Plotkin}}{{Aller}
  et~al.}{2003}]{Aller:03}
{Aller} H.~D.,  {Aller} M.~F.,   {Plotkin} R.~M.,  2003, \mn@doi [\apss]
  {10.1023/B:ASTR.0000004990.22510.53}, \href
  {http://adsabs.harvard.edu/abs/2003Ap%26SS.288...17A} {288, 17}

\bibitem[\protect\citeauthoryear{{Amiri}, {Vlemmings}, {Kemball}  \& {van
  Langevelde}}{{Amiri} et~al.}{2012}]{Amiri:12}
{Amiri} N.,  {Vlemmings} W.~H.~T.,  {Kemball} A.~J.,   {van Langevelde} H.~J.,
  2012, \mn@doi [\aap] {10.1051/0004-6361/201117452}, \href
  {http://adsabs.harvard.edu/abs/2012A%26A...538A.136A} {538, A136}

\bibitem[\protect\citeauthoryear{{Asensio Ramos}, {Landi Degl'Innocenti}  \&
  {Trujillo Bueno}}{{Asensio Ramos} et~al.}{2005}]{Asensio_Ramos:05}
{Asensio Ramos} A.,  {Landi Degl'Innocenti} E.,   {Trujillo Bueno} J.,  2005,
  \mn@doi [\apj] {10.1086/429719}, \href
  {http://adsabs.harvard.edu/abs/2005ApJ...625..985A} {625, 985}

\bibitem[\protect\citeauthoryear{{Assaf}, {Diamond}, {Richards}  \&
  {Gray}}{{Assaf} et~al.}{2013}]{Assaf:13}
{Assaf} K.~A.,  {Diamond} P.~J.,  {Richards} A.~M.~S.,   {Gray} M.~D.,  2013,
  \mn@doi [\mnras] {10.1093/mnras/stt242}, \href
  {http://adsabs.harvard.edu/abs/2013MNRAS.431.1077A} {431, 1077}

\bibitem[\protect\citeauthoryear{{Auri{\`e}re} et~al.,}{{Auri{\`e}re}
  et~al.}{2008}]{Auriere:08}
{Auri{\`e}re} M.,  et~al., 2008, \mn@doi [\aap] {10.1051/0004-6361:200810502},
  \href {http://adsabs.harvard.edu/abs/2008A%26A...491..499A} {491, 499}

\bibitem[\protect\citeauthoryear{{Auri{\`e}re}, {Donati},
  {Konstantinova-Antova}, {Perrin}, {Petit}  \& {Roudier}}{{Auri{\`e}re}
  et~al.}{2010}]{Auriere:10}
{Auri{\`e}re} M.,  {Donati} J.,  {Konstantinova-Antova} R.,  {Perrin} G.,
  {Petit} P.,   {Roudier} T.,  2010, \mn@doi [\aap]
  {10.1051/0004-6361/201014925}, \href
  {http://adsabs.harvard.edu/abs/2010A%26A...516L...2A} {516, L2}

\bibitem[\protect\citeauthoryear{{Bains}, {Gledhill}, {Yates}  \&
  {Richards}}{{Bains} et~al.}{2003}]{Bains:03}
{Bains} I.,  {Gledhill} T.~M.,  {Yates} J.~A.,   {Richards} A.~M.~S.,  2003,
  \mn@doi [\mnras] {10.1046/j.1365-8711.2003.06071.x}, \href
  {http://adsabs.harvard.edu/abs/2003MNRAS.338..287B} {338, 287}

\bibitem[\protect\citeauthoryear{{Bains}, {Richards}, {Gledhill}  \&
  {Yates}}{{Bains} et~al.}{2004}]{Bains:04}
{Bains} I.,  {Richards} A.~M.~S.,  {Gledhill} T.~M.,   {Yates} J.~A.,  2004,
  \mn@doi [\mnras] {10.1111/j.1365-2966.2004.08209.x}, \href
  {http://adsabs.harvard.edu/abs/2004MNRAS.354..529B} {354, 529}

\bibitem[\protect\citeauthoryear{{Barvainis} \& {Predmore}}{{Barvainis} \&
  {Predmore}}{1985}]{Barvainis:85}
{Barvainis} R.,  {Predmore} C.~R.,  1985, \apj, \href
  {http://adsabs.harvard.edu/cgi-bin/nph-bib_query?bibcode=1985ApJ...288..694B&amp;db_key=AST}
  {288, 694}

\bibitem[\protect\citeauthoryear{{Barvainis}, {McIntosh}  \&
  {Predmore}}{{Barvainis} et~al.}{1987}]{Barvainis:87}
{Barvainis} R.,  {McIntosh} G.,   {Predmore} C.~R.,  1987, \mn@doi [\nat]
  {10.1038/329613a0}, \href {http://adsabs.harvard.edu/abs/1987Natur.329..613B}
  {329, 613}

\bibitem[\protect\citeauthoryear{{Benson} \& {Mutel}}{{Benson} \&
  {Mutel}}{1982}]{Benson:82}
{Benson} J.~M.,  {Mutel} R.~L.,  1982, \apj, \href
  {http://adsabs.harvard.edu/cgi-bin/nph-bib_query?bibcode=1982ApJ...253..199B&db_key=AST}
  {253, 199}

\bibitem[\protect\citeauthoryear{{Boboltz}}{{Boboltz}}{1997}]{Boboltz:thesis}
{Boboltz} D.~A.~J.,  1997, PhD thesis, Virginia Polytechnic Institute and State
  University

\bibitem[\protect\citeauthoryear{{Briggs}}{{Briggs}}{1995}]{Briggs:thesis}
{Briggs} D.~S.,  1995, PhD thesis, New Mexico Institute of Mining and
  Technology

\bibitem[\protect\citeauthoryear{{Bujarrabal} \& {Nguyen-Q-Rieu}}{{Bujarrabal}
  \& {Nguyen-Q-Rieu}}{1981}]{Bujarrabal:81}
{Bujarrabal} V.,  {Nguyen-Q-Rieu} M.,  1981, \aap, \href
  {http://adsabs.harvard.edu/cgi-bin/nph-bib_query?bibcode=1981A%26A...102...65B&db_key=AST}
  {102, 65}

\bibitem[\protect\citeauthoryear{{Cernicharo}, {Bujarrabal}  \&
  {Santaren}}{{Cernicharo} et~al.}{1993}]{Cernicharo:93}
{Cernicharo} J.,  {Bujarrabal} V.,   {Santaren} J.~L.,  1993, \apjl, \href
  {http://adsabs.harvard.edu/cgi-bin/nph-bib_query?bibcode=1993ApJ...407L..33C&db_key=AST}
  {407, L33}

\bibitem[\protect\citeauthoryear{{Chapman} \& {Cohen}}{{Chapman} \&
  {Cohen}}{1986}]{Chapman:86}
{Chapman} J.~M.,  {Cohen} R.~J.,  1986, \mnras, \href
  {http://adsabs.harvard.edu/cgi-bin/nph-bib_query?bibcode=1986MNRAS.220..513C&db_key=AST}
  {220, 513}

\bibitem[\protect\citeauthoryear{{Clark}, {Troland}  \& {Johnson}}{{Clark}
  et~al.}{1982}]{Clark:82}
{Clark} F.~O.,  {Troland} T.~H.,   {Johnson} D.~R.,  1982, \mn@doi [\apj]
  {10.1086/160367}, \href {http://adsabs.harvard.edu/abs/1982ApJ...261..569C}
  {261, 569}

\bibitem[\protect\citeauthoryear{{Clark}, {Troland}, {Pepper}  \&
  {Johnson}}{{Clark} et~al.}{1984}]{Clark:84}
{Clark} F.~O.,  {Troland} T.~H.,  {Pepper} G.~H.,   {Johnson} D.~R.,  1984,
  \mn@doi [\apj] {10.1086/161646}, \href
  {http://adsabs.harvard.edu/abs/1984ApJ...276..572C} {276, 572}

\bibitem[\protect\citeauthoryear{{Cohen}}{{Cohen}}{1989}]{Cohen:89}
{Cohen} R.~J.,  1989, \mn@doi [Reports on Progress in Physics]
  {10.1088/0034-4885/52/8/001}, \href
  {http://adsabs.harvard.edu/abs/1989RPPh...52..881C} {52, 881}

\bibitem[\protect\citeauthoryear{{Cohen}, {Downs}, {Emerson}, {Grimm},
  {Gulkis}, {Stevens}  \& {Tarter}}{{Cohen} et~al.}{1987}]{Cohen:87b}
{Cohen} R.~J.,  {Downs} G.,  {Emerson} R.,  {Grimm} M.,  {Gulkis} S.,
  {Stevens} G.,   {Tarter} J.,  1987, \mnras, \href
  {http://adsabs.harvard.edu/abs/1987MNRAS.225..491C} {225, 491}

\bibitem[\protect\citeauthoryear{{Cotton} et~al.,}{{Cotton}
  et~al.}{2006}]{Cotton:06}
{Cotton} W.~D.,  et~al., 2006, \mn@doi [\aap] {10.1051/0004-6361:20065134},
  \href {http://adsabs.harvard.edu/abs/2006A%26A...456..339C} {456, 339}

\bibitem[\protect\citeauthoryear{{Cotton}, {Ragland}  \& {Danchi}}{{Cotton}
  et~al.}{2011}]{Cotton:11}
{Cotton} W.~D.,  {Ragland} S.,   {Danchi} W.~C.,  2011, \mn@doi [\apj]
  {10.1088/0004-637X/736/2/96}, \href
  {http://adsabs.harvard.edu/abs/2011ApJ...736...96C} {736, 96}

\bibitem[\protect\citeauthoryear{{Davison} \& {Hinkley}}{{Davison} \&
  {Hinkley}}{1997}]{bk:Davison}
{Davison} A.~C.,  {Hinkley} D.~V.,  1997, Bootstrap Methods and Their
  Application.
Cambridge University Press

\bibitem[\protect\citeauthoryear{{Desmurs}, {Bujarrabal}, {Colomer}  \&
  {Alcolea}}{{Desmurs} et~al.}{2000}]{Desmurs:00}
{Desmurs} J.~F.,  {Bujarrabal} V.,  {Colomer} F.,   {Alcolea} J.,  2000, \aap,
  \href
  {http://adsabs.harvard.edu/cgi-bin/nph-bib_query?bibcode=2000A%26A...360..189D&db_key=AST}
  {360, 189}

\bibitem[\protect\citeauthoryear{{Diamond} \& {Kemball}}{{Diamond} \&
  {Kemball}}{2003}]{Diamond:03}
{Diamond} P.~J.,  {Kemball} A.~J.,  2003, \mn@doi [\apj] {10.1086/379347},
  \href {http://adsabs.harvard.edu/abs/2003ApJ...599.1372D} {599, 1372}

\bibitem[\protect\citeauthoryear{{Diamond}, {Kemball}, {Junor}, {Zensus},
  {Benson}  \& {Dhawan}}{{Diamond} et~al.}{1994}]{Diamond:94}
{Diamond} P.~J.,  {Kemball} A.~J.,  {Junor} W.,  {Zensus} A.,  {Benson} J.,
  {Dhawan} V.,  1994, \apjl, \href
  {http://adsabs.harvard.edu/cgi-bin/nph-bib_query?bibcode=1994ApJ...430L..61D&db_key=AST}
  {430, L61}

\bibitem[\protect\citeauthoryear{{Donati}, {Semel}, {Carter}, {Rees}  \&
  {Collier Cameron}}{{Donati} et~al.}{1997}]{Donati:97}
{Donati} J.,  {Semel} M.,  {Carter} B.~D.,  {Rees} D.~E.,   {Collier Cameron}
  A.,  1997, \mnras, \href {http://adsabs.harvard.edu/abs/1997MNRAS.291..658D}
  {291, 658}

\bibitem[\protect\citeauthoryear{{Draine}}{{Draine}}{2011}]{Draine:11}
{Draine} B.~T.,  2011, {Physics of the Interstellar and Intergalactic Medium}.
Princeton University Press

\bibitem[\protect\citeauthoryear{{Elitzur}}{{Elitzur}}{1991}]{Elitzur:91}
{Elitzur} M.,  1991, \mn@doi [\apj] {10.1086/169827}, \href
  {http://adsabs.harvard.edu/cgi-bin/nph-bib_query?bibcode=1991ApJ...370..407E&db_key=AST}
  {370, 407}

\bibitem[\protect\citeauthoryear{{Elitzur}}{{Elitzur}}{1992}]{bk:Elit}
{Elitzur} M.,  1992, Astronomical Masers.
Kluwer Academic Publishers

\bibitem[\protect\citeauthoryear{{Elitzur}}{{Elitzur}}{1996}]{Elitzur:96}
{Elitzur} M.,  1996, \mn@doi [\apj] {10.1086/176741}, \href
  {http://adsabs.harvard.edu/cgi-bin/nph-bib_query?bibcode=1996ApJ...457..415E&db_key=AST}
  {457, 415}

\bibitem[\protect\citeauthoryear{{Elitzur}}{{Elitzur}}{2002}]{Elitzur:02}
{Elitzur} M.,  2002, in {Migenes} V.,  {Reid} M.~J.,  eds,  IAU Symposium Vol.
  206, Cosmic Masers: From Proto-Stars to Black Holes. Astronomical Society of
  the Pacific, p.~452

\bibitem[\protect\citeauthoryear{{Etoka} \& {Diamond}}{{Etoka} \&
  {Diamond}}{2004}]{Etoka:04}
{Etoka} S.,  {Diamond} P.,  2004, \mn@doi [\mnras]
  {10.1111/j.1365-2966.2004.07370.x}, \href
  {http://adsabs.harvard.edu/abs/2004MNRAS.348...34E} {348, 34}

\bibitem[\protect\citeauthoryear{{Garc{\'{\i}}a-Segura}, {L{\'o}pez}  \&
  {Franco}}{{Garc{\'{\i}}a-Segura} et~al.}{2005}]{Garcia-Segura:05}
{Garc{\'{\i}}a-Segura} G.,  {L{\'o}pez} J.~A.,   {Franco} J.,  2005, \mn@doi
  [\apj] {10.1086/426110}, \href
  {http://adsabs.harvard.edu/abs/2005ApJ...618..919G} {618, 919}

\bibitem[\protect\citeauthoryear{{Glassgold} \& {Huggins}}{{Glassgold} \&
  {Huggins}}{1983}]{Glassgold:83}
{Glassgold} A.~E.,  {Huggins} P.~J.,  1983, \mn@doi [\mnras]
  {10.1093/mnras/203.2.517}, \href
  {http://adsabs.harvard.edu/abs/1983MNRAS.203..517G} {203, 517}

\bibitem[\protect\citeauthoryear{{Glenn}, {Jewell}, {Fourre}  \&
  {Miaja}}{{Glenn} et~al.}{2003}]{Glenn:03}
{Glenn} J.,  {Jewell} P.~R.,  {Fourre} R.,   {Miaja} L.,  2003, \apj, \href
  {http://adsabs.harvard.edu/cgi-bin/nph-bib_query?bibcode=2003ApJ...588..478G&db_key=AST}
  {588, 478}

\bibitem[\protect\citeauthoryear{{Goldreich} \& {Kylafis}}{{Goldreich} \&
  {Kylafis}}{1981}]{Goldreich:81}
{Goldreich} P.,  {Kylafis} N.~D.,  1981, \mn@doi [\apjl] {10.1086/183446},
  \href {http://adsabs.harvard.edu/abs/1981ApJ...243L..75G} {243, L75}

\bibitem[\protect\citeauthoryear{{Goldreich}, {Keeley}  \& {Kwan}}{{Goldreich}
  et~al.}{1973}]{GKK}
{Goldreich} P.,  {Keeley} D.~A.,   {Kwan} J.~Y.,  1973, \apj, \href
  {http://adsabs.harvard.edu/cgi-bin/nph-bib_query?bibcode=1973ApJ...179..111G&db_key=AST}
  {179, 111}

\bibitem[\protect\citeauthoryear{{G{\'o}mez}, {Tafoya}, {Anglada}, {Miranda},
  {Torrelles}, {Patel}  \& {Hern{\'a}ndez}}{{G{\'o}mez}
  et~al.}{2009}]{Gomez:09}
{G{\'o}mez} Y.,  {Tafoya} D.,  {Anglada} G.,  {Miranda} L.~F.,  {Torrelles}
  J.~M.,  {Patel} N.~A.,   {Hern{\'a}ndez} R.~F.,  2009, \mn@doi [\apj]
  {10.1088/0004-637X/695/2/930}, \href
  {http://adsabs.harvard.edu/abs/2009ApJ...695..930G} {695, 930}

\bibitem[\protect\citeauthoryear{{Gray}}{{Gray}}{2003}]{Gray:03}
{Gray} M.~D.,  2003, \mn@doi [\mnras] {10.1046/j.1365-8711.2003.06836.x}, \href
  {http://adsabs.harvard.edu/cgi-bin/nph-bib_query?bibcode=2003MNRAS.343L..33G&db_key=AST}
  {343, L33}

\bibitem[\protect\citeauthoryear{{Gray}}{{Gray}}{2012}]{Gray:12}
{Gray} M.,  2012, {Maser Sources in Astrophysics}.
Cambridge, UK: Cambridge University Press

\bibitem[\protect\citeauthoryear{{Gray} \& {Field}}{{Gray} \&
  {Field}}{1995}]{Gray:95c}
{Gray} M.~D.,  {Field} D.,  1995, \aap, \href
  {http://adsabs.harvard.edu/abs/1995A%26A...298..243G} {298, 243}

\bibitem[\protect\citeauthoryear{{Greaves}}{{Greaves}}{2002}]{Greaves:02}
{Greaves} J.~S.,  2002, \mn@doi [\aap] {10.1051/0004-6361:20021002}, \href
  {http://adsabs.harvard.edu/abs/2002A%26A...392L...1G} {392, L1}

\bibitem[\protect\citeauthoryear{{Grunhut}, {Wade}, {Hanes}  \&
  {Alecian}}{{Grunhut} et~al.}{2010}]{Grunhut:10}
{Grunhut} J.~H.,  {Wade} G.~A.,  {Hanes} D.~A.,   {Alecian} E.,  2010, \mn@doi
  [\mnras] {10.1111/j.1365-2966.2010.17275.x}, \href
  {http://adsabs.harvard.edu/abs/2010MNRAS.408.2290G} {408, 2290}

\bibitem[\protect\citeauthoryear{{Gustafsson} \& {H{\"o}fner}}{{Gustafsson} \&
  {H{\"o}fner}}{2004}]{bk:Gustafsson}
{Gustafsson} B.,  {H{\"o}fner} S.,  2004, Atmospheres of AGB Stars. In H. J.
  Habing \& H. Olofsson, eds. Asymptotic Giant Branch Stars. Astronomy and
  Astrophysics Library.
Springer New York

\bibitem[\protect\citeauthoryear{{Hall}, {Allen}, {Troup}, {Wark}  \&
  {Wright}}{{Hall} et~al.}{1990}]{Hall:90}
{Hall} P.~J.,  {Allen} D.~A.,  {Troup} E.~R.,  {Wark} R.~M.,   {Wright} A.~E.,
  1990, \mnras, \href {http://adsabs.harvard.edu/abs/1990MNRAS.243..480H} {243,
  480}

\bibitem[\protect\citeauthoryear{{Harper} \& {Brown}}{{Harper} \&
  {Brown}}{2006}]{Harper:06}
{Harper} G.~M.,  {Brown} A.,  2006, \mn@doi [\apj] {10.1086/505073}, \href
  {http://adsabs.harvard.edu/abs/2006ApJ...646.1179H} {646, 1179}

\bibitem[\protect\citeauthoryear{{Harper}, {Brown}  \& {Lim}}{{Harper}
  et~al.}{2001}]{Harper:01}
{Harper} G.~M.,  {Brown} A.,   {Lim} J.,  2001, \mn@doi [\apj]
  {10.1086/320215}, \href {http://adsabs.harvard.edu/abs/2001ApJ...551.1073H}
  {551, 1073}

\bibitem[\protect\citeauthoryear{{Hartquist} \& {Dyson}}{{Hartquist} \&
  {Dyson}}{1997}]{Hartquist:97}
{Hartquist} T.~W.,  {Dyson} J.~E.,  1997, \aap, \href
  {http://adsabs.harvard.edu/abs/1997A%26A...319..589H} {319, 589}

\bibitem[\protect\citeauthoryear{{Harwit} et~al.,}{{Harwit}
  et~al.}{2010}]{Harwit:10}
{Harwit} M.,  et~al., 2010, \mn@doi [\aap] {10.1051/0004-6361/201015042}, \href
  {http://adsabs.harvard.edu/abs/2010A%26A...521L..51H} {521, L51}

\bibitem[\protect\citeauthoryear{{Hayya}, {Armstrong}  \& {Gressis}}{{Hayya}
  et~al.}{1975}]{Hayya:75}
{Hayya} J.,  {Armstrong} D.,   {Gressis} D.,  1975, Management Science, 21,
  1338

\bibitem[\protect\citeauthoryear{{Heger}, {Fryer}, {Woosley}, {Langer}  \&
  {Hartmann}}{{Heger} et~al.}{2003}]{Heger:03}
{Heger} A.,  {Fryer} C.~L.,  {Woosley} S.~E.,  {Langer} N.,   {Hartmann} D.~H.,
   2003, \mn@doi [\apj] {10.1086/375341}, \href
  {http://adsabs.harvard.edu/abs/2003ApJ...591..288H} {591, 288}

\bibitem[\protect\citeauthoryear{{Herpin}, {Baudry}, {Thum}, {Morris}  \&
  {Wiesemeyer}}{{Herpin} et~al.}{2006}]{Herpin:06}
{Herpin} F.,  {Baudry} A.,  {Thum} C.,  {Morris} D.,   {Wiesemeyer} H.,  2006,
  \mn@doi [\aap] {10.1051/0004-6361:20054255}, \href
  {http://adsabs.harvard.edu/abs/2006A%26A...450..667H} {450, 667}

\bibitem[\protect\citeauthoryear{{Homan} \& {Lister}}{{Homan} \&
  {Lister}}{2006}]{Homan:06}
{Homan} D.~C.,  {Lister} M.~L.,  2006, \mn@doi [\aj] {10.1086/500256}, \href
  {http://adsabs.harvard.edu/abs/2006AJ....131.1262H} {131, 1262}

\bibitem[\protect\citeauthoryear{{Homan} \& {Wardle}}{{Homan} \&
  {Wardle}}{2003}]{Homan:03}
{Homan} D.~C.,  {Wardle} J.~F.~C.,  2003, \mn@doi [\apss]
  {10.1023/B:ASTR.0000004991.81026.2f}, \href
  {http://adsabs.harvard.edu/abs/2003Ap%26SS.288...29H} {288, 29}

\bibitem[\protect\citeauthoryear{{Homan}, {Attridge}  \& {Wardle}}{{Homan}
  et~al.}{2001}]{Homan:01}
{Homan} D.~C.,  {Attridge} J.~M.,   {Wardle} J.~F.~C.,  2001, \mn@doi [\apj]
  {10.1086/321568}, \href {http://adsabs.harvard.edu/abs/2001ApJ...556..113H}
  {556, 113}

\bibitem[\protect\citeauthoryear{{Houde}}{{Houde}}{2014}]{Houde:14}
{Houde} M.,  2014, \mn@doi [\apj] {10.1088/0004-637X/795/1/27}, \href
  {http://adsabs.harvard.edu/abs/2014ApJ...795...27H} {795, 27}

\bibitem[\protect\citeauthoryear{{Iben} \& {Renzini}}{{Iben} \&
  {Renzini}}{1983}]{Iben:83}
{Iben} I.,  {Renzini} A.,  1983, \araa, \href
  {http://adsabs.harvard.edu/cgi-bin/nph-bib_query?bibcode=1983ARA%26A..21..271I&db_key=AST}
  {21, 271}

\bibitem[\protect\citeauthoryear{{Ireland}, {Scholz}  \& {Wood}}{{Ireland}
  et~al.}{2011}]{Ireland:11}
{Ireland} M.~J.,  {Scholz} M.,   {Wood} P.~R.,  2011, \mn@doi [\mnras]
  {10.1111/j.1365-2966.2011.19469.x}, \href
  {http://adsabs.harvard.edu/abs/2011MNRAS.418..114I} {418, 114}

\bibitem[\protect\citeauthoryear{{Kemball}}{{Kemball}}{1992}]{Kemball:thesis}
{Kemball} A.~J.,  1992, PhD thesis, Rhodes University

\bibitem[\protect\citeauthoryear{{Kemball}}{{Kemball}}{1999}]{Kemball:99}
{Kemball} A.~J.,  1999, in {Taylor} G.~B.,  {Carilli} C.~L.,   {Perley} R.~A.,
  eds,  Astronomical Society of the Pacific Conference Series Vol. 180,
  Synthesis Imaging in Radio Astronomy II. Astronomical Society of the Pacific,
  p.~499

\bibitem[\protect\citeauthoryear{{Kemball} \& {Diamond}}{{Kemball} \&
  {Diamond}}{1997}]{Kemball:97}
{Kemball} A.~J.,  {Diamond} P.~J.,  1997, \apjl, \href
  {http://adsabs.harvard.edu/cgi-bin/nph-bib_query?bibcode=1997ApJ...481L.111K&db_key=AST}
  {481, L111}

\bibitem[\protect\citeauthoryear{{Kemball} \& {Richter}}{{Kemball} \&
  {Richter}}{2011}]{Kemball:11}
{Kemball} A.~J.,  {Richter} L.,  2011, \mn@doi [\aap]
  {10.1051/0004-6361/201117337}, \href
  {http://adsabs.harvard.edu/abs/2011A%26A...533A..26K} {533, A26}

\bibitem[\protect\citeauthoryear{{Kemball}, {Diamond}  \& {Cotton}}{{Kemball}
  et~al.}{1995}]{Kemball:95}
{Kemball} A.~J.,  {Diamond} P.~J.,   {Cotton} W.~D.,  1995, \aaps, \href
  {http://adsabs.harvard.edu/cgi-bin/nph-bib_query?bibcode=1995A%26AS..110..383K&amp;db_key=AST}
  {110, 383}

\bibitem[\protect\citeauthoryear{{Kemball}, {Diamond}, {Gonidakis}, {Mitra},
  {Yim}, {Pan}  \& {Chiang}}{{Kemball} et~al.}{2009}]{Kemball:09}
{Kemball} A.~J.,  {Diamond} P.~J.,  {Gonidakis} I.,  {Mitra} M.,  {Yim} K.,
  {Pan} K.,   {Chiang} H.,  2009, \mn@doi [\apj]
  {10.1088/0004-637X/698/2/1721}, \href
  {http://adsabs.harvard.edu/abs/2009ApJ...698.1721K} {698, 1721}

\bibitem[\protect\citeauthoryear{{Kemball}, {Diamond}, {Richter}, {Gonidakis}
  \& {Xue}}{{Kemball} et~al.}{2011}]{Kemball:11b}
{Kemball} A.~J.,  {Diamond} P.~J.,  {Richter} L.,  {Gonidakis} I.,   {Xue} R.,
  2011, \mn@doi [\apj] {10.1088/0004-637X/743/1/69}, \href
  {http://adsabs.harvard.edu/abs/2011ApJ...743...69K} {743, 69}

\bibitem[\protect\citeauthoryear{{Knapp}, {Bowers}, {Young}  \&
  {Phillips}}{{Knapp} et~al.}{1995}]{Knapp:95}
{Knapp} G.~R.,  {Bowers} P.~F.,  {Young} K.,   {Phillips} T.~G.,  1995, \apj,
  \href
  {http://adsabs.harvard.edu/cgi-bin/nph-bib_query?bibcode=1995ApJ...455..293K&db_key=AST}
  {455, 293}

\bibitem[\protect\citeauthoryear{{Konstantinova-Antova}, {Auri{\`e}re},
  {Schr{\"o}der}  \& {Petit}}{{Konstantinova-Antova}
  et~al.}{2009}]{Konstantinova-Antova:09}
{Konstantinova-Antova} R.,  {Auri{\`e}re} M.,  {Schr{\"o}der} K.,   {Petit} P.,
   2009, in {K.~G.~Strassmeier, A.~G.~Kosovichev, \& J.~E.~Beckman} ed.,  IAU
  Symposium Vol. 259, Cosmic Magnetic Fields: From Planets, to Stars and
  Galaxies. Cambridge University Press, p.~433,
  \mn@doi{10.1017/S1743921309031020}

\bibitem[\protect\citeauthoryear{{Leal-Ferreira}, {Vlemmings}, {Kemball}  \&
  {Amiri}}{{Leal-Ferreira} et~al.}{2013}]{Leal-Ferreira:13}
{Leal-Ferreira} M.~L.,  {Vlemmings} W.~H.~T.,  {Kemball} A.,   {Amiri} N.,
  2013, \mn@doi [\aap] {10.1051/0004-6361/201321218}, \href
  {http://adsabs.harvard.edu/abs/2013A%26A...554A.134L} {554, A134}

\bibitem[\protect\citeauthoryear{{L{\`e}bre}, {Auri{\`e}re}, {Fabas}, {Gillet},
  {Herpin}, {Konstantinova-Antova}  \& {Petit}}{{L{\`e}bre}
  et~al.}{2014}]{Lebre:14}
{L{\`e}bre} A.,  {Auri{\`e}re} M.,  {Fabas} N.,  {Gillet} D.,  {Herpin} F.,
  {Konstantinova-Antova} R.,   {Petit} P.,  2014, \mn@doi [\aap]
  {10.1051/0004-6361/201322826}, \href
  {http://adsabs.harvard.edu/abs/2014A%26A...561A..85L} {561, A85}

\bibitem[\protect\citeauthoryear{{Lipscy}, {Jura}  \& {Reid}}{{Lipscy}
  et~al.}{2005}]{Lipscy:05}
{Lipscy} S.~J.,  {Jura} M.,   {Reid} M.~J.,  2005, \mn@doi [\apj]
  {10.1086/429900}, \href {http://adsabs.harvard.edu/abs/2005ApJ...626..439L}
  {626, 439}

\bibitem[\protect\citeauthoryear{{Matt}, {Balick}, {Winglee}  \&
  {Goodson}}{{Matt} et~al.}{2000}]{Matt:00}
{Matt} S.,  {Balick} B.,  {Winglee} R.,   {Goodson} A.,  2000, \apj, \href
  {http://adsabs.harvard.edu/cgi-bin/nph-bib_query?bibcode=2000ApJ...545..965M&db_key=AST}
  {545, 965}

\bibitem[\protect\citeauthoryear{{McIntosh} \& {Predmore}}{{McIntosh} \&
  {Predmore}}{1991}]{McIntosh:91}
{McIntosh} A.~C.,  {Predmore} C.~R.,  1991, in {Haschick} A.~D.,  {Ho}
  P.~T.~P.,  eds,  Astronomical Society of the Pacific Conference Series Vol.
  16, Atoms, Ions and Molecules: New Results in Spectral Line Astrophysics.
  Astronomical Society of the Pacific, p.~83

\bibitem[\protect\citeauthoryear{{McIntosh} \& {Predmore}}{{McIntosh} \&
  {Predmore}}{1993}]{McIntosh:93}
{McIntosh} G.~C.,  {Predmore} C.~R.,  1993, \apjl, \href
  {http://adsabs.harvard.edu/cgi-bin/nph-bib_query?bibcode=1993ApJ...404L..71M&db_key=AST}
  {404, L71}

\bibitem[\protect\citeauthoryear{{McIntosh}, {Predmore}, {Moran}, {Greenhill},
  {Rogers}  \& {Barvainis}}{{McIntosh} et~al.}{1989}]{McIntosh:89}
{McIntosh} G.~C.,  {Predmore} C.~R.,  {Moran} J.~M.,  {Greenhill} L.~J.,
  {Rogers} A.~E.~E.,   {Barvainis} R.,  1989, \apj, \href
  {http://adsabs.harvard.edu/cgi-bin/nph-bib_query?bibcode=1989ApJ...337..934M&db_key=AST}
  {337, 934}

\bibitem[\protect\citeauthoryear{{McIntosh}, {Predmore}  \& {Patel}}{{McIntosh}
  et~al.}{1994}]{McIntosh:94}
{McIntosh} G.~C.,  {Predmore} C.~R.,   {Patel} N.~A.,  1994, \apjl, \href
  {http://adsabs.harvard.edu/cgi-bin/nph-bib_query?bibcode=1994ApJ...428L..29M&db_key=AST}
  {428, L29}

\bibitem[\protect\citeauthoryear{{Monnier} et~al.,}{{Monnier}
  et~al.}{2004}]{Monnier:04}
{Monnier} J.~D.,  et~al., 2004, \mn@doi [\apj] {10.1086/382218}, \href
  {http://adsabs.harvard.edu/abs/2004ApJ...605..436M} {605, 436}

\bibitem[\protect\citeauthoryear{{M{\"u}ller}, {Schl{\"o}der}, {Stutzki}  \&
  {Winnewisser}}{{M{\"u}ller} et~al.}{2005}]{Muller:05}
{M{\"u}ller} H.~S.~P.,  {Schl{\"o}der} F.,  {Stutzki} J.,   {Winnewisser} G.,
  2005, \mn@doi [Journal of Molecular Structure]
  {10.1016/j.molstruc.2005.01.027}, \href
  {http://adsabs.harvard.edu/abs/2005JMoSt.742..215M} {742, 215}

\bibitem[\protect\citeauthoryear{{Nedoluha} \& {Watson}}{{Nedoluha} \&
  {Watson}}{1990a}]{Nedoluha:90a}
{Nedoluha} G.~E.,  {Watson} W.~D.,  1990a, \mn@doi [\apj] {10.1086/168723},
  \href
  {http://adsabs.harvard.edu/cgi-bin/nph-bib_query?bibcode=1990ApJ...354..660N&db_key=AST}
  {354, 660}

\bibitem[\protect\citeauthoryear{{Nedoluha} \& {Watson}}{{Nedoluha} \&
  {Watson}}{1990b}]{Nedoluha:90b}
{Nedoluha} G.~E.,  {Watson} W.~D.,  1990b, \mn@doi [\apjl] {10.1086/185825},
  \href
  {http://adsabs.harvard.edu/cgi-bin/nph-bib_query?bibcode=1990ApJ...361L..53N&db_key=AST}
  {361, L53}

\bibitem[\protect\citeauthoryear{{Nedoluha} \& {Watson}}{{Nedoluha} \&
  {Watson}}{1993}]{Nedoluha:93}
{Nedoluha} G.~E.,  {Watson} W.~D.,  1993, in {Clegg} A.~W.,  {Nedoluha} G.~E.,
  eds,  Lecture Notes in Physics Vol. 412, Astrophysical Masers. Springer
  Berlin / Heidelberg, p.~47

\bibitem[\protect\citeauthoryear{{Nedoluha} \& {Watson}}{{Nedoluha} \&
  {Watson}}{1994}]{Nedoluha:94}
{Nedoluha} G.~E.,  {Watson} W.~D.,  1994, \mn@doi [\apj] {10.1086/173816},
  \href
  {http://adsabs.harvard.edu/cgi-bin/nph-bib_query?bibcode=1994ApJ...423..394N&db_key=AST}
  {423, 394}

\bibitem[\protect\citeauthoryear{{Perley} \& {Taylor}}{{Perley} \&
  {Taylor}}{2003}]{bk:VLAmanual}
{Perley} R.,  {Taylor} G.,  2003, The VLA Calibrator Manual

\bibitem[\protect\citeauthoryear{{Reid}}{{Reid}}{2007}]{Reid:07}
{Reid} M.~J.,  2007, in {J.~M.~Chapman \& W.~A.~Baan} ed.,  IAU Symposium Vol.
  242, Astrophysical Masers \& their Environments. Cambridge University Press,
  p.~522, \mn@doi{10.1017/S1743921307013701}

\bibitem[\protect\citeauthoryear{{Reid} \& {Menten}}{{Reid} \&
  {Menten}}{1997}]{Reid:97}
{Reid} M.~J.,  {Menten} K.~M.,  1997, \apj, \href
  {http://adsabs.harvard.edu/abs/1997ApJ...476..327R} {476, 327}

\bibitem[\protect\citeauthoryear{{Reid} \& {Moran}}{{Reid} \&
  {Moran}}{1981}]{Reid:81b}
{Reid} M.~J.,  {Moran} J.~M.,  1981, \mn@doi [\araa]
  {10.1146/annurev.aa.19.090181.001311}, \href
  {http://adsabs.harvard.edu/abs/1981ARA%26A..19..231R} {19, 231}

\bibitem[\protect\citeauthoryear{{Reid}, {Moran}, {Leach}, {Ball}, {Johnston},
  {Spencer}  \& {Swenson}}{{Reid} et~al.}{1979}]{Reid:79}
{Reid} M.~J.,  {Moran} J.~M.,  {Leach} R.~W.,  {Ball} J.~A.,  {Johnston} K.~J.,
   {Spencer} J.~H.,   {Swenson} G.~W.,  1979, \mn@doi [\apjl] {10.1086/182873},
  \href {http://adsabs.harvard.edu/abs/1979ApJ...227L..89R} {227, L89}

\bibitem[\protect\citeauthoryear{{Richards} et~al.,}{{Richards}
  et~al.}{2004}]{Richards:04}
{Richards} A.~M.~S.,  et~al., 2004, in {R.~Bachiller, F.~Colomer,
  J.-F.~Desmurs, \& P.~de Vicente} ed., Proceedings of the 7th EVN Symposium.
  Observatorio Astronomico Nacional, p.~209 (\mn@eprint {}
  {arXiv:astro-ph/0501028})

\bibitem[\protect\citeauthoryear{{Richter}, {Kemball}  \& {Jonas}}{{Richter}
  et~al.}{2013}]{Richter:13}
{Richter} L.,  {Kemball} A.,   {Jonas} J.,  2013, preprint, \href
  {http://adsabs.harvard.edu/abs/2013arXiv1309.2260R} {} (\mn@eprint {arXiv}
  {1309.2260})

\bibitem[\protect\citeauthoryear{{Sabin}, {Zijlstra}  \& {Greaves}}{{Sabin}
  et~al.}{2007}]{Sabin:07}
{Sabin} L.,  {Zijlstra} A.~A.,   {Greaves} J.~S.,  2007, \mn@doi [\mnras]
  {10.1111/j.1365-2966.2007.11445.x}, \href
  {http://adsabs.harvard.edu/abs/2007MNRAS.376..378S} {376, 378}

\bibitem[\protect\citeauthoryear{{Sabin}, {Wade}  \& {L{\`e}bre}}{{Sabin}
  et~al.}{2015}]{Sabin:15}
{Sabin} L.,  {Wade} G.~A.,   {L{\`e}bre} A.,  2015, \mn@doi [\mnras]
  {10.1093/mnras/stu2227}, \href
  {http://adsabs.harvard.edu/abs/2015MNRAS.446.1988S} {446, 1988}

\bibitem[\protect\citeauthoryear{{Soker}}{{Soker}}{2000}]{Soker:00}
{Soker} N.,  2000, \mnras, \href
  {http://adsabs.harvard.edu/cgi-bin/nph-bib_query?bibcode=2000MNRAS.312..217S&db_key=AST}
  {312, 217}

\bibitem[\protect\citeauthoryear{{Soker} \& {Clayton}}{{Soker} \&
  {Clayton}}{1999}]{Soker:99a}
{Soker} N.,  {Clayton} G.~C.,  1999, \mnras, \href
  {http://adsabs.harvard.edu/cgi-bin/nph-bib_query?bibcode=1999MNRAS.307..993S&db_key=AST}
  {307, 993}

\bibitem[\protect\citeauthoryear{{Szymczak} \& {G{\'e}rard}}{{Szymczak} \&
  {G{\'e}rard}}{2004}]{Szymczak:04}
{Szymczak} M.,  {G{\'e}rard} E.,  2004, \mn@doi [\aap]
  {10.1051/0004-6361:20040461}, \href
  {http://adsabs.harvard.edu/abs/2004A%26A...423..209S} {423, 209}

\bibitem[\protect\citeauthoryear{{Szymczak}, {Cohen}  \& {Richards}}{{Szymczak}
  et~al.}{1998}]{Szymczak:98}
{Szymczak} M.,  {Cohen} R.~J.,   {Richards} A.~M.~S.,  1998, \mn@doi [\mnras]
  {10.1046/j.1365-8711.1998.01601.x}, \href
  {http://adsabs.harvard.edu/abs/1998MNRAS.297.1151S} {297, 1151}

\bibitem[\protect\citeauthoryear{{Thompson}, {Moran}  \& {Swenson}}{{Thompson}
  et~al.}{2004}]{Thompson:04}
{Thompson} A.,  {Moran} J.,   {Swenson} G.~J.,  2004, Interferometry and
  Synthesis in Radio Astronomy, second edn.
WILEY-VCH Verlag GmbH \& Co. KGaA

\bibitem[\protect\citeauthoryear{{Troland}, {Heiles}, {Johnson}  \&
  {Clark}}{{Troland} et~al.}{1979}]{Troland:79}
{Troland} T.~H.,  {Heiles} C.,  {Johnson} D.~R.,   {Clark} F.~O.,  1979,
  \mn@doi [\apj] {10.1086/157273}, \href
  {http://adsabs.harvard.edu/abs/1979ApJ...232..143T} {232, 143}

\bibitem[\protect\citeauthoryear{{Vitrishchak}, {Gabuzda}, {Algaba},
  {Rastorgueva}, {O'Sullivan}  \& {O'Dowd}}{{Vitrishchak}
  et~al.}{2008}]{Vitrishchak:08}
{Vitrishchak} V.~M.,  {Gabuzda} D.~C.,  {Algaba} J.~C.,  {Rastorgueva} E.~A.,
  {O'Sullivan} S.~P.,   {O'Dowd} A.,  2008, \mn@doi [\mnras]
  {10.1111/j.1365-2966.2008.13919.x}, \href
  {http://adsabs.harvard.edu/abs/2008MNRAS.391..124V} {391, 124}

\bibitem[\protect\citeauthoryear{{Vlemmings}}{{Vlemmings}}{2007}]{Vlemmings:07}
{Vlemmings} W.~H.~T.,  2007, in {J.~M.~Chapman \& W.~A.~Baan} ed.,  IAU
  Symposium Vol. 242, Astrophysical Masers \& their Environments. Cambridge
  University Press, p.~37, \mn@doi{10.1017/S1743921307012549}

\bibitem[\protect\citeauthoryear{{Vlemmings} \& {van Langevelde}}{{Vlemmings}
  \& {van Langevelde}}{2008}]{Vlemmings:08}
{Vlemmings} W.~H.~T.,  {van Langevelde} H.~J.,  2008, \mn@doi [\aap]
  {10.1051/0004-6361:200810205}, \href
  {http://adsabs.harvard.edu/abs/2008A%26A...488..619V} {488, 619}

\bibitem[\protect\citeauthoryear{{Vlemmings}, {Diamond}  \& {van
  Langevelde}}{{Vlemmings} et~al.}{2002}]{Vlemmings:02a}
{Vlemmings} W.~H.~T.,  {Diamond} P.~J.,   {van Langevelde} H.~J.,  2002, \aap,
  \href
  {http://adsabs.harvard.edu/cgi-bin/nph-bib_query?bibcode=2002A%26A...394..589V&amp;db_key=AST}
  {394, 589}

\bibitem[\protect\citeauthoryear{{Vlemmings}, {van Langevelde}  \&
  {Diamond}}{{Vlemmings} et~al.}{2005}]{Vlemmings:05}
{Vlemmings} W.~H.~T.,  {van Langevelde} H.~J.,   {Diamond} P.~J.,  2005,
  \mn@doi [\aap] {10.1051/0004-6361:20042488}, \href
  {http://adsabs.harvard.edu/abs/2005A%26A...434.1029V} {434, 1029}

\bibitem[\protect\citeauthoryear{{Vlemmings}, {Diamond}  \& {Imai}}{{Vlemmings}
  et~al.}{2006}]{Vlemmings:06d}
{Vlemmings} W.~H.~T.,  {Diamond} P.~J.,   {Imai} H.,  2006, \mn@doi [\nat]
  {10.1038/nature04466}, \href
  {http://adsabs.harvard.edu/abs/2006Natur.440...58V} {440, 58}

\bibitem[\protect\citeauthoryear{{Vlemmings}, {Ramstedt}, {Rao}  \&
  {Maercker}}{{Vlemmings} et~al.}{2012}]{Vlemmings:12}
{Vlemmings} W.~H.~T.,  {Ramstedt} S.,  {Rao} R.,   {Maercker} M.,  2012,
  \mn@doi [\aap] {10.1051/0004-6361/201218897}, \href
  {http://adsabs.harvard.edu/abs/2012A%26A...540L...3V} {540, L3}

\bibitem[\protect\citeauthoryear{{Wallin} \& {Watson}}{{Wallin} \&
  {Watson}}{1997}]{Wallin:97}
{Wallin} B.~K.,  {Watson} W.~D.,  1997, \mn@doi [\apj] {10.1086/304079}, \href
  {http://adsabs.harvard.edu/cgi-bin/nph-bib_query?bibcode=1997ApJ...481..832W&db_key=AST}
  {481, 832}

\bibitem[\protect\citeauthoryear{{Wardle} \& {Kronberg}}{{Wardle} \&
  {Kronberg}}{1974}]{Wardle:74}
{Wardle} J.~F.~C.,  {Kronberg} P.~P.,  1974, \apj, \href
  {http://adsabs.harvard.edu/cgi-bin/nph-bib_query?bibcode=1974ApJ...194..249W&db_key=AST}
  {194, 249}

\bibitem[\protect\citeauthoryear{{Watson}}{{Watson}}{1994}]{Watson:94}
{Watson} W.~D.,  1994, \mn@doi [\apjl] {10.1086/187269}, \href
  {http://adsabs.harvard.edu/abs/1994ApJ...424L..37W} {424, L37}

\bibitem[\protect\citeauthoryear{{Watson}}{{Watson}}{2002}]{Watson:02}
{Watson} W.~D.,  2002, in {V.~Migenes \& M.~J.~Reid} ed.,  IAU Symposium Vol.
  206, Cosmic Masers: From Proto-Stars to Black Holes. Astronomical Society of
  the Pacific, p.~464 (\mn@eprint {} {arXiv:astro-ph/0107572})

\bibitem[\protect\citeauthoryear{{Watson}}{{Watson}}{2009}]{Watson:09}
{Watson} W.~D.,  2009, in {A.~Esquivel, J.~Franco, G.~Garcia-Segura, E.~de
  Gouveia Dal Pino, A.~Lazarian, S.~Lizano \& A.~Raga} ed.,  Revista Mexicana
  de Astronomia y Astrofisica Conference Series Vol. 36, Magnetic Fields in the
  Universe II: from Laboratory and Stars to the Primordial Universe. Instituto
  de Astronomia, p.~113 (\mn@eprint {arXiv} {0811.1292})

\bibitem[\protect\citeauthoryear{{Watson} \& {Wyld}}{{Watson} \&
  {Wyld}}{2001}]{Watson:01}
{Watson} W.~D.,  {Wyld} H.~W.,  2001, \mn@doi [\apjl] {10.1086/323513}, \href
  {http://adsabs.harvard.edu/cgi-bin/nph-bib_query?bibcode=2001ApJ...558L..55W&db_key=AST}
  {558, L55}

\bibitem[\protect\citeauthoryear{{Western} \& {Watson}}{{Western} \&
  {Watson}}{1983}]{Western:83c}
{Western} L.~R.,  {Watson} W.~D.,  1983, \apj, \href
  {http://adsabs.harvard.edu/cgi-bin/nph-bib_query?bibcode=1983ApJ...275..195W&db_key=AST}
  {275, 195}

\bibitem[\protect\citeauthoryear{{Western} \& {Watson}}{{Western} \&
  {Watson}}{1984}]{Western:84}
{Western} L.~R.,  {Watson} W.~D.,  1984, \mn@doi [\apj] {10.1086/162487}, \href
  {http://adsabs.harvard.edu/cgi-bin/nph-bib_query?bibcode=1984ApJ...285..158W&db_key=AST}
  {285, 158}

\bibitem[\protect\citeauthoryear{{Whiting}}{{Whiting}}{2012}]{Whiting:12}
{Whiting} M.~T.,  2012, \mn@doi [\mnras] {10.1111/j.1365-2966.2012.20548.x},
  \href {http://adsabs.harvard.edu/abs/2012MNRAS.421.3242W} {421, 3242}

\bibitem[\protect\citeauthoryear{{Wiebe} \& {Watson}}{{Wiebe} \&
  {Watson}}{1998}]{Wiebe:98}
{Wiebe} D.~S.,  {Watson} W.~D.,  1998, \apjl, \href
  {http://adsabs.harvard.edu/cgi-bin/nph-bib_query?bibcode=1998ApJ...503L..71W&db_key=AST}
  {503, L71}

\bibitem[\protect\citeauthoryear{{Wittkowski} et~al.,}{{Wittkowski}
  et~al.}{2011}]{Wittkowski:11}
{Wittkowski} M.,  et~al., 2011, \mn@doi [\aap] {10.1051/0004-6361/201117411},
  \href {http://adsabs.harvard.edu/abs/2011A%26A...532L...7W} {532, L7}

\bibitem[\protect\citeauthoryear{{Wittkowski}, {Hauschildt}, {Arroyo-Torres}
  \& {Marcaide}}{{Wittkowski} et~al.}{2012}]{Wittkowski:12}
{Wittkowski} M.,  {Hauschildt} P.~H.,  {Arroyo-Torres} B.,   {Marcaide} J.~M.,
  2012, \mn@doi [\aap] {10.1051/0004-6361/201219126}, \href
  {http://adsabs.harvard.edu/abs/2012A%26A...540L..12W} {540, L12}

\bibitem[\protect\citeauthoryear{{Wong}, {Kami{\'n}ski}, {Menten}  \&
  {Wyrowski}}{{Wong} et~al.}{2016}]{Wong:16}
{Wong} K.~T.,  {Kami{\'n}ski} T.,  {Menten} K.~M.,   {Wyrowski} F.,  2016,
  preprint, \href {http://adsabs.harvard.edu/abs/2016arXiv160303371W} {}
  (\mn@eprint {arXiv} {1603.03371})

\bibitem[\protect\citeauthoryear{{Zhang}, {Reid}, {Menten}  \& {Zheng}}{{Zhang}
  et~al.}{2012}]{Zhang:12}
{Zhang} B.,  {Reid} M.~J.,  {Menten} K.~M.,   {Zheng} X.~W.,  2012, \mn@doi
  [\apj] {10.1088/0004-637X/744/1/23}, \href
  {http://adsabs.harvard.edu/abs/2012ApJ...744...23Z} {744, 23}

\makeatother
\end{thebibliography}


\appendix

\section{Maser features}
\label{appendixA}

The maser features detected in the v=1 J=2-1, \mbox{v=1 J=1-0} and v=2 J=1-0 SiO maser transitions are listed in Tables~\ref{tab:BR123A-components}
to \ref{tab:BR123D1-components}. In each table, the columns from left to right are: 
Velocity $v$, full velocity extent $\Delta v$ of the feature, position offsets $\Delta\alpha$ and $\Delta\delta$ relative to the aligned image 
centre, Stokes $I$, error in Stokes $I$ $\sigma_I$, Stokes $V$, error in Stokes $V$ $\sigma_V$, Stokes $Q$, error in Stokes $Q$ $\sigma_Q$, 
Stokes $U$, error in Stokes $U$ $\sigma_U$, fractional circular polarisation $m_c$, error in fractional circular polarisation $\sigma_c$, 
fractional linear polarisation $m_l$, error in fractional linear polarisation $\sigma_l$, polarisation position angle $\chi$, error in polarisation 
position angle $\sigma_{\chi}$. The velocity and Stokes parameters of each feature are measured at the position of the peak Stokes $I$. The units 
of the Stokes parameters and their uncertainties are Janskys per synthesized beam (Jy/beam). The uncertainties in the Stokes parameters have been 
empirically broadened as described in the text, and the linear polarisation calculation includes a correction for Ricean bias. The fractional 
linear and circular polarisation values are presented if they exceed the detection threshold described in the text.


\begin{table*}
\caption[List of maser features detected in the v=1 J=2-1 SiO maser transition.]
{\footnotesize List of maser features detected in the v=1 J=2-1 SiO maser transition.
The column descriptions are given in the text.}
\tiny
\begin{tabular}{llllllllllllllllllll}
\hline
$v$ & $\Delta v$ & $\Delta\alpha$ & $\Delta\delta$ & I & $\sigma_I$ & Q & $\sigma_Q$ & U & $\sigma_U$ & V & $\sigma_V$ &  $m_c$ & $\sigma_c$ & $m_l$ & $\sigma_l$ & $\chi$ & $\sigma_{\chi}$ \\
$[$km/s$]$ & $[$km/s$]$ & $[$mas$]$ & $[$mas$]$ & $[$Jy/b$]$ & $[$Jy/b$]$ & $[$Jy/b$]$ & $[$Jy/b$]$ & $[$Jy/b$]$ & $[$Jy/b$]$ & $[$Jy/b$]$ & $[$Jy/b$]$ & $[\%]$ & $[\%]$ & $[\%]$ & $[\%]$ & $[$deg$]$ & $[$deg$]$ \\
\hline
-1.99 & 0.44 & -6.33 & -20.64 & 1.311 & 0.055 & 0.014 & 0.052 & 0.062 & 0.051 & -0.027 & 0.053 &   &  &   &   &   &   \\
3.23 & 3.48 & -18.93 & -3.93 & 11.113 & 0.063 & 0.071 & 0.058 & 0.023 & 0.061 & -0.131 & 0.059 & -1.18 & 0.53 &   &   &   &   \\
3.23 & 2.61 & -19.20 & 5.73 & 3.308 & 0.063 & -0.040 & 0.058 & 0.048 & 0.061 & -0.046 & 0.059 &   &  &   &   &   &   \\
4.53 & 0.44 & -13.95 & -34.44 & 1.935 & 0.051 & 0.005 & 0.051 & 0.083 & 0.050 & -0.020 & 0.049 &   &  &   &   &   &   \\
5.40 & 1.74 & 15.30 & -0.30 & 4.117 & 0.050 & -0.026 & 0.050 & 0.003 & 0.049 & -0.050 & 0.051 &   &  &   &   &   &   \\
7.14 & 1.30 & -18.39 & -19.92 & 2.171 & 0.048 & 0.042 & 0.049 & 0.004 & 0.048 & -0.102 & 0.050 & -4.71 & 2.29 &   &   &   &   \\
7.14 & 0.87 & -24.12 & 0.00 & 1.897 & 0.048 & -0.101 & 0.049 & 0.035 & 0.048 & 0.015 & 0.050 &   &  &   &   &   &   \\
7.57 & 2.61 & -18.69 & 6.45 & 2.554 & 0.047 & -0.004 & 0.050 & 0.010 & 0.047 & 0.014 & 0.050 &   &  &   &   &   &   \\
8.87 & 0.87 & -19.35 & -20.88 & 1.875 & 0.056 & 0.007 & 0.053 & 0.026 & 0.052 & -0.043 & 0.053 &   &  &   &   &   &   \\
8.87 & 0.87 & -26.25 & -0.36 & 1.808 & 0.056 & 0.016 & 0.053 & -0.002 & 0.052 & 0.042 & 0.053 &   &  &   &   &   &   \\
9.74 & 1.30 & -16.95 & 7.11 & 5.497 & 0.051 & -0.035 & 0.050 & -0.021 & 0.051 & -0.049 & 0.048 &   &  &   &   &   &   \\
9.74 & 0.87 & -18.63 & -2.55 & 1.900 & 0.051 & -0.021 & 0.050 & 0.062 & 0.051 & 0.017 & 0.048 &   &  &   &   &   &   \\
10.18 & 0.87 & -17.76 & -3.54 & 5.643 & 0.045 & -0.014 & 0.048 & 0.047 & 0.049 & -0.019 & 0.046 &   &  &   &   &   &   \\
12.35 & 1.30 & -18.24 & -2.94 & 8.091 & 0.045 & 0.059 & 0.046 & -0.061 & 0.046 & -0.142 & 0.043 & -1.76 & 0.53 &   &   &   &   \\
13.65 & 0.87 & -18.78 & -1.95 & 2.703 & 0.055 & 0.008 & 0.054 & -0.026 & 0.053 & -0.052 & 0.053 &   &  &   &   &   &   \\
14.09 & 0.44 & 33.45 & 0.36 & 2.903 & 0.078 & 0.031 & 0.075 & 0.065 & 0.074 & -0.057 & 0.073 &   &  &   &   &   &   \\
14.09 & 0.44 & -31.86 & 3.63 & 2.255 & 0.078 & -0.064 & 0.075 & 0.050 & 0.074 & -0.034 & 0.073 &   &  &   &   &   &   \\
14.09 & 0.44 & -32.01 & -1.11 & 3.829 & 0.078 & -0.046 & 0.075 & -0.072 & 0.074 & -0.021 & 0.073 &   &  &   &   &   &   \\
19.30 & 0.87 & 20.46 & -21.06 & 1.537 & 0.061 & 0.072 & 0.058 & 0.065 & 0.060 & 0.004 & 0.060 &   &  &   &   &   &   \\
19.30 & 0.44 & 16.65 & -23.58 & 1.046 & 0.061 & -0.069 & 0.058 & 0.007 & 0.060 & -0.028 & 0.060 &   &  &   &   &   &   \\
19.74 & 0.44 & 19.14 & -24.12 & 2.575 & 0.102 & 0.034 & 0.101 & -0.011 & 0.102 & -0.015 & 0.101 &   &  &   &   &   &   \\
20.61 & 1.30 & 20.43 & -24.42 & 5.166 & 0.107 & 0.268 & 0.101 & -0.045 & 0.104 & -0.061 & 0.102 &   &  &   &   &   &   \\
20.61 & 2.17 & 18.33 & -23.79 & 5.595 & 0.107 & -0.026 & 0.101 & 0.238 & 0.104 & -0.037 & 0.102 &   &  &   &   &   &   \\
20.61 & 0.44 & 0.24 & -22.02 & 1.568 & 0.107 & -0.033 & 0.101 & -0.008 & 0.104 & 0.018 & 0.102 &   &  &   &   &   &   \\
21.04 & 1.30 & 19.86 & -25.98 & 12.601 & 0.185 & -0.626 & 0.176 & -0.099 & 0.182 & -0.193 & 0.187 &   &  & 4.83 & 1.14 & -85.50 & 8.94 \\
21.48 & 0.44 & 22.08 & -24.30 & 3.641 & 0.200 & -0.018 & 0.194 & -0.117 & 0.192 & -0.092 & 0.187 &   &  &   &   &   &   \\
21.91 & 0.87 & 21.51 & -26.58 & 16.183 & 0.156 & -2.407 & 0.163 & -0.422 & 0.169 & -0.268 & 0.157 &   &  & 15.07 & 1.58 & -85.03 & 1.99 \\
22.78 & 0.44 & 19.23 & -30.21 & 8.730 & 0.057 & -0.101 & 0.055 & 0.045 & 0.054 & -0.128 & 0.053 & -1.46 & 0.61 &   &   &   &   \\
23.22 & 0.44 & 25.20 & -22.17 & 1.731 & 0.063 & 0.002 & 0.063 & -0.067 & 0.063 & -0.046 & 0.060 &   &  &   &   &   &   \\
24.95 & 0.44 & 27.15 & -20.76 & 4.136 & 0.046 & -0.058 & 0.045 & -0.056 & 0.046 & 0.007 & 0.046 &   &  &   &   &   &   \\
26.69 & 0.44 & -8.70 & 12.69 & 1.797 & 0.050 & 0.023 & 0.048 & 0.011 & 0.049 & 0.014 & 0.046 &   &  &   &   &   &   \\
27.13 & 0.44 & -2.46 & 6.69 & 1.192 & 0.071 & -0.019 & 0.070 & 0.048 & 0.069 & 0.014 & 0.064 &   &  &   &   &   &   \\
27.13 & 0.44 & -8.67 & 9.96 & 1.072 & 0.071 & -0.050 & 0.070 & 0.018 & 0.069 & -0.054 & 0.064 &   &  &   &   &   &   \\
27.13 & 0.44 & -23.79 & -23.43 & 3.003 & 0.071 & -0.056 & 0.070 & -0.018 & 0.069 & -0.065 & 0.064 &   &  &   &   &   &   \\
27.56 & 0.87 & -8.73 & 8.76 & 5.864 & 0.056 & -0.009 & 0.056 & 0.088 & 0.056 & 0.034 & 0.058 &   &  &   &   &   &   \\
28.43 & 0.44 & 1.11 & -44.55 & 1.999 & 0.059 & -0.005 & 0.060 & 0.009 & 0.058 & -0.055 & 0.057 &   &  &   &   &   &   \\
29.30 & 0.44 & 12.18 & -10.17 & 1.175 & 0.060 & -0.033 & 0.059 & -0.009 & 0.059 & -0.006 & 0.062 &   &  &   &   &   &   \\
29.73 & 1.74 & -3.69 & 5.64 & 6.151 & 0.067 & 0.104 & 0.065 & 0.054 & 0.067 & -0.016 & 0.069 &   &  &   &   &   &   \\
30.17 & 0.44 & -32.49 & -19.62 & 3.033 & 0.058 & 0.058 & 0.057 & -0.213 & 0.055 & -0.034 & 0.055 &   &  & 7.05 & 0.88 & -37.42 & 7.89 \\
31.04 & 2.61 & -3.93 & 3.96 & 7.506 & 0.142 & -0.149 & 0.145 & 0.019 & 0.138 & -0.021 & 0.142 &   &  &   &   &   &   \\
31.04 & 8.26 & -8.37 & -25.26 & 53.815 & 0.142 & -1.694 & 0.145 & -3.012 & 0.138 & -1.187 & 0.142 & -2.21 & 0.26 & 6.42 & 0.48 & -59.68 & 1.19 \\
32.78 & 1.30 & 20.64 & -14.40 & 6.326 & 0.063 & -1.866 & 0.061 & -0.081 & 0.065 & 0.004 & 0.062 &   &  & 29.51 & 1.35 & -88.75 & 1.01 \\
33.64 & 0.44 & -4.92 & 4.47 & 1.294 & 0.056 & -0.067 & 0.059 & 0.036 & 0.060 & 0.015 & 0.058 &   &  &   &   &   &   \\
34.51 & 0.44 & 18.45 & -14.22 & 2.751 & 0.055 & 0.380 & 0.056 & 0.793 & 0.055 & -0.119 & 0.057 & -4.31 & 2.06 & 31.88 & 1.98 & 32.20 & 1.83 \\
34.95 & 0.44 & 3.51 & -18.03 & 1.624 & 0.051 & -0.076 & 0.046 & 0.041 & 0.046 & 0.033 & 0.049 &   &  &   &   &   &   \\
36.25 & 0.87 & -9.99 & -28.35 & 2.632 & 0.051 & -0.017 & 0.051 & -0.107 & 0.052 & 0.084 & 0.051 &   &  &   &   &   &   \\
37.99 & 0.44 & -10.08 & -27.09 & 1.890 & 0.042 & -0.060 & 0.041 & -0.185 & 0.040 & -0.031 & 0.041 &   &  & 10.05 & 0.97 & -53.98 & 6.26 \\
40.16 & 0.44 & -9.12 & -18.99 & 1.796 & 0.048 & -0.032 & 0.049 & -0.032 & 0.049 & -0.037 & 0.050 &   &  &   &   &   &   \\

\hline
\label{tab:BR123A-components}
\end{tabular}
\begin{flushleft}
\end{flushleft}
\end{table*}

\begin{table*}
\caption[List of maser features detected in the v=1 J=1-0 SiO maser transition.]
{\footnotesize List of maser features detected in the v=1 J=1-0 SiO maser transition.
The column descriptions are given in the text.}
\tiny
\begin{tabular}{llllllllllllllllllll}
\hline
$v$ & $\Delta v$ & $\Delta\alpha$ & $\Delta\delta$ & I & $\sigma_I$ & Q & $\sigma_Q$ & U & $\sigma_U$ & V & $\sigma_V$ &  $m_c$ & $\sigma_c$ & $m_l$ & $\sigma_l$ & $\chi$ & $\sigma_{\chi}$ \\
$[$km/s$]$ & $[$km/s$]$ & $[$mas$]$ & $[$mas$]$ & $[$Jy/b$]$ & $[$Jy/b$]$ & $[$Jy/b$]$ & $[$Jy/b$]$ & $[$Jy/b$]$ & $[$Jy/b$]$ & $[$Jy/b$]$ & $[$Jy/b$]$ & $[\%]$ & $[\%]$ & $[\%]$ & $[\%]$ & $[$deg$]$ & $[$deg$]$ \\
\hline   
3.22 & 1.74 & -14.52 & -34.77 & 2.280 & 0.020 & 0.093 & 0.021 & -0.005 & 0.019 & -0.010 & 0.020 &   &  & 4.01 & 0.94 & -8.36 & 6.21 \\
4.09 & 0.87 & -13.95 & -34.68 & 2.792 & 0.021 & 0.008 & 0.020 & -0.259 & 0.020 & -0.057 & 0.020 & -2.03 & 0.72 & 9.25 & 0.73 & -50.86 & 2.26 \\
5.40 & 0.44 & 26.76 & 36.30 & 1.023 & 0.018 & 0.008 & 0.018 & -0.005 & 0.018 & 0.015 & 0.017 &   &  &   &   &   &   \\
5.40 & 0.44 & 20.79 & 43.23 & 0.740 & 0.018 & -0.008 & 0.018 & 0.012 & 0.018 & 0.011 & 0.017 &   &  &   &   &   &   \\
5.83 & 0.44 & 27.96 & 38.37 & 0.922 & 0.031 & -0.003 & 0.029 & 0.006 & 0.028 & -0.011 & 0.027 &   &  &   &   &   &   \\
6.27 & 1.74 & 28.68 & 37.20 & 3.211 & 0.036 & -0.011 & 0.034 & 0.012 & 0.033 & 0.017 & 0.034 &   &  &   &   &   &   \\
6.27 & 0.44 & -23.28 & -11.22 & 0.698 & 0.036 & -0.002 & 0.034 & 0.008 & 0.033 & 1.8e-4 & 0.034 &   &  &   &   &   &   \\
6.70 & 1.74 & 29.91 & 39.33 & 1.622 & 0.032 & -0.028 & 0.030 & 0.013 & 0.030 & 0.013 & 0.030 &   &  &   &   &   &   \\
6.70 & 2.61 & -16.14 & 7.02 & 5.258 & 0.032 & -0.009 & 0.030 & -0.008 & 0.030 & 0.145 & 0.030 & 2.75 & 0.57 &   &   &   &   \\
8.44 & 2.17 & 23.73 & 29.19 & 1.755 & 0.021 & -0.004 & 0.020 & 0.037 & 0.020 & -0.011 & 0.019 &   &  &   &   &   &   \\
9.74 & 1.30 & 23.10 & 27.39 & 2.040 & 0.018 & -0.096 & 0.018 & 0.064 & 0.018 & -0.014 & 0.018 &   &  & 5.57 & 0.88 & 66.40 & 4.52 \\
9.74 & 1.30 & 20.40 & 2.25 & 2.042 & 0.018 & -0.094 & 0.018 & -0.030 & 0.018 & 0.014 & 0.018 &   &  & 4.73 & 0.87 & -87.92 & 5.37 \\
12.35 & 1.30 & 25.47 & 0.15 & 1.402 & 0.018 & 0.004 & 0.018 & 0.122 & 0.018 & 0.002 & 0.018 &   &  & 8.63 & 1.32 & 37.27 & 4.31 \\
12.79 & 0.44 & 26.37 & -0.63 & 0.719 & 0.016 & -0.013 & 0.015 & 0.010 & 0.015 & 0.001 & 0.015 &   &  &   &   &   &   \\
13.65 & 1.30 & 13.68 & 20.28 & 1.058 & 0.020 & 0.013 & 0.019 & 0.013 & 0.021 & -0.002 & 0.019 &   &  &   &   &   &   \\
14.09 & 0.87 & 33.42 & 0.18 & 7.551 & 0.024 & -0.108 & 0.024 & -0.500 & 0.024 & 0.139 & 0.023 & 1.84 & 0.31 & 6.77 & 0.31 & -57.91 & 1.33 \\
14.52 & 0.44 & 5.10 & 21.45 & 0.852 & 0.021 & 0.002 & 0.022 & -0.019 & 0.024 & 0.004 & 0.023 &   &  &   &   &   &   \\
14.96 & 0.44 & 11.34 & 37.74 & 0.759 & 0.025 & 0.101 & 0.026 & -0.145 & 0.024 & 0.008 & 0.026 &   &  & 23.12 & 3.39 & -34.34 & 4.17 \\
15.39 & 3.48 & 10.32 & 17.82 & 1.763 & 0.027 & -0.005 & 0.026 & -0.015 & 0.027 & 0.061 & 0.027 & 3.48 & 1.51 &   &   &   &   \\
15.83 & 0.87 & 24.48 & -43.68 & 1.011 & 0.034 & -0.350 & 0.035 & 0.008 & 0.035 & -0.007 & 0.034 &   &  & 34.47 & 3.66 & 82.52 & 2.89 \\
15.83 & 2.17 & 12.24 & 16.71 & 1.523 & 0.034 & 0.007 & 0.035 & -0.003 & 0.035 & 0.025 & 0.034 &   &  &   &   &   &   \\
15.83 & 0.44 & 10.05 & 16.50 & 0.655 & 0.034 & 0.014 & 0.035 & 0.011 & 0.035 & 0.030 & 0.034 &   &  &   &   &   &   \\
16.26 & 1.30 & 9.63 & 19.29 & 0.616 & 0.052 & -0.014 & 0.052 & 0.014 & 0.053 & 0.026 & 0.053 &   &  &   &   &   &   \\
16.26 & 0.87 & -28.62 & -0.18 & 5.903 & 0.052 & 0.314 & 0.052 & -0.061 & 0.053 & -0.145 & 0.053 & -2.45 & 0.89 & 5.35 & 0.90 & -12.28 & 4.85 \\
17.13 & 0.87 & 14.01 & 44.04 & 2.161 & 0.034 & 0.402 & 0.035 & -0.006 & 0.036 & 0.013 & 0.036 &   &  & 18.55 & 1.66 & -7.23 & 2.60 \\
17.13 & 2.17 & 3.42 & 22.08 & 1.832 & 0.034 & -0.003 & 0.035 & -0.065 & 0.036 & 0.001 & 0.036 &   &  &   &   &   &   \\
17.57 & 1.30 & 24.03 & 26.58 & 1.223 & 0.037 & 0.150 & 0.037 & 0.011 & 0.036 & 0.017 & 0.035 &   &  & 11.91 & 3.13 & -4.78 & 7.25 \\
17.57 & 0.44 & 13.44 & 39.96 & 0.624 & 0.037 & 0.006 & 0.037 & -0.024 & 0.036 & -0.003 & 0.035 &   &  &   &   &   &   \\
17.57 & 0.87 & 12.84 & 43.11 & 0.753 & 0.037 & 0.020 & 0.037 & 0.007 & 0.036 & -0.002 & 0.035 &   &  &   &   &   &   \\
18.00 & 0.44 & 14.91 & 36.84 & 0.651 & 0.034 & 0.013 & 0.037 & -0.008 & 0.034 & -0.009 & 0.036 &   &  &   &   &   &   \\
18.00 & 0.44 & 14.04 & 36.81 & 0.860 & 0.034 & 0.022 & 0.037 & 4.1e-4 & 0.034 & -0.007 & 0.036 &   &  &   &   &   &   \\
18.00 & 0.44 & 11.40 & 42.72 & 0.690 & 0.034 & 0.006 & 0.037 & -0.007 & 0.034 & 0.008 & 0.036 &   &  &   &   &   &   \\
18.00 & 4.35 & 8.43 & 18.30 & 1.324 & 0.034 & 0.004 & 0.037 & 6.4e-5 & 0.034 & 0.006 & 0.036 &   &  &   &   &   &   \\
18.43 & 0.44 & 7.68 & 18.78 & 0.776 & 0.045 & 0.020 & 0.046 & 0.005 & 0.048 & 0.019 & 0.046 &   &  &   &   &   &   \\
18.87 & 0.44 & 19.86 & -21.51 & 0.979 & 0.048 & 0.008 & 0.051 & 0.017 & 0.049 & 0.009 & 0.048 &   &  &   &   &   &   \\
18.87 & 0.87 & 7.71 & 37.98 & 2.262 & 0.048 & 0.226 & 0.051 & -0.072 & 0.049 & -0.002 & 0.048 &   &  & 10.26 & 2.29 & -15.56 & 6.15 \\
18.87 & 0.44 & 7.38 & 37.02 & 0.753 & 0.048 & 0.140 & 0.051 & -0.018 & 0.049 & 0.010 & 0.048 &   &  &   &   &   &   \\
19.30 & 0.44 & 21.54 & -22.65 & 0.933 & 0.077 & -0.001 & 0.070 & 0.003 & 0.069 & -0.024 & 0.071 &   &  &   &   &   &   \\
19.30 & 0.44 & 20.91 & -22.68 & 0.667 & 0.077 & -0.001 & 0.070 & 0.019 & 0.069 & -0.009 & 0.071 &   &  &   &   &   &   \\
19.30 & 6.52 & 15.99 & 14.16 & 5.225 & 0.077 & -0.023 & 0.070 & 0.015 & 0.069 & -0.194 & 0.071 & -3.72 & 1.36 &   &   &   &   \\
19.30 & 1.30 & 9.48 & 40.98 & 2.219 & 0.077 & 0.017 & 0.070 & -0.023 & 0.069 & -0.004 & 0.071 &   &  &   &   &   &   \\
19.30 & 0.87 & 8.43 & 38.01 & 0.916 & 0.077 & 0.008 & 0.070 & -0.039 & 0.069 & -0.022 & 0.071 &   &  &   &   &   &   \\
19.30 & 0.44 & 7.32 & 19.08 & 0.658 & 0.077 & 0.005 & 0.070 & 0.002 & 0.069 & 0.015 & 0.071 &   &  &   &   &   &   \\
19.30 & 0.44 & 1.35 & 30.93 & 0.744 & 0.077 & 0.011 & 0.070 & -0.004 & 0.069 & 0.012 & 0.071 &   &  &   &   &   &   \\
19.74 & 1.30 & 13.95 & 15.87 & 0.823 & 0.066 & 0.004 & 0.061 & 0.003 & 0.058 & 0.014 & 0.063 &   &  &   &   &   &   \\
20.17 & 0.44 & 5.88 & 18.00 & 0.739 & 0.041 & -0.007 & 0.043 & -0.008 & 0.042 & -0.004 & 0.045 &   &  &   &   &   &   \\
20.61 & 0.44 & 22.92 & -27.78 & 0.660 & 0.075 & 0.002 & 0.071 & -0.014 & 0.068 & -3.1e-4 & 0.066 &   &  &   &   &   &   \\
20.61 & 0.44 & 21.51 & -27.24 & 0.554 & 0.075 & -0.003 & 0.071 & -0.008 & 0.068 & 0.017 & 0.066 &   &  &   &   &   &   \\
20.61 & 0.87 & 20.97 & -26.88 & 0.819 & 0.075 & -0.002 & 0.071 & -0.016 & 0.068 & 0.006 & 0.066 &   &  &   &   &   &   \\
20.61 & 0.44 & 8.10 & 14.40 & 0.813 & 0.075 & 0.006 & 0.071 & 0.003 & 0.068 & 0.008 & 0.066 &   &  &   &   &   &   \\
20.61 & 0.44 & 6.48 & 36.75 & 1.462 & 0.075 & 0.042 & 0.071 & -0.119 & 0.068 & 0.012 & 0.066 &   &  &   &   &   &   \\
20.61 & 0.44 & 6.45 & 11.94 & 0.981 & 0.075 & 0.005 & 0.071 & 0.005 & 0.068 & 0.008 & 0.066 &   &  &   &   &   &   \\
20.61 & 0.44 & 6.00 & 35.67 & 0.696 & 0.075 & 0.022 & 0.071 & -0.013 & 0.068 & 0.006 & 0.066 &   &  &   &   &   &   \\
20.61 & 0.44 & 4.98 & 17.46 & 0.994 & 0.075 & -0.009 & 0.071 & -0.013 & 0.068 & -0.005 & 0.066 &   &  &   &   &   &   \\
21.04 & 0.87 & 22.41 & -27.48 & 1.082 & 0.087 & -0.009 & 0.074 & -0.003 & 0.069 & 0.024 & 0.072 &   &  &   &   &   &   \\
21.04 & 0.87 & 18.72 & -20.85 & 0.905 & 0.087 & -0.005 & 0.074 & 0.001 & 0.069 & 0.023 & 0.072 &   &  &   &   &   &   \\
21.04 & 0.44 & 18.36 & -20.58 & 0.652 & 0.087 & -0.002 & 0.074 & 0.005 & 0.069 & 0.013 & 0.072 &   &  &   &   &   &   \\
21.04 & 0.44 & 10.77 & 10.71 & 0.679 & 0.087 & -0.010 & 0.074 & 0.005 & 0.069 & 0.005 & 0.072 &   &  &   &   &   &   \\
21.04 & 0.44 & 6.03 & 12.27 & 0.509 & 0.087 & -0.009 & 0.074 & -0.008 & 0.069 & 0.001 & 0.072 &   &  &   &   &   &   \\
21.04 & 0.44 & 4.98 & 34.11 & 0.629 & 0.087 & -0.001 & 0.074 & -0.080 & 0.069 & -0.012 & 0.072 &   &  &   &   &   &   \\
21.48 & 3.48 & 21.48 & -24.66 & 16.871 & 0.089 & -0.029 & 0.066 & 0.275 & 0.062 & 0.026 & 0.065 &   &  & 1.60 & 0.38 & 41.26 & 7.22 \\
21.48 & 0.44 & 12.24 & 12.12 & 0.828 & 0.089 & 0.003 & 0.066 & -0.003 & 0.062 & 0.003 & 0.065 &   &  &   &   &   &   \\
21.48 & 4.35 & 8.70 & 10.53 & 3.235 & 0.089 & 0.001 & 0.066 & 0.027 & 0.062 & 0.026 & 0.065 &   &  &   &   &   &   \\
21.91 & 0.44 & 15.84 & 3.03 & 0.526 & 0.059 & 0.006 & 0.048 & 0.051 & 0.049 & 0.027 & 0.047 &   &  &   &   &   &   \\
21.91 & 3.91 & 12.84 & 10.02 & 7.306 & 0.059 & 0.017 & 0.048 & 0.242 & 0.049 & 0.410 & 0.047 & 5.62 & 0.65 & 3.25 & 0.69 & 36.15 & 5.94 \\
21.91 & 0.44 & 7.44 & 12.96 & 0.687 & 0.059 & -0.001 & 0.048 & 0.014 & 0.049 & 0.022 & 0.047 &   &  &   &   &   &   \\
21.91 & 0.44 & 4.74 & 11.70 & 0.778 & 0.059 & -0.002 & 0.048 & -0.010 & 0.049 & -0.008 & 0.047 &   &  &   &   &   &   \\
22.35 & 0.87 & 10.41 & 5.76 & 0.932 & 0.037 & 0.009 & 0.035 & 0.022 & 0.034 & 0.026 & 0.034 &   &  &   &   &   &   \\
22.78 & 0.44 & 16.38 & 31.05 & 0.610 & 0.034 & -0.019 & 0.033 & 0.024 & 0.031 & 0.014 & 0.032 &   &  &   &   &   &   \\
23.65 & 0.44 & 16.14 & 16.56 & 0.653 & 0.044 & -0.011 & 0.036 & -0.009 & 0.040 & -0.007 & 0.041 &   &  &   &   &   &   \\
23.65 & 0.44 & 3.42 & 20.55 & 1.020 & 0.044 & 0.012 & 0.036 & -0.001 & 0.040 & -0.007 & 0.041 &   &  &   &   &   &   \\
24.08 & 0.44 & 2.91 & 23.94 & 1.295 & 0.068 & 0.003 & 0.063 & -0.018 & 0.065 & -0.031 & 0.065 &   &  &   &   &   &   \\
24.08 & 0.87 & -4.56 & 17.55 & 1.372 & 0.068 & -0.011 & 0.063 & -0.023 & 0.065 & -0.003 & 0.065 &   &  &   &   &   &   \\
24.52 & 1.30 & 1.80 & 21.12 & 5.650 & 0.062 & 0.029 & 0.049 & -0.134 & 0.052 & -0.044 & 0.050 &   &  &   &   &   &   \\
24.52 & 3.91 & 0.90 & 9.12 & 11.387 & 0.062 & 0.415 & 0.049 & 0.361 & 0.052 & -0.239 & 0.050 & -2.10 & 0.44 & 4.81 & 0.44 & 13.70 & 2.66 \\
25.39 & 4.35 & 13.17 & 5.22 & 9.455 & 0.063 & -0.013 & 0.055 & -0.011 & 0.056 & 0.030 & 0.056 &   &  &   &   &   &   \\
25.39 & 0.44 & -0.30 & 20.94 & 0.945 & 0.063 & -0.007 & 0.055 & -0.001 & 0.056 & -0.010 & 0.056 &   &  &   &   &   &   \\
25.39 & 0.44 & -1.53 & 22.14 & 0.959 & 0.063 & 0.006 & 0.055 & 0.005 & 0.056 & -0.006 & 0.056 &   &  &   &   &   &   \\
25.39 & 0.87 & -5.97 & 17.28 & 1.379 & 0.063 & -0.009 & 0.055 & -0.010 & 0.056 & 0.010 & 0.056 &   &  &   &   &   &   \\
25.82 & 1.30 & -3.06 & 7.50 & 2.621 & 0.033 & 0.054 & 0.030 & 0.001 & 0.029 & 0.012 & 0.031 &   &  &   &   &   &   \\
25.82 & 0.87 & -6.36 & 14.64 & 2.127 & 0.033 & 0.015 & 0.030 & -0.008 & 0.029 & 0.005 & 0.031 &   &  &   &   &   &   \\
26.26 & 0.44 & -2.31 & 20.04 & 0.735 & 0.027 & 0.019 & 0.025 & -0.011 & 0.025 & -0.003 & 0.026 &   &  &   &   &   &   \\
26.26 & 1.30 & -3.69 & 19.29 & 1.203 & 0.027 & 0.025 & 0.025 & 0.020 & 0.025 & 0.003 & 0.026 &   &  &   &   &   &   \\
26.26 & 0.44 & -4.50 & 18.39 & 0.871 & 0.027 & -5.2e-5 & 0.025 & -0.008 & 0.025 & 0.004 & 0.026 &   &  &   &   &   &   \\
26.69 & 0.87 & -1.89 & 11.88 & 2.741 & 0.025 & 0.274 & 0.025 & 0.024 & 0.025 & -0.050 & 0.024 & -1.82 & 0.88 & 9.99 & 0.94 & -4.25 & 2.62 \\
28.00 & 1.74 & 1.35 & -45.03 & 4.416 & 0.028 & -0.099 & 0.026 & 0.590 & 0.027 & -0.011 & 0.027 &   &  & 13.54 & 0.61 & 42.99 & 1.25 \\
28.86 & 0.87 & 0.54 & 5.58 & 0.858 & 0.020 & -0.011 & 0.020 & 0.005 & 0.020 & -0.004 & 0.019 &   &  &   &   &   &   \\
29.30 & 4.35 & 1.14 & 4.74 & 3.700 & 0.022 & 0.020 & 0.021 & 0.014 & 0.022 & 0.012 & 0.022 &   &  &   &   &   &   \\
30.60 & 0.44 & 24.33 & 0.63 & 1.289 & 0.023 & 0.384 & 0.022 & 0.054 & 0.023 & -0.001 & 0.023 &   &  & 30.00 & 1.78 & -2.81 & 1.68 \\
30.60 & 4.78 & -0.12 & 4.74 & 4.387 & 0.023 & 0.099 & 0.022 & 0.230 & 0.023 & 0.069 & 0.023 & 1.58 & 0.54 & 5.68 & 0.52 & 26.51 & 2.53 \\
31.04 & 0.44 & 13.47 & -33.12 & 1.932 & 0.022 & 0.024 & 0.020 & -0.466 & 0.021 & -0.051 & 0.021 & -2.65 & 1.10 & 24.15 & 1.11 & -50.33 & 1.26 \\
31.47 & 0.44 & -8.64 & -24.81 & 1.925 & 0.024 & -0.113 & 0.023 & 0.108 & 0.023 & 0.113 & 0.023 & 5.88 & 1.21 & 8.06 & 1.23 & 61.37 & 4.36 \\
31.91 & 0.44 & 14.49 & -31.41 & 0.787 & 0.022 & -0.002 & 0.021 & -0.100 & 0.022 & 0.020 & 0.022 &   &  & 12.37 & 2.94 & -52.32 & 6.37 \\
31.91 & 0.44 & 14.07 & -30.78 & 0.620 & 0.022 & 0.089 & 0.021 & -0.028 & 0.022 & 0.016 & 0.022 &   &  & 14.66 & 3.56 & -15.35 & 7.24 \\
32.34 & 0.44 & -3.87 & 4.89 & 0.568 & 0.031 & 0.005 & 0.029 & -2.8e-4 & 0.029 & 0.021 & 0.028 &   &  &   &   &   &   \\
32.34 & 0.44 & -13.68 & -15.30 & 0.752 & 0.031 & 0.027 & 0.029 & 0.013 & 0.029 & 4.8e-4 & 0.028 &   &  &   &   &   &   \\
32.78 & 10.43 & 20.58 & -14.55 & 22.149 & 0.061 & -1.233 & 0.050 & -0.333 & 0.048 & -0.723 & 0.050 & -3.26 & 0.23 & 5.76 & 0.23 & -89.24 & 1.08 \\
32.78 & 0.87 & 11.55 & -29.73 & 1.476 & 0.061 & 0.651 & 0.050 & -0.225 & 0.048 & -0.008 & 0.050 &   &  & 46.52 & 3.91 & -16.32 & 2.02 \\
33.64 & 0.44 & 11.07 & -28.35 & 1.835 & 0.034 & 0.742 & 0.032 & -0.154 & 0.032 & 0.007 & 0.033 &   &  & 41.25 & 1.89 & -12.65 & 1.20 \\
33.64 & 2.17 & -8.28 & -26.04 & 1.088 & 0.034 & 0.020 & 0.032 & 0.005 & 0.032 & 0.004 & 0.033 &   &  &   &   &   &   \\
35.38 & 0.44 & 16.02 & -15.54 & 0.722 & 0.021 & -0.009 & 0.019 & 0.005 & 0.019 & 0.012 & 0.019 &   &  &   &   &   &   \\
38.42 & 0.44 & -15.60 & -27.18 & 0.614 & 0.020 & 0.012 & 0.020 & 0.003 & 0.020 & 0.007 & 0.019 &   &  &   &   &   &   \\
39.29 & 3.04 & -9.18 & -19.23 & 5.232 & 0.033 & 0.328 & 0.030 & 0.019 & 0.031 & -0.008 & 0.029 &   &  & 6.24 & 0.58 & -5.15 & 2.76 \\

\hline
\label{tab:BR123D2-components}
\end{tabular}
\end{table*}

\begin{table*}
\caption[List of maser features detected in the v=2 J=1-0 SiO maser transition.]
{\footnotesize List of maser features detected in the v=2 J=1-0 SiO maser transition.
The column descriptions are given in the text.} 
\tiny
\begin{tabular}{llllllllllllllllllll}
\hline
$v$ & $\Delta v$ & $\Delta\alpha$ & $\Delta\delta$ & I & $\sigma_I$ & Q & $\sigma_Q$ & U & $\sigma_U$ & V & $\sigma_V$ &  $m_c$ & $\sigma_c$ & $m_l$ & $\sigma_l$ & $\chi$ & $\sigma_{\chi}$ \\
$[$km/s$]$ & $[$km/s$]$ & $[$mas$]$ & $[$mas$]$ & $[$Jy/b$]$ & $[$Jy/b$]$ & $[$Jy/b$]$ & $[$Jy/b$]$ & $[$Jy/b$]$ & $[$Jy/b$]$ & $[$Jy/b$]$ & $[$Jy/b$]$ & $[\%]$ & $[\%]$ & $[\%]$ & $[\%]$ & $[$deg$]$ & $[$deg$]$ \\
\hline  
6.18 & 0.44 & -16.27 & 7.02 & 3.187 & 0.018 & -0.002 & 0.018 & -0.010 & 0.018 & 0.139 & 0.018 & 4.36 & 0.56 &   &   &   &   \\
15.37 & 0.44 & 9.74 & 17.40 & 0.998 & 0.027 & 0.003 & 0.026 & -0.005 & 0.025 & 8.9e-5 & 0.025 &   &  &   &   &   &   \\
15.37 & 0.44 & 9.47 & 16.74 & 1.054 & 0.027 & 0.010 & 0.026 & -0.019 & 0.025 & 0.002 & 0.025 &   &  &   &   &   &   \\
15.81 & 0.88 & 16.85 & 17.34 & 1.475 & 0.030 & -0.029 & 0.030 & 4.0e-4 & 0.030 & 0.011 & 0.029 &   &  &   &   &   &   \\
16.25 & 0.44 & 8.72 & 16.68 & 1.229 & 0.036 & -0.002 & 0.036 & 0.002 & 0.037 & -0.010 & 0.036 &   &  &   &   &   &   \\
16.69 & 0.44 & 11.27 & 14.97 & 2.028 & 0.035 & 0.028 & 0.032 & -0.015 & 0.033 & 0.015 & 0.034 &   &  &   &   &   &   \\
17.12 & 0.88 & 7.97 & 15.72 & 1.690 & 0.069 & 0.016 & 0.067 & 0.001 & 0.060 & 0.018 & 0.065 &   &  &   &   &   &   \\
17.56 & 0.44 & 7.91 & 14.40 & 0.898 & 0.070 & -0.012 & 0.069 & -0.010 & 0.069 & -0.006 & 0.072 &   &  &   &   &   &   \\
18.00 & 2.63 & 14.03 & 14.01 & 3.338 & 0.039 & 0.068 & 0.039 & -0.013 & 0.040 & -0.012 & 0.038 &   &  &   &   &   &   \\
18.88 & 0.44 & 14.75 & 14.19 & 0.957 & 0.070 & 0.007 & 0.064 & -0.018 & 0.065 & 0.007 & 0.065 &   &  &   &   &   &   \\
18.88 & 0.44 & 13.01 & 12.66 & 1.424 & 0.070 & 0.008 & 0.064 & 0.008 & 0.065 & 0.019 & 0.065 &   &  &   &   &   &   \\
19.31 & 3.94 & 15.08 & 12.60 & 5.120 & 0.066 & -0.005 & 0.059 & -0.005 & 0.061 & 0.021 & 0.059 &   &  &   &   &   &   \\
19.31 & 0.44 & 11.09 & 11.85 & 1.302 & 0.066 & 0.006 & 0.059 & -0.026 & 0.061 & -0.007 & 0.059 &   &  &   &   &   &   \\
19.31 & 1.31 & 9.62 & 11.82 & 0.958 & 0.066 & -0.006 & 0.059 & 0.005 & 0.061 & 0.005 & 0.059 &   &  &   &   &   &   \\
19.75 & 0.44 & 17.48 & -22.77 & 1.377 & 0.082 & -0.036 & 0.071 & 0.018 & 0.070 & -0.015 & 0.073 &   &  &   &   &   &   \\
20.63 & 2.19 & 19.43 & -23.01 & 6.353 & 0.138 & 0.007 & 0.127 & -0.006 & 0.123 & 0.104 & 0.125 &   &  &   &   &   &   \\
20.63 & 2.19 & 11.48 & 10.62 & 4.847 & 0.138 & -0.013 & 0.127 & -0.013 & 0.123 & 0.017 & 0.125 &   &  &   &   &   &   \\
21.06 & 1.75 & 21.35 & -24.66 & 12.669 & 0.079 & -0.207 & 0.076 & -0.139 & 0.070 & -0.004 & 0.073 &   &  & 1.88 & 0.61 & -163.15 & 9.08 \\
21.06 & 1.75 & 12.68 & 10.17 & 6.566 & 0.079 & 0.006 & 0.076 & -0.030 & 0.070 & 0.293 & 0.073 & 4.46 & 1.12 &   &   &   &   \\
21.06 & 0.88 & 8.36 & 10.50 & 3.047 & 0.079 & -0.084 & 0.076 & 0.003 & 0.070 & 0.059 & 0.073 &   &  &   &   &   &   \\
21.06 & 0.44 & 4.64 & 11.94 & 1.250 & 0.079 & 0.004 & 0.076 & -0.031 & 0.070 & -0.006 & 0.073 &   &  &   &   &   &   \\
21.50 & 0.88 & 21.92 & -23.91 & 6.219 & 0.102 & 0.101 & 0.090 & 0.028 & 0.091 & 0.009 & 0.093 &   &  &   &   &   &   \\
21.50 & 0.88 & 13.25 & 9.30 & 6.429 & 0.102 & -0.013 & 0.090 & -0.016 & 0.091 & 0.003 & 0.093 &   &  &   &   &   &   \\
21.94 & 0.44 & 15.89 & 10.92 & 2.467 & 0.057 & 0.006 & 0.054 & -0.006 & 0.052 & 0.017 & 0.052 &   &  &   &   &   &   \\
22.38 & 0.44 & 14.03 & 8.97 & 2.420 & 0.041 & -0.004 & 0.040 & -0.005 & 0.039 & -0.011 & 0.040 &   &  &   &   &   &   \\
22.81 & 0.44 & 7.73 & 7.62 & 0.877 & 0.039 & -0.020 & 0.036 & -0.002 & 0.035 & 0.005 & 0.037 &   &  &   &   &   &   \\
23.25 & 1.75 & 14.45 & 7.08 & 8.069 & 0.054 & -0.019 & 0.048 & -0.048 & 0.047 & 0.246 & 0.048 & 3.05 & 0.59 &   &   &   &   \\
23.69 & 3.50 & 1.46 & 9.33 & 6.535 & 0.046 & -0.014 & 0.037 & -0.004 & 0.035 & -0.144 & 0.039 & -2.20 & 0.60 &   &   &   &   \\
24.13 & 0.44 & 1.67 & 21.12 & 3.879 & 0.044 & -0.013 & 0.040 & 0.042 & 0.038 & -0.065 & 0.039 &   &  &   &   &   &   \\
24.13 & 0.44 & -4.78 & 17.70 & 1.410 & 0.044 & 0.013 & 0.040 & 0.022 & 0.038 & 0.028 & 0.039 &   &  &   &   &   &   \\
24.56 & 0.88 & 12.95 & 5.31 & 2.278 & 0.034 & 0.006 & 0.033 & -0.020 & 0.032 & -0.017 & 0.031 &   &  &   &   &   &   \\
24.56 & 0.44 & 1.76 & 17.94 & 1.525 & 0.034 & 0.015 & 0.033 & -0.036 & 0.032 & -0.016 & 0.031 &   &  &   &   &   &   \\
24.56 & 0.44 & -0.97 & 8.94 & 1.338 & 0.034 & 0.003 & 0.033 & -0.015 & 0.032 & -0.095 & 0.031 & -7.13 & 2.36 &   &   &   &   \\
25.00 & 0.44 & -1.66 & 22.17 & 1.752 & 0.029 & -0.007 & 0.029 & -0.012 & 0.029 & 0.004 & 0.030 &   &  &   &   &   &   \\
25.00 & 0.88 & -5.95 & 16.92 & 2.346 & 0.029 & -0.005 & 0.029 & 0.017 & 0.029 & -0.004 & 0.030 &   &  &   &   &   &   \\
25.44 & 0.88 & -3.85 & 19.38 & 1.448 & 0.030 & -0.017 & 0.028 & -0.005 & 0.027 & 0.014 & 0.027 &   &  &   &   &   &   \\
25.44 & 0.44 & -6.52 & 14.70 & 2.593 & 0.030 & -0.027 & 0.028 & -0.004 & 0.027 & -0.004 & 0.027 &   &  &   &   &   &   \\
25.88 & 0.88 & -2.08 & 11.67 & 1.951 & 0.032 & -0.024 & 0.030 & -0.001 & 0.031 & -0.005 & 0.031 &   &  &   &   &   &   \\
28.50 & 1.75 & 1.10 & 4.95 & 3.739 & 0.022 & -0.026 & 0.021 & 0.039 & 0.021 & -0.003 & 0.021 &   &  &   &   &   &   \\
29.82 & 0.88 & -0.13 & 5.10 & 1.147 & 0.018 & 0.012 & 0.018 & 0.011 & 0.018 & 0.020 & 0.018 &   &  &   &   &   &   \\
32.00 & 0.44 & -0.28 & 3.33 & 1.053 & 0.017 & -0.014 & 0.017 & -0.003 & 0.017 & -0.023 & 0.017 &   &  &   &   &   &   \\
32.44 & 0.88 & 20.42 & -14.49 & 3.294 & 0.019 & 0.471 & 0.019 & -0.003 & 0.019 & -0.189 & 0.020 & -5.75 & 0.60 & 14.28 & 0.60 & -90.20 & 1.18 \\
34.63 & 0.44 & -1.69 & 2.16 & 0.976 & 0.016 & 0.003 & 0.015 & -0.009 & 0.015 & -0.004 & 0.015 &   &  &   &   &   &   \\
39.01 & 2.19 & 12.62 & -13.26 & 12.563 & 0.028 & 0.873 & 0.028 & -0.166 & 0.028 & 0.334 & 0.027 & 2.66 & 0.22 & 7.07 & 0.22 & -95.43 & 0.90 \\

\hline
\label{tab:BR123D1-components}
\end{tabular}
\end{table*}


\end{document}